\documentclass[10pt]{article}
\pdfoutput=1
\usepackage[utf8]{inputenc}

\usepackage{bbold}
\usepackage{amssymb}
\usepackage{amsmath}
\usepackage[dvips]{graphicx}
\usepackage{setspace}
\usepackage{slashed}
\usepackage{mathtools}
\usepackage{amsfonts}
\usepackage{fancyhdr}
\usepackage{dsfont}
\usepackage[dvipsnames]{xcolor}
\usepackage{graphicx}
\usepackage{rotating}
\usepackage{comment}
\usepackage{color} 
\usepackage{subcaption}
\usepackage[percent]{overpic}
\usepackage{cite}
\usepackage{braket}
\usepackage{standalone}
\definecolor{darkgreen}{rgb}{0,0.5,0}
\definecolor{darkblue}{rgb}{0,0,0.6}
\definecolor{purple}{rgb}{0.4,.2,0.7}

\newcommand{\h}{\theta}

\newcommand{\be}{\begin{equation}}
\newcommand{\ee}{\end{equation}}

\usepackage[colorlinks=true,citecolor=darkgreen,linkcolor=black,urlcolor=purple]{hyperref}

\usepackage{pdfsync}

\makeatletter
\newcommand*{\defeq}{\mathrel{\rlap{%
                     \raisebox{0.3ex}{$\m@th\cdot$}}%
                     \raisebox{-0.3ex}{$\m@th\cdot$}}%
                     =} 
\makeatother

\DeclareMathOperator{\Tr}{Tr}
\def\be{\begin{eqnarray}}
\def\ee{\end{eqnarray}}

\newcommand{\tr}{\textrm{Tr}\,}

\newcommand{\bea}{\begin{eqnarray}}
\newcommand{\eea}{\end{eqnarray}}
\def\ben{\begin{equation}}
\def\een{\end{equation}}

 \let\h=\eta  
     \let\r=v

\let\w=\omega

\def\be{\begin{equation}}
\def\ee{\end{equation}}
\def\ba{\begin{eqnarray}}
\def\ea{\end{eqnarray}}

\def\bal#1\eal{\begin{align}#1\end{align}}
\def\bs#1\es{\begin{split}#1\end{split}}

\def\mygreen{green!60!brown}
\def\mypurple{magenta!60!black}
\def\myred{red!70}
\def\myblue{blue!70!white!80!green}
\def\myorange{orange!70!brown}

\def\AColor{blue!70!white!80!green}
\def\BColor{red!70}
\def\CColor{green!60!brown}
\def\aColor{blue!65!red!80!magenta!90}
\def\bColor{magenta!30!red!75!blue}
\def\cColor{orange!90!brown!80}
\def\minSurfaceColor{VioletRed!80!WildStrawberry}

\allowdisplaybreaks  

\numberwithin{equation}{section}

\def\bw{\bar{W}}

\def\w{{\omega}}

\def\hp{\hat{\phi}}

\def\ch{{\chi}}

\def\be{\begin{equation}}
\def\ee{\end{equation}}
\def\ba{\begin{eqnarray}}
\def\ea{\end{eqnarray}}
\def\bal#1\eal{\begin{align}#1\end{align}}

\def\r{\rightarrow}

\def\r{\right}

\def\w{\bar{w}}

\usepackage{tikz}
\usetikzlibrary{positioning,arrows}
\usetikzlibrary{decorations.pathmorphing}
\usetikzlibrary{decorations.markings}
\tikzset{
particle/.style={postaction={decorate}},
graviton/.style={decorate, decoration={snake, amplitude=0.8 mm, segment length=1.5 mm, pre length=0.8 mm, post length=0.8 mm}},
photon/.style={
        decoration={complete sines, amplitude=0.15cm, segment length=0.2cm},
        decorate    
    },
gluon/.style={
        decoration={coil, aspect=0.75, mirror, segment length=1.5mm},
        decorate
    }
}
 
\def \be {\begin{equation}}
\def \ee {\end{equation}}

\renewcommand{\min}{{\rm min}}
\newcommand{\bh}{\bar{h}}

\usepackage{framed}

\begin{document}
\onehalfspacing

\begin{center}

~
\vskip5mm

{\LARGE  {
Coarse-Grained Fixed-Point Tensor Networks and Holographic Reflected Entropy in 3D Gravity
\\
\ \\
}}

\vskip 5mm

Ning Bao${}^{1,2}$, Jinwei Chu${}^{3}$, Yikun Jiang${}^{1}$ and Jacob March${}^{1}$

\vskip5mm

\it{${}^1$ Department of Physics, Northeastern University, Boston, MA 02115, USA.}\\
\it{${}^2$ Computational Sciences Initiative, Brookhaven National Laboratory, Upton, NY 11973, USA.}\\
\it{${}^3$ Department of Physics, University of Chicago, Chicago, IL 60637, USA}

\vskip5mm
{\tt ningbao75@gmail.com, 
jinweichu@uchicago.edu, phys.yk.jiang@gmail.com, 
march.j@northeastern.edu}

\vskip5mm

\end{center}

\vspace{4mm}

\begin{abstract}
\noindent

We use the framework of \textit{fixed-point BCFT tensor networks} to present a microscopic CFT derivation of the correspondence between reflected entropy ($\text{RE}$) and entanglement wedge cross section ($\text{EW}$) in AdS$_3$/CFT$_2$, for both bipartite and multipartite settings. These fixed-point tensor networks, obtained by triangulating Euclidean CFT path integrals, allow us to explicitly construct the canonical purification via cutting-and-gluing CFT path integrals. Employing modular flow in the large-$c$ limit, we demonstrate that these intrinsic CFT manipulations reproduce bulk geometric prescriptions, without assuming the AdS/CFT dictionary. The emergence of bulk geometry is traced to coarse-graining over heavy states in the large-$c$ limit. Universal coarse-grained BCFT data for compact 2D CFTs, through the relation to Liouville theory with ZZ boundary conditions, yields hyperbolic geometry on the Cauchy slice. The corresponding averaged replica partition functions reproduce all candidate $\text{EW}$s, arising from different averaging patterns, with the dominant one providing the correct $\text{RE}$ and $\text{EW}$. In this way, many heuristic tensor-network intuitions in toy models are made precise and established directly from intrinsic CFT data.

 \end{abstract}

\pagebreak
\pagestyle{plain}

\setcounter{tocdepth}{3}
{}
\vfill
\tableofcontents

\newpage

\date{}

\section{Introduction}\label{sec:intro}

One of the most important insights in the study of the AdS/CFT correspondence is the idea that quantum entanglement plays a crucial role in building up the holographic dual spacetime \cite{VanRaamsdonk:2010pw}. The most famous relation capturing this idea is the Ryu–Takayanagi (RT) formula \cite{Ryu:2006bv, Ryu:2006ef}, which states that the von Neumann entropy $S(A)$ of the reduced density matrix $\rho_A$ for a subregion $A$ in the CFT is, at leading order in $G_N$, proportional to the smallest area among all minimal surfaces $\gamma_A$ in the bulk dual that are homologous to $A$:
\be
\label{RT}
S(A)\equiv -\tr(\rho_A \ln \rho_A)=\min_{\gamma_A} \frac{\text{Area}(\gamma_A)}{4G_N}\ .
\ee
Given a pure state $|\psi\rangle_{AB}$ on two complementary subsystems $A$ and $B$, the von Neumann entropy of either subsystem defines the entanglement entropy
\be
\label{EE}
{\rm EE}(A:B)_{|\psi\rangle_{AB}}\equiv S(A)=S(B)\ .
\ee

Assuming the holographic dictionary, the RT formula (\ref{RT}) was proven in \cite{Lewkowycz:2013nqa} by mapping the Euclidean path integral, constructed by the CFT replica trick \cite{Callan:1994py, Calabrese:2004eu} for computing $S(A)$, to an on-shell gravitational action. This result demystified part of the connection between bulk geometric areas and quantum entanglement, but it still leaves unanswered the deeper question of how macroscopic bulk geometries emerge from the microscopic algebraic degrees of freedom of the CFT, since the derivation assumes the holographic dictionary from the outset. 

Progress in this direction has been motivated by the observation that the RT formula shares many similarities with tensor network representations of quantum states in condensed matter systems \cite{Vidal:2007hda, Swingle:2009bg}. A variety of tensor network toy models have been proposed to capture key aspects of the holographic mapping between boundary and bulk, including RT-like formulas for the von Neumann entropy, most notably in \cite{Pastawski:2015qua, Hayden:2016cfa}. However, it remains unclear how these models, on the one hand, give rise to \textit{geometric areas} in the bulk that appear in RT-like formulas—often interpreted as counting cuts—and capture the exact entanglement phase structures, and, on the other hand, connect directly to the \textit{microscopic algebraic data} of the boundary CFT.

On a different note, entanglement entropy (\ref{EE}) is certainly not the only quantum information quantity of interest for CFT states. In particular, it is important to characterize entanglement properties that are intrinsic to mixed states \cite{Takayanagi:2017knl, Nguyen:2017yqw}. On the bulk side, one may also look for dual geometric objects beyond boundary-homologous minimal surfaces. One such dual pair has be proposed for the reflected entropy (RE) \cite{Dutta:2019gen},\footnote{See also e.g., \cite{Jeong:2019xdr,Bao:2019zqc,Kusuki:2019evw, BabaeiVelni:2019pkw, Chu:2019etd,Akers:2019gcv,Chandrasekaran:2020qtn,Li:2020ceg,Berthiere:2020ihq,Bueno:2020fle,Hayden:2021gno,Li:2021dmf,Ling:2021vxe,Akers:2021pvd, Kudler-Flam:2020url, Wen:2021qgx, Basu:2023jtf, Basu:2022nds,Chen:2022fte,Harper:2019lff,Chen:2021lnq,Hayden:2023yij} for recent developments.} defined via the canonical purification of mixed states.

\begin{figure}
    \centering
\includegraphics[width=1\linewidth]{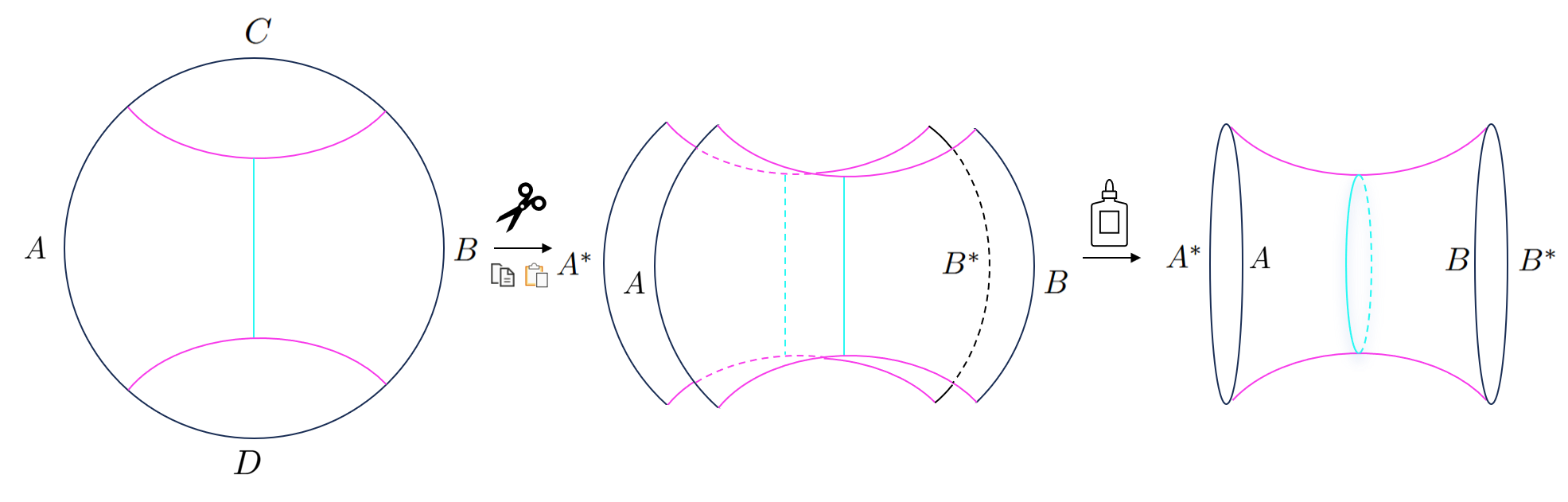}
    \caption{\small{An illustration of $\text{RE}=\text{EW}$: The entanglement wedge cross section $\text{EW}(A:B)$ is represented by the cyan line in the left panel. As will be reviewed in Sec~\ref{purificationreview}, from the AdS bulk point of view the canonical purification is obtained by cutting the manifold open along the pink RT surfaces and gluing it to a CPT-conjugate copy. The entanglement entropy of the resulting pure state, which defines the reflected entropy $\text{RE}(A:B)$, is dual to the RT surface highlighted in cyan. Clearly, it is two times the size of $\text{EW}(A:B)$.}}
    \label{crosssection}
\end{figure}

Similar to the proof of the RT formula in \cite{Lewkowycz:2013nqa}, by assuming the holographic dictionary, it was shown that, at leading order in $G_N$, this quantity is proportional to the area of the entanglement wedge cross section (EW)~\cite{Takayanagi:2017knl, Nguyen:2017yqw}. Specifically,
\be \label{RE=2EW}
\text{RE}(A:B)=\frac{\text{2EW}(A:B)}{4G_N}\ .
\ee
The detailed definitions will be reviewed in Sec.~\ref{purificationreview}. An illustration of (\ref{RE=2EW}) is sketched in Fig.~\ref{crosssection}, where the mixed state is associated to two disconnected subregions $A$ and $B$ on the CFT boundary. For brevity, we will denote the relation (\ref{RE=2EW}) as ``${\rm RE}={\rm EW}$''. This correspondence has also been supported by tensor network toy model constructions \cite{Akers:2021pvd, Akers:2022zxr}, similar to the RT-like formula for entanglement entropy \cite{Pastawski:2015qua, Hayden:2016cfa}.

\begin{figure}
    \centering
\includegraphics[width=0.4\linewidth]{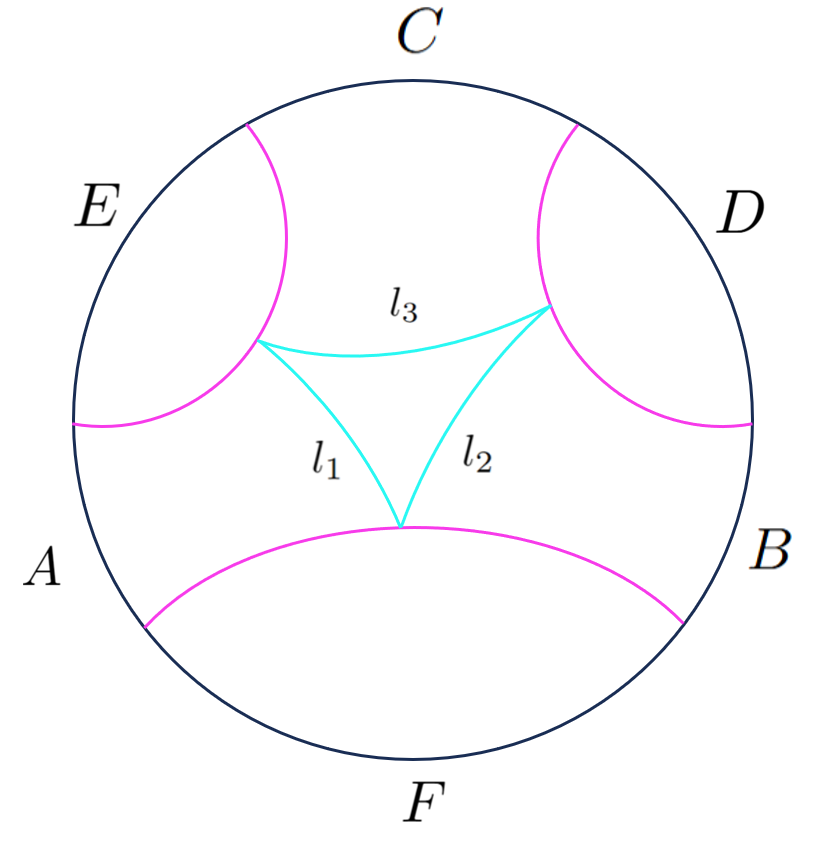}
    \caption{\small{The tripartite entanglement wedge cross section $\text{EW}_3(A:B:C)$ is defined by minimizing $l_{1}+l_{2}+l_{3}$, where the geodesics $l_i$'s are anchored on the RT surface of $ABC$ and form a closed cycle.}}
    \label{EW}
\end{figure}

This dual pair unveils new aspects of the holographic correspondence, particularly for mixed states. However, bipartite entanglement is not the most general structure for mixed-state correlations. Motivated by this, multipartite extensions of the reflected entropy were proposed in \cite{Bao:2019zqc, Chu:2019etd} and shown to be dual to the multipartite entanglement wedge cross section introduced in \cite{Umemoto:2018jpc}, see Fig.~\ref{EW} for a representative example. These developments open the door to a broader family of quantum information quantities with holographic duals.

Based on the above considerations, it is natural to ask: is there a microscopic CFT derivation of the correspondence between (multipartite) reflected entropies and entanglement wedge cross sections, without assuming the holographic dictionary? Can we see directly how such a CFT derivation is related to the holographic one? In particular, can this correspondence be realized through tensor networks constructed directly from microscopic algebraic CFT data, producing genuine geometric areas in emergent hyperbolic space and capturing the full entanglement phase structure?

\begin{figure}
    \centering
\includegraphics[width=0.4\linewidth]{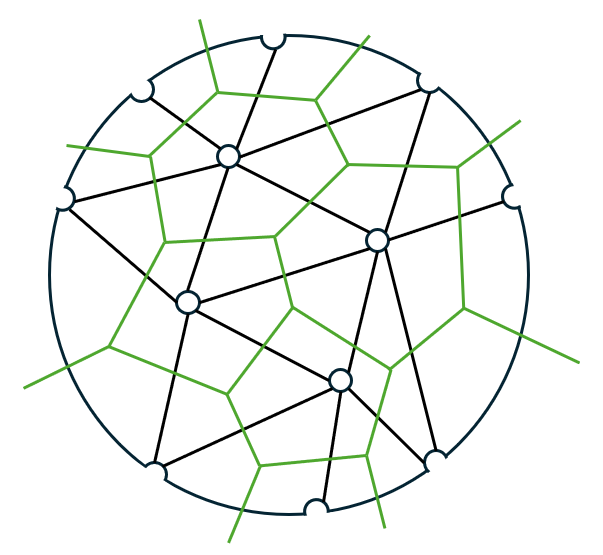}
    \caption{\small{The general BCFT tensor network representation of a CFT quantum state prepared by a Euclidean path integral is obtained by introducing tiny holes as regulators on the state-preparation manifold and performing the OPE block decomposition; the resulting network is defined on the dual graph, indicated by the green lines.}}
    \label{TNtriangulation}
\end{figure}

For the RT formula (\ref{RT}), recent progress in \cite{Bao:2025plr, Geng:2025efs} has provided an answer in the context of the AdS$_3$/CFT$_2$ correspondence. In particular, it was shown that 2D CFT states prepared by a Euclidean path integral can be represented as fixed-point tensor networks obtained by \textit{triangulating} the 2D CFT state-preparation manifolds. These tensor networks are constructed by introducing tiny CFT boundaries as regulators, which naturally lead to boundary conformal field theory (BCFT) building blocks. The dual graph of the triangulation encodes the BCFT conformal block decomposition, and in doing so naturally provides a tensor network representation of the state, as illustrated by the green lines in Fig.~\ref{TNtriangulation}. In this way, earlier constructions of exact fixed-point spacetime tensor networks, namely state sum representations for 2D CFT partition functions and correlation functions\cite{Chen:2022wvy, Cheng:2023kxh, Chen:2024unp, Hung:2024gma, Bao:2024ixc, Brehm:2021wev, Brehm:2024zun} are extended to the level of \textit{quantum states prepared by Euclidean path integrals}. We refer to such exact discrete representation of CFT path integrals as \textit{BCFT tensor networks}. The tensors in these networks are built entirely from intrinsic CFT data—namely, BCFT operator product expansion (OPE) coefficients and conformal blocks—allowing exact computation of entanglement entropy from these building blocks. Importantly, the results are independent of the choice of triangulation, since the building blocks satisfy all the intricate (B)CFT self-consistency relations \cite{CARDY1991274, Lewellen:1991tb, Geng:2025efs, Bao:2024ixc}.

However, to establish the holographic correspondence with bulk geometric objects, it is not necessary to retain all the microscopic details of the CFT. The asymptotic CFT data for heavy states with conformal dimension $h \gg c$ exhibits universality enforced by conformal bootstrap constraints \cite{Cardy:1986ie, Collier:2019weq, Numasawa:2022cni}. This universality manifests in the density of states through the Cardy formula \cite{Cardy:1986ie}, and for OPE coefficients—matrix elements between heavy eigenstates—it has been demonstrated in both CFT and BCFT \cite{Cardy:2017qhl, Collier:2019weq, Numasawa:2022cni}. Guided by this universality, the OPE coefficients are modeled as random tensors encoding the statistics of coarse-grained heavy CFT data, which amounts to a direct generalization of the Eigenstate Thermalization Hypothesis (ETH) \cite{Lashkari:2016vgj, Collier:2019weq, Belin:2020hea}. 

In the large-$c$ limit, these universal behaviors are proposed to be further enhanced, extending all the way down to the states at the black hole threshold $h=(c-1)/24$ \cite{Hartman:2014oaa, Chandra:2022bqq, Dey:2024nje}, thereby capturing the universal coarse-grained behavior of chaotic 3D black hole microstates. The resulting \textit{coarse-grained tensor networks} are therefore \textit{BCFT random tensor networks}.

This coarse-graining over the heavy state CFT data in the large-$c$ limit provides the bridge to macroscopic emergent geometries. In particular, semiclassical spacetime wormhole solutions \cite{Maldacena:2004rf} in 3D gravity\footnote{Precursors of this result appeared in \cite{Saad:2019pqd, Stanford:2020wkf}, in the context of 2D JT gravity.} have been shown to emerge from coarse-graining the CFT data using an ETH-type ansatz \cite{Cotler:2020ugk, Belin:2020hea, Chandra:2022bqq}, see also recent developments in \cite{Belin:2023efa, Jafferis:2025vyp, DiUbaldo:2023qli, deBoer:2024mqg, Belin:2021ryy, deBoer:2023vsm, Sasieta:2022ksu, Chandra:2025fef, Collier:2024mgv, deBoer:2025rct, Boruch:2025ilr, Yan:2023rjh, Yan:2025usw, Anous:2021caj}. Extending this idea, a similar coarse-graining procedure \cite{Hung:2025vgs, Wang:2025bcx, Jafferis:2025yxt} within BCFT tensor networks demonstrates that the mechanism is not restricted to spacetime wormhole geometries. In fact, 3D hyperbolic geometries with \textit{a single asymptotic boundary} naturally arise through coarse-graining, both in the CFT setting \cite{Chua:2023ios, Bao:2025plr} and within BCFT tensor networks \cite{Geng:2025efs}. When applied to replica partition functions, this provides precisely the ingredients that yield the entanglement entropy as the areas of minimal surfaces predicted by the RT formula \cite{Bao:2025plr, Geng:2025efs}. Remarkably, the phase transitions in the multi-interval RT formula—a long-standing challenge for tensor network toy models of holography—also appear naturally in this framework, arising from distinct Gaussian averaging patterns in the statistics of OPE coefficients. 

For cases with a single asymptotic boundary, the averaging procedure yields Liouville theory with ZZ boundary conditions \cite{Zamolodchikov:2001ah, Chua:2023ios, Bao:2025plr, Geng:2025efs, Chandra:2023dgq},\footnote{It should be emphasized that Liouville theory is \textit{not} itself a holographic CFT. Rather, it arises as the universal outcome in the coarse-grained norm computation of compact holographic CFTs, which in turn furnishes the bridge to geometry and 3D gravity.} which describes quantization of the geometry on the bulk constant-time Cauchy slice. This relates BCFT tensor networks to conventional tensor networks associated with bulk Cauchy slices \cite{Swingle:2009bg, Pastawski:2015qua, Hayden:2016cfa}, through a mechanism reminiscent of the path-integral optimization interpretation of holographic tensor networks \cite{Caputa:2017urj, Caputa:2017yrh, VanRaamsdonk:2018zws}. This paradigm, which extracts emergent local geometry through CFT path-integral discretization combined with quantum-chaos–based averaging, is referred to as ``\textit{It from ETH}” in \cite{Geng:2025efs}.

In this paper, we extend this framework beyond the derivation of the RT formula \cite{Geng:2025efs} and establish ${\rm RE}={\rm EW}$ for the vacuum state in AdS$_3$/CFT$_2$ directly from BCFT tensor networks. Once triangulated into BCFT building blocks, the cutting and gluing of Euclidean path integrals can be implemented exactly. As anticipated in \cite{Marolf:2019zoo} from the consideration of “fixed-area states” \cite{Akers:2018fow, Dong:2018seb}, the boundary dual of AdS cut-and-paste operations is realized through the sewing of CFT path integrals. We make this proposal exact by formulating it entirely in terms of intrinsic BCFT building blocks, thereby elevating intuitions from tensor-network toy models into precise statements about 2D CFTs. 

In the context of reflected entropy, our central observation is that the $m\to 1$ analytic continuation proposed in \cite{Dutta:2019gen} can be carried out explicitly in the CFT, allowing us to construct the canonical purification within the BCFT tensor network framework. Key elements in this construction are the correspondence between the single-interval modular Hamiltonian and the BCFT Hamiltonian \cite{Ohmori:2014eia, Cardy:2016fqc}, together with the factorization of reduced density matrices at large $c$.

Finally, as we will review, the reflected entropy is defined as the entanglement entropy of the canonical purification. Hence, having constructed its path-integral and BCFT tensor network representation, this directly reduces to the RT formula derivation in \cite{Bao:2025plr, Geng:2025efs}. Coarse-graining over heavy states subsequently induces an averaging over OPE coefficients in the norm and replica partition functions, which gives rise to emergent hyperbolic bulk geometries and also establishes the correspondence with entanglement wedge cross sections. Compared with the twist-operator derivation in AdS$_3$/CFT$_2$ \cite{Dutta:2019gen, Kusuki:2019evw}, the BCFT tensor network connects the computation more directly to holographic reasoning and tensor-network intuitions, while using intrinsic CFT data and furnishing a mechanism for direct geometric emergence.

This paper is organized as follows. In Sec.~\ref{sec2}, we review the canonical purification and reflected entropy, and the holographic argument relating them to the entanglement wedge cross section. We also explain how the CFT state-preparation manifold relates to bulk Cauchy slices via hyperbolic slicing and coarse-graining. In Sec.~\ref{sec3}, we employ the BCFT tensor network framework to derive the $\text{RE}=\text{EW}$ relation. In particular, we construct the canonical purification using CFT cutting and gluing, and the modular Hamiltonian in the large-$c$ limit. The reflected entropy is then computed by coarse-graining BCFT data in the replica trick, yielding areas of entanglement wedge cross sections. The reflected entropy phase transitions are shown to arise from different averaging patterns. We present three representative examples before generalizing the result to arbitrary cases. In Sec.~\ref{multipartitesection}, we extend the framework to multipartite settings. We conclude in Sec.~\ref{conclusionsec} with a summary and future directions.

\section{Preliminaries} \label{sec2}

In this section, we set up the preliminaries for deriving the $\text{RE}=\text{EW}$ correspondence (\ref{RE=2EW}) from BCFT tensor networks. In Sec.~\ref{purificationreview}, we review the definition of the canonical purification and reflected entropy, and then summarize the argument that assumes the holographic dictionary to establish its correspondence with the entanglement wedge cross section. In Sec.~\ref{prep=cauchy}, we explain how, for a large class of quantum states, the boundary CFT state-preparation manifold is related to the bulk Cauchy slice via hyperbolic slicing. We further outline how this geometrical relation can emerge from coarse-graining the algebraic (B)CFT data, thereby connecting BCFT tensor networks to tensor networks defined on Cauchy slices.

\subsection{Reflected Entropy and Its Holographic Dual} \label{purificationreview}
We begin by reviewing the reflected entropy and its connection to the entanglement wedge cross section \cite{Dutta:2019gen}.

First, for a density matrix $\rho_{AB}$ defined on the union system $A \cup B$, its canonical purification is a state $\ket{\sqrt{\rho_{AB}}}$ in a doubled Hilbert space $\left(\mathcal{H}_A \otimes \mathcal{H}_B\right) \otimes \left(\mathcal{H}_{A^*} \otimes \mathcal{H}_{B^*}\right)$, where the auxiliary systems $A^*$ and $B^*$ are copies of $A$ and $B$, respectively. That is,
\be
\label{cp}
\ket{\sqrt{\rho_{AB}}} \in \left( \mathcal{H}_{A} \otimes \mathcal{H}_{A^*}\right) \otimes \left(\mathcal{H}_{B} \otimes \mathcal{H}_{B^*} \right)=\mathcal{H}_{AA^*BB^*}
\ee
satisfies the purification condition:
\be
\tr_{A^* B^*} \ket{\sqrt{\rho_{AB}}} \bra{\sqrt{\rho_{AB}}} =\rho_{AB}~.
\ee
The square root indicates that for the density matrix $\rho_{AB}$ with eigenvalues $p_i$ and eigenvectors $|p_i\rangle_{AB}$, the canonical purification can be written explicitly as
\be
\ket{\sqrt{\rho_{AB}}}=\sum_i\sqrt{p_i}|p_i\rangle_{AB}|p_i\rangle_{A^*B^*}\ .
\ee
The simplest example was illustrated in the familiar setup of the thermofield double state in \cite{Dutta:2019gen}.

The reflected entropy is then defined as the entanglement entropy of the canonical purification $\ket{\sqrt{\rho_{AB}}}$~\cite{Dutta:2019gen}:
\be \label{reflected def}
\text{RE}(A:B)\equiv \text{EE}(AA^*:BB^*)_{\ket{\sqrt{\rho_{AB}}}}\ .
\ee
In holographic CFTs, this quantity has attracted significant attention due to its proposed duality with a geometric quantity in the bulk: the entanglement wedge cross section $\text{EW}(A:B)$. The latter is defined as the area of the minimal surface within the entanglement wedge of $AB$ that bifurcates it into two parts homologous to subsystems $A$ and $B$, respectively. See, for example, the left panel of Fig.~\ref{crosssection}, where $\text{EW}(A:B)$ is represented by the cyan line. At leading order in $G_N$, the relation between the reflected entropy and the entanglement wedge cross section is given by \eqref{RE=2EW}, a formula whose structure directly mimics the RT formula \eqref{RT}.

\begin{figure}

    \includegraphics[width=\linewidth]{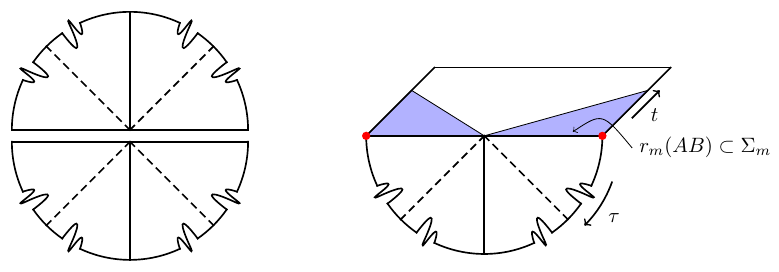}
   
    \caption{\small{Left: Euclidean path integral for computing $\tr \rho^4_{AB}$ in the boundary theory and its dual bulk saddle. Each wedge (bounded by solid lines) represents a path integral that prepares $\rho_{AB}$. Right: slicing open the bulk path integral for $\tr \rho_{AB}^4$ produces the analogue of the Hartle–Hawking state $\ket{\rho_{AB}^2}$ in the bulk, with the dual CFT state defined on the red dots.}}
    \label{bulkreplica}
\end{figure}

We now briefly review the holographic derivation of \eqref{RE=2EW} presented in \cite{Dutta:2019gen}, which translates the replica trick computation in CFT into a gravitational path integral, and follows a line of reasoning similar to that used in the proof of the RT formula \cite{Lewkowycz:2013nqa}. Specifically, the holographic dictionary allows us to compute the $m$-th replica partition function $Z_m = \mathrm{Tr}(\rho_{AB}^m)$ by filling in smooth gravitational solutions with boundary conditions determined by the dual CFT. The dominant contributions are assumed to preserve the $\mathbb{Z}_m$ replica symmetry \cite{Lewkowycz:2013nqa}, as illustrated in the left panel of Fig.~\ref{bulkreplica}.

For the case where $m$ is even, one can slice open the gravitational dual solution in the bulk along the $\mathbb{Z}_2$ Euclidean time-reflection symmetric Cauchy slice, denoted by $\Sigma_m$, and get an analogue of the Hartle–Hawking state, denoted $\ket{\Psi_m}$, associated to the $m$-replicated geometry, such that\footnote{We adopt a different convention from \cite{Dutta:2019gen}, in which the Hartle–Hawking state is unnormalized.} 
\be
Z_m=\bra{\Psi_m}\Psi_m \rangle\ .
\ee
From the boundary CFT point of view, $|\Psi_m \rangle$ is given by the path integral on $m/2$ replica. By definition, this is the state $\ket{\rho_{AB}^{m/2}}$. For $m=4$, this is illustrated in Fig.~\ref{bulkreplica}: the Hartle–Hawking state in gravity is defined on the time-reflection symmetric Cauchy slice $\Sigma_m$ at $\tau=0$. The state in the dual CFT lives on the boundary of $\Sigma_m$, and contains one copy of $\mathcal{H}_{AB}$ on the left end and another on the right, indicated by the red dots. The Lorentzian section for the time evolution of this initial condition at $\tau=t=0$ is indicated in blue.

One can then apply the Ryu–Takayanagi prescription (\ref{RT}) to the subsystem $AA^*$ of the state $\ket{\rho_{AB}^{m/2}}$ to compute its von Neumann entropy. This is given by the area of the homologous minimal surface in the bulk region $\Sigma_m=r_m(AB)\cup r_m(A^*B^*)$, divided by $4G_N$. Here, $r_m(AB)$ denotes the $m$-th generalization of the entanglement wedge $r_m(AB)$—that is, the bulk region bounded by the boundary region $AB$ and the bulk $\mathbb{Z}_m$ replica-symmetric fixed-point surface.
 
We are interested in the $m \to 1$ limit, where the state $|\Psi_m\rangle=\ket{\rho_{AB}^{m/2}}$ reduces to the canonical purification $\ket{\rho_{AB}^{1/2}}$. To analytically continue $m$ away from even integers, one can take the $\mathbb{Z}_m$ quotient of the gravitational replica solution \cite{Lewkowycz:2013nqa}. Subsequently, as $m$ goes to 1, the $\mathbb{Z}_m$ fixed-point locus reduces to the RT surface of $AB$, and $r_m(AB)$ becomes the entanglement wedge. This implies that the Cauchy slice $\Sigma_1$ consists of two copies of the entanglement wedge glued together along this RT surface \cite{Dutta:2019gen, Faulkner:2018faa, Engelhardt:2017aux, Engelhardt:2018kcs} as in Fig.~\ref{crosssection}. 

Furthermore, on the slice $\Sigma_1$, the minimal surface homologous to $AA^*$—such as the cyan circle in Fig.~\ref{crosssection}—is composed of two copies of the entanglement wedge cross section $\text{EW}(A:B)$. Therefore, combining the RT prescription (\ref{RT}) and the definition of reflected entropy (\ref{reflected def}) leads to the relation (\ref{RE=2EW}) stated earlier.

\subsection{CFT State-Preparation Manifold/Bulk Cauchy Slice Correspondence: Hyperbolic Slicing and Averaging}\label{prep=cauchy}

\begin{figure}
    \centering
\includegraphics[width=0.8\linewidth]{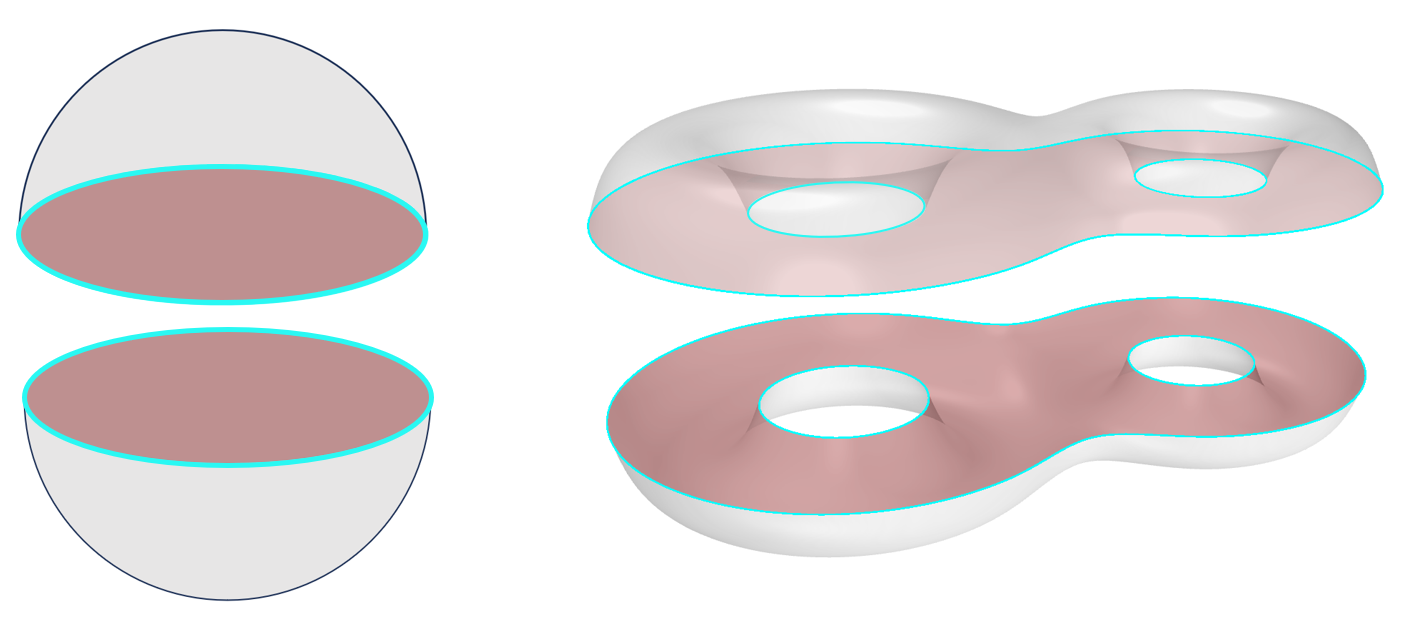}
    \caption{\small{We depict the vacuum AdS$_3$ solution together with a ``three-boundary black hole'' solution in hyperbolic slicing. The manifolds are cut open along the $\tau=0$ Cauchy slice, shown in red. On this slice, the cyan curves mark the $1D$ asymptotic boundaries of the Cauchy slice, which are described by ZZ boundary conditions on the Liouville field. The grey surface corresponds to the CFT state-preparation manifold.}}
    \label{hyperbolic}
\end{figure}

The discussion in Sec.~\ref{purificationreview} was based on cutting and gluing operations on the bulk dual Cauchy slice. In this section, we explain how, in many instances of the AdS$_3$/CFT$_2$ correspondence, the geometry of the \textit{bulk Cauchy slice} is naturally related to the \textit{boundary CFT state-preparation manifold} through hyperbolic slicing \eqref{slicing}. We further explain how this correspondence can arise from coarse-graining the underlying CFT data, and thus emerges directly within the CFT without assuming the holographic dictionary. These considerations motivate us to perform analogous cutting and gluing operations directly in the CFT in Sec.~\ref{sec3}.

The class of CFT quantum states with the above property can be prepared by Euclidean path integrals on 2D Riemann surfaces with boundaries. Two representative examples—a disk with one boundary and a genus-zero surface with three boundaries—are illustrated in Fig.~\ref{hyperbolic}. The CFT state-preparation manifolds are shown as the grey surfaces in the lower half of the figure. The resulting quantum states $\ket{\Psi}$, obtained from the Euclidean path integrals on these grey surfaces, are defined on the 1D cyan curves. Computing the norm $\bra{\Psi}\Psi\rangle$ amounts to gluing the state-preparation manifold to its orientation reversal along the cyan curves. In the examples of Fig.~\ref{hyperbolic}, this procedure yields the sphere and a genus-two manifold, indicated in grey.

The holographic dual of the norm computation corresponds to 3D bulk geometries that fill the boundary grey surfaces. Over a large region of moduli space, the dominant three-dimensional solutions are handlebodies \cite{Yin:2007gv}, shown as the bulk regions enclosed by the grey surfaces in Fig.~\ref{hyperbolic}. By virtue of the $\mathbb{Z}_2$ symmetry of these solutions, the bulk path integral can be sliced open, yielding Hartle–Hawking states $\ket{\text{HH}}$ \cite{Maldacena:2001kr} on the red surfaces, which are the holographic dual of the CFT states $\ket{\Psi}$.

The metrics of these 3D solutions can be expressed in terms of hyperbolic slicing \cite{Maldacena:2001kr, Maldacena:2004rf, Balasubramanian:2014hda, Verlindeunpublished, Verlindetalk, Verlinde:2022xkw, Chandra:2023dgq, Chua:2023ios, Bao:2025plr, Geng:2025efs},
\be \label{slicing}
ds^2=d\tau^2+\cosh^2(\tau) e^{\phi(z,
\bar{z})} dz d\bar{z}
\ee
where the Liouville field $\phi(z,\bar{z})$ solves the Liouville equation,
\be
\partial \bar{\partial}\phi=\frac{e^\phi}{2}~.
\ee
Thus $e^{\phi(z,
\bar{z})} dz d\bar{z}$ denotes a 2D hyperbolic metric, and the 3D hyperbolic solutions are foliated by such 2D hyperbolic manifolds. In this foliation, the CFT state-preparation manifold lies at $\tau \to -\infty$, while the bulk Cauchy slice is located at $\tau=0$. Different slices share the same conformal geometry—namely the same metric up to a Weyl factor $\cosh^2(\tau)$—and points on different slicings admit a natural \textit{one-to-one correspondence}.

In \cite{Belin:2020hea, Chandra:2022bqq}, 3D solutions of the form \eqref{slicing} were applied to 2D hyperbolic spaces without boundaries (but with punctures or higher genus). In these cases, \eqref{slicing} yields 3D spacetime wormhole solutions with \textit{two} asymptotic spacetime boundaries at $\tau \to \pm \infty$ \cite{Maldacena:2004rf, Belin:2020hea, Chandra:2022bqq}.

On the other hand, it was observed in \cite{Chua:2023ios, Bao:2025plr, Geng:2025efs, Chandra:2023dgq} that when a 2D hyperbolic space with 1D asymptotic boundaries is substituted into \eqref{slicing}, the resulting 3D bulk geometry possesses only one \textit{single} asymptotic boundary. This is because the \textit{two} 2D asymptotic spacetime boundaries at $\tau \to \pm \infty$ are \textit{glued together} along their 1D asymptotic boundaries, indicated by the cyan curves in Fig.~\ref{hyperbolic}. This feature was already pointed out in \cite{Maldacena:2004rf}, and can be readily verified in the familiar cases of vacuum AdS$_3$ and BTZ solutions.

In terms of the Liouville field $\phi(z,\bar{z})$, the asymptotic boundaries correspond to imposing ZZ boundary conditions \cite{Zamolodchikov:2001ah, Chua:2023ios, Bao:2025plr, Geng:2025efs, Chandra:2023dgq},\footnote{In \cite{Chandra:2024vhm}, ZZ boundary conditions were also used to investigate thin-shell black holes in 3D gravity.}
\be \label{zzbc}
\phi \to -2 \ln( \text{Im}(z)), \qquad \text{Im}(z) \to 0\ .
\ee
Under these conditions, the 3D gravitational path integrals on the hyperbolic manifolds \eqref{slicing} can be evaluated, and are found to match precisely with the quantization of 2D Liouville CFT with ZZ boundary conditions \cite{Chandra:2022bqq, Chua:2023ios, Collier:2023fwi}, yielding 
\be
Z_{\text{grav,3D}}=Z_{\text{Liouville,2D}}^{\text{ZZ}}\ .
\ee

The preceding discussion shows how, once the 3D solutions \eqref{slicing} are given, the $\tau=0$ and $\tau \to -\infty$ slices are related, but these 3D solutions were assumed a priori. In fact, the emergence of 3D geometry can be \textit{derived} directly from the coarse-graining framework within the CFT itself \cite{Bao:2025plr, Geng:2025efs, Chua:2023ios}, as we will illustrate with explicit examples in Sec.~\ref{EEsec}. Here we briefly sketch the logic.

The idea is that the computation of the norm $\bra{\Psi}\Psi\rangle$, for example in Fig.~\ref{hyperbolic}, can be carried out in the CFT using conformal block decomposition, which involves OPE coefficients and conformal blocks. Upon coarse-graining over heavy states and replacing the density of states and OPE coefficients by the universal large-$c$ averaged (B)CFT data fixed by the conformal bootstrap \cite{Cardy:1986ie, Collier:2019weq, Numasawa:2022cni}, the result precisely reproduces the partition function of Liouville CFT with ZZ boundary conditions \cite{Chua:2023ios, Bao:2025plr, Geng:2025efs}, i.e.,
\be
\overline{\bra{\Psi}\Psi \rangle}=Z_{\text{Liouville,2D}}^{\text{ZZ}}~.
\ee
This Liouville theory with ZZ boundaries, in turn, describes quantization of the 2D hyperbolic metric on the corresponding Cauchy slice of \eqref{slicing}.

We emphasize that $\ket{\Psi}$ is a state in the compact holographic CFT, not in Liouville theory; the latter emerges only after universal coarse-graining over heavy states, thereby serving as a bridge to geometry and 3D gravity.

A canonical example is the thermofield state prepared on a cylinder. Replacing the density of states in the norm computation with the Cardy density for all heavy states above the black hole threshold \cite{Cardy:1986ie, Hartman:2014oaa} reproduces the BTZ black hole partition function \cite{Maloney:2007ud}. This coincides with the Liouville partition function with two ZZ boundaries \cite{Bao:2025plr, Chua:2023ios, Zamolodchikov:2001ah}, which quantizes 2D hyperbolic cylinder geometries.\footnote{Note the manifolds involved: the holographic CFT lives on a 2D torus, the bulk dual is a 3D solid torus, while the effective Liouville theory is defined on a cylinder with two specific boundary conditions.}

As a final remark, the mechanism we identified links the BCFT tensor network defined at $\tau \to -\infty$ to conventional tensor networks proposed to describe discretized geometries on the Cauchy slice $\tau=0$ \cite{Swingle:2009bg, Pastawski:2015qua, Hayden:2016cfa}, via its connection to Liouville theory with ZZ boundary conditions under coarse-graining. In this way, our mechanism naturally realizes and generalizes the path-integral optimization interpretation of tensor networks \cite{Caputa:2017urj, Caputa:2017yrh, VanRaamsdonk:2018zws}.

\section{$\text{RE}=\text{EW}$ from BCFT Random Tensor Networks}\label{sec3}

In this section, we use BCFT tensor networks to construct the canonical purification (\ref{cp}) and compute the reflected entropy (\ref{reflected def}). We will show that this calculation leads to the entanglement wedge cross section, thereby proving the $\text{RE}=\text{EW}$ relation (\ref{RE=2EW}). 

Our strategy is as follows. We first analyze the state $\ket{\rho_{AB}^{m/2}}$ from the BCFT perspective, where the tensor network representation can be written explicitly for even $m$. Using the relation between the modular Hamiltonian and BCFT Hamiltonians, we continue to $m=1$, obtaining the canonical purification (\ref{cp}) as a BCFT path integral. With this     representation in hand, the problem reduces to an analogue of the setup used in the derivation of the RT formula \cite{Bao:2025plr, Geng:2025efs}. This allows us to apply the same coarse-graining procedure on OPE coefficients within the BCFT tensor network to compute the von Neumann entropy of $AA^*$ (\ref{reflected def}), and establish its equality with the area of the entanglement wedge cross section. 

Throughout the derivation, we perform the cutting and gluing operations directly on the CFT state-preparation manifold, rather than on the bulk Cauchy slice as in \cite{Dutta:2019gen}. Nevertheless, the correspondence between the two manifolds via \eqref{slicing} ensures that these boundary operations map to the bulk ones, making our construction the explicit boundary dual of the bulk construction in \cite{Dutta:2019gen}.

We will illustrate the idea in three representative cases with different features, which can then be straightforwardly generalized to arbitrary cases.

\subsection{Case I: $\text{RE}(A:B)$ for Adjacent $A$ and $B$}

\subsubsection{Construction of Canonical Purification}
\begin{figure}
    \centering
\includegraphics[width=0.6\linewidth]{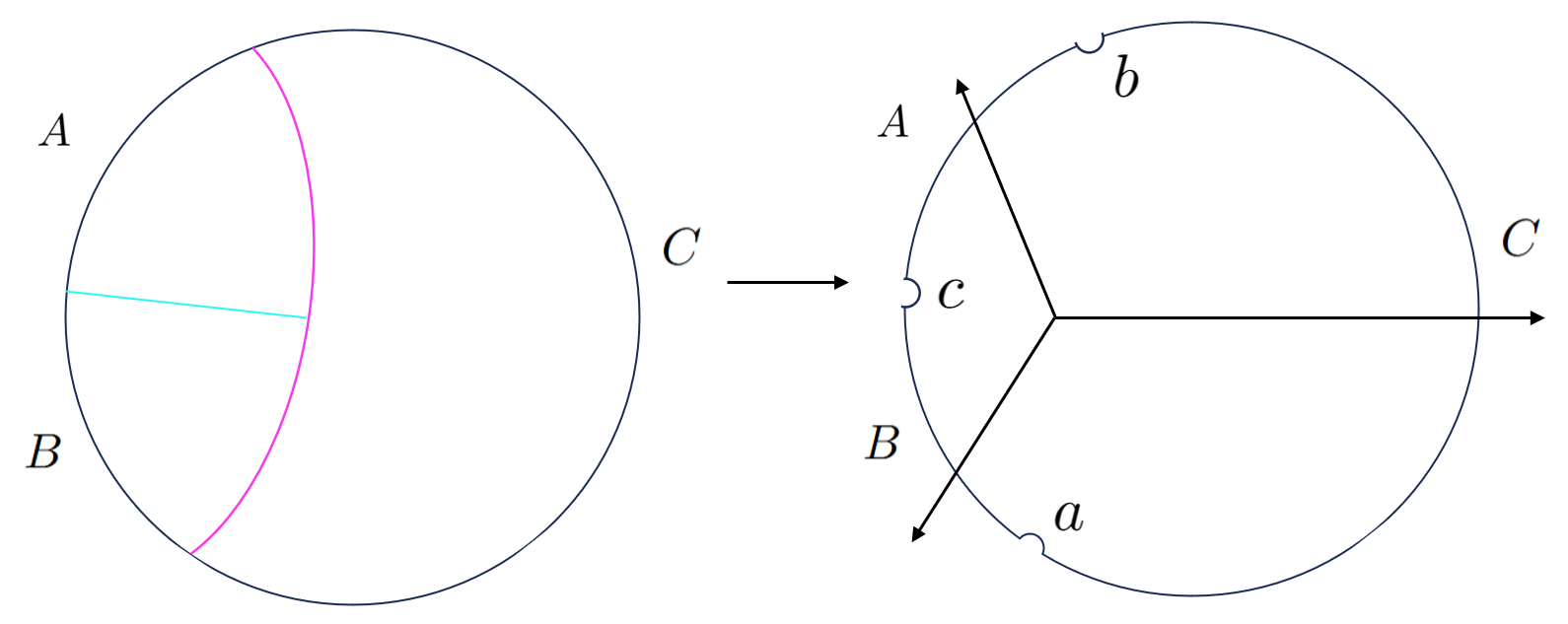}
    \caption{\small Left: The vacuum state is defined on $ABC$. The pink curve denotes the RT surface for $AB$, while the cyan curve represents the entanglement wedge cross section. Right: The CFT vacuum state on $ABC$ is prepared by a Euclidean path integral on a disk; by introducing tiny conformal boundaries $abc$ as regulators, we obtain an entangled BCFT state.}
    \label{3CONNECTED}
\end{figure}

The first example we study is illustrated in the left panel of Fig.~\ref{3CONNECTED}, where the vacuum state is defined on the three adjacent single intervals $A$, $B$ and $C$. On the zero time Cauchy slice, the RT surface of $AB$ and the entanglement wedge cross section are again colored in pink and cyan, respectively. 

The vacuum state of the CFT on $ABC$ is prepared by a Euclidean path integral on a disk (see the right panel of Fig.~\ref{3CONNECTED}). We begin by introducing tiny conformal boundaries \cite{Cardy:2004hm, Geng:2025efs} into the path integral, thereby turning it into a state defined within the BCFT framework. Since the regulator holes are shrinkable and the resulting quantum states are independent of the triangulation, we will throughout this paper adopt the simplest triangulation; in the present setup, this reduces to a single triangle—an \textit{open pair of pants}—whose dual graph provides the tensor network representation. More explicitly, we regulate the path integral by introducing three tiny conformal boundaries $a,b,c$, thereby expressing the vacuum as an entangled state in three BCFT Hilbert spaces associated with the regulated subregions $A$, $B$, and $C$, respectively.

\begin{figure}
    \centering
\includegraphics[width=0.45\linewidth]{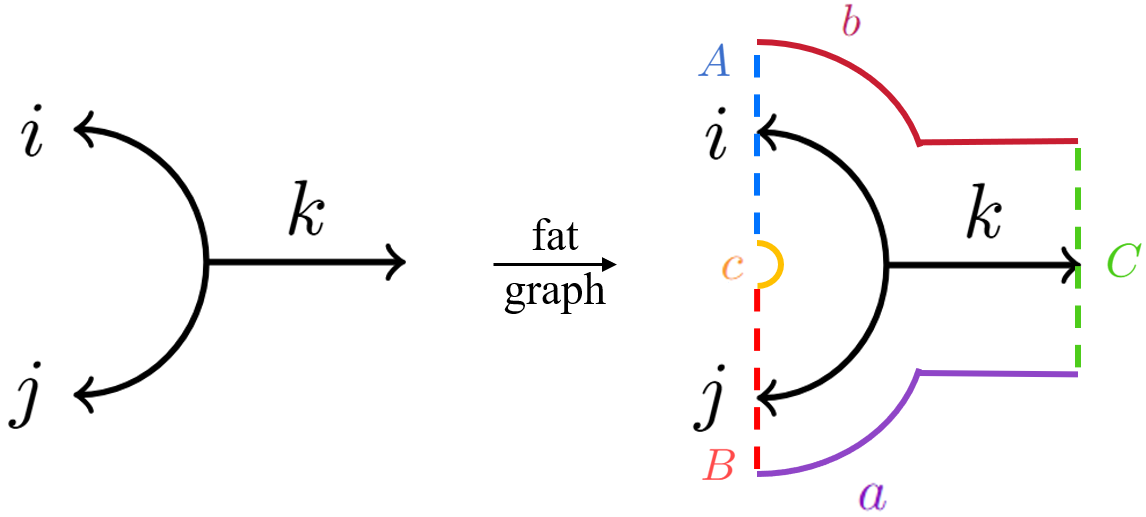}
    \caption{\small{A fat graph depiction of (\ref{eq:2BCFT1CFTchiral}), making its connection to Fig.~\ref{3CONNECTED} manifest.}}
    \label{fat}
\end{figure}

Using the ``OPE block'' formalism \cite{Czech:2016xec, Chandra:2023dgq, Bao:2025plr, Geng:2025efs}, this state can be expressed as\footnote{We adopt the convention that, for the BCFT OPE coefficients $C^{abc}_{ijk}$'s, the conformal boundary on the opposite side is placed above the primary label. For instance, the state $\ket{i}$ on $A$ has an opposite boundary $a$, so we write $a$ on top of $i$.}

\begin{equation}
\begin{aligned}
\ket{\Psi}^{abc}_{ABC}=\sum_{\text{primaries}}  C^{abc}_{ijk}  \;\mathcal{B}\left[
 \vcenter{\hbox{\begin{tikzpicture}[scale=0.6]
 \draw[<->,thick,black!!40] (0,1) arc (90:-90:1);
  \draw[->, thick, black!!40] (1,0) to (2.5,0);
  \node[left] at (0,1) {\textcolor{black}{$i$}};
  \node[left] at (0,-1) {\textcolor{black}{$j$}};
  \node[above] at (1.75,0) {\textcolor{black}{$k$}};
 \end{tikzpicture}}} \right] \ket{i} \ket{j} \ket{k}
 ~,
 \end{aligned}\label{eq:2BCFT1CFTchiral}
\end{equation}
where $\ket{i}, \ket{j}, \ket{k}$ denote the boundary primary states propagating on intervals $A$, $B$, and $C$, with boundary conditions $bc$, $ca$, and $ab$ at their respective ends. The connection to Fig.~\ref{3CONNECTED} is made manifest through the associated fat graph shown in Fig.~\ref{fat}. $C^{abc}_{ijk}$ denote the BCFT operator product expansion (OPE) coefficients, while $\mathcal{B}[\cdots]$ represents the corresponding OPE block, which packages the full contribution of all descendant operators associated with the chosen primary sectors.

It is also useful to rewrite this in more conventional tensor network notation by expanding the descendant contributions explicitly, and expressing the result as a tensor $\mathcal{T}_{ijk,IJK}^{abc}$:

\be
\ket{\Psi}^{abc}_{ABC}=\sum_{\text{primaries}, I,J,K} C_{ijk}^{abc} \gamma_{ijk}^{IJK} \ket{i,j,k,I,J,K}=\sum_{\text{primaries}, I,J,K} \mathcal{T}_{ijk,IJK}^{abc}\ket{i,j,k,I,J,K}~.
\ee
Here $I,J,K$ label the descendant states, and $\gamma_{ijk}^{IJK}$ is determined by mapping the regulated curly triangle onto the upper half-plane \cite{Cheng:2023kxh, Brehm:2021wev, Brehm:2024zun}.\footnote{Both $\gamma_{ijk}^{IJK}$ and $\mathcal{T}_{ijk,IJK}^{abc}$ depend on the shape and metric of the curly triangle, though this dependence is omitted in our notation.} The resulting object is a nine-index\footnote{In this paper, we fix the conformal boundary conditions as in \cite{Hung:2025vgs, Geng:2025efs}, so that the boundary condition labels reduce to delta-function matching constraints, and each tensor effectively carries only six uncontracted legs, represented by the BCFT Hilbert space labeled by primary and descendant labels.} tensor with infinite bond dimensions. In this language, the sewing rules of (B)CFT—arising from summing over a complete basis of internal states—are naturally represented as tensor contractions in tensor networks.

We now construct the canonical purification within the BCFT framework, where the method is inspired by the original holographic argument of \cite{Dutta:2019gen}. Here, the relation between the modular Hamiltonian of a single interval and the BCFT Hamiltonian \cite{Ohmori:2014eia, Cardy:2016fqc} allows for an explicit analytic continuation in the construction of $\ket{\sqrt{\rho_{AB}}}$.

First, using (\ref{eq:2BCFT1CFTchiral}), the reduced density matrix for $AB$ is given by
\begin{equation}
\begin{aligned}
\rho_{AB}=\tr_{C} \ket{\Psi}^{abc}_{ABC} {}^{abc}_{ABC} \bra{\Psi}=\sum_{\text{primaries}}  \frac{C^{abc}_{ijk}   C^{*abc}_{lmk}}{\sqrt{g_a g_b}}  \;\mathcal{B}\left[
 \vcenter{\hbox{\begin{tikzpicture}[scale=0.6]
 \draw[<->,thick,black!!40] (0,1) arc (90:-90:1);
  \draw[<->,thick,black!!40] (4,1) arc (90:270:1);
  \draw[-, thick, black!!40] (1,0) to (3,0);
  \node[left] at (0,1) {\textcolor{black}{$i$}};
  \node[left] at (0,-1) {\textcolor{black}{$j$}};
  \node[right] at (4,-1) {\textcolor{black}{$m$}};
  \node[right] at (4,1) {\textcolor{black}{$l$}};
  \node[above] at (2,0) {$k$};
 \end{tikzpicture}}} \right] \ket{i} \ket{j} \bra{l} \bra{m}
 ~,
 \end{aligned} \label{densitymatrix}
\end{equation} 
where we include the normalization factor $1/\sqrt{g_ag_b}$ from BCFT two-point function for the $k$ primaries \cite{Numasawa:2022cni, Hung:2025vgs, Geng:2025efs}, determined by the boundary $g$-factors of the $a$ and $b$ boundary conditions \cite{Affleck:1991tk}.

Correspondingly, the replica partition function is

\be
\begin{aligned}
&Z_{AB,m}=\tr \rho_{AB}^{m}=\sum_{\text{primaries}}\frac{C_{ijk}^{abc} C^{*abc}_{lmk} C^{abc}_{lmn}C_{pqn}^{*abc}C^{abc}_{pqo} C^{*abc}_{rso} C^{abc}_{rst}\cdots C^{*abc}_{iju}}{(g_a g_b g_c)^m} \\
&\vcenter{\hbox{

\begin{tikzpicture}[scale=0.75]
	\draw[thick] (0,1) arc (90:-90:1)
    node[pos=0, left] {$i$}
    node[pos=1, left] {$j$};
	\draw[thick] (1,0) -- ++(2,0)
    node[pos=0.5, above] {$k$};
  \draw[thick] (4,0) circle (1)
	  node[below] at ++(0,1) {$l$}
	  node[above] at ++(0,-1) {$m$};
	\draw[thick] (5,0) -- ++(2,0)
    node[pos=0.5, above] {$n$};
	\draw[thick] (8,0) circle (1)
	  node[below] at ++(0,1) {$p$}
	  node[above] at ++(0,-1) {$q$};
  \draw[thick] (9,0) -- ++(2,0)
    node[pos=0.5, above] {$o$};
  \draw[thick] (12,0) circle (1)
	  node[below] at ++(0,1) {$r$}
	  node[above] at ++(0,-1) {$s$};
  \draw[thick] (13,0) -- (14,0)
    node[pos=0.7, above] {$t$} ;
  \node[above] at (15,-0.3) {$\cdots$};
  \draw[thick] (16,0) -- ++(1,0)
    node[pos=0.3, above] {$u$};
  \draw[thick] (18,1) arc (90:270:1)
    node[pos=0, right] {$i$}
    node[pos=1, right] {$j$};

  \def\w{1.4}
  \def\h{2}
  \draw[thick,Mulberry, rounded corners=2pt,dashed]
    (4-\w, -\h) rectangle ++(2*\w,2*\h);
  \draw[thick, Mulberry, rounded corners=2pt,dashed]
    (8-\w,-2) rectangle ++(2*\w,2*\h);
  \draw[thick, Mulberry, rounded corners=2pt,dashed]
    (12-\w,-2) rectangle ++(2*\w,2*\h);

  \draw[dashed,thick] (0,-1) -- ++(0,-0.5) -- ++(18,0) -- ++(0,0.5);
  \draw[dashed,thick] (0,1) -- ++(0,0.5) -- ++(18,0) -- ++(0,-0.5);
\end{tikzpicture}

	}}~,
\end{aligned}
\ee 
where the additional $g$-factors again arise from the normalization of BCFT primary states—for instance, the $l$ primaries contribute a factor of $1/\sqrt{g_b g_c}$. The expression on the second line represents the conformal block obtained by gluing OPE blocks, and we have introduced several purple dashed boxes to highlight the repeating segments of the conformal block. As the replica index $m$ increases, the number of such boxes grows correspondingly. Each purple box, together with the OPE coefficients, corresponds precisely to the reduced density matrix of the complement region, $\rho_C$, by definition. For later use, we also introduce the following notation:

\be
\begin{tikzpicture}
	\draw[thick] (0.5,0) -- ++(1.5,0)
    node[pos=0.4, above] {$k$};
  \draw[thick] (3,0) circle (1)
	  node[below] at ++(0,1) {$l$}
	  node[above] at ++(0,-1) {$m$};
	\draw[thick] (4,0) -- ++(1.5,0)
    node[pos=0.6, above] {$n$};

  \def\w{1.5}
  \def\h{1.5}
  \draw[thick, Mulberry, rounded corners=2pt, dashed]
    (3-\w, -\h) rectangle ++(2*\w,2*\h);

  \node at (6.25,0) {$=$};

	\draw[thick] (7,0) -- ++(1,0)
    node[pos=0.5, above] {$k$};
  \draw[thick] (8,-0.5) rectangle ++(1.5,1)
    node[below] at ++(-0.75,-1) {$\rho_C$};
	\draw[thick] (9.5,0) -- ++(1,0)
    node[pos=0.5, above] {$n$};
\end{tikzpicture} \label{bubbleresistor}
\ee
For notational simplicity, we will omit the OPE coefficients and $g$-factors in what follows unless explicitly stated. For example, the left-hand side of the equation above should also include the factors $\frac{C^{abc}_{lmk} C^{*abc}_{lmn}}{\sqrt{g_a g_b g_c^2}}$, together with the sum over primaries $l,m$. 

The replica partition function can then be represented as, 

\be
\begin{aligned}
Z_{AB,m}&=\vcenter{\hbox{\begin{tikzpicture}[scale=0.75]
	\draw[thick] (0,1) arc (90:-90:1)
    node[pos=0, left] {$i$}
    node[pos=1, left] {$j$};
	\draw[thick] (1,0) -- ++(1.5,0)
    node[pos=0.5, above] {$k$};
  \draw[thick] (2.5,-0.5) rectangle ++(1.5,1)
    node[below] at ++(-0.75,-1) {$\rho_C$};
	\draw[thick] (4,0) -- ++(1,0)
    node[pos=0.5, above] {$n$};
  \draw[thick] (5,-0.5) rectangle ++(1.5,1)
    node[below] at ++(-0.75,-1) {$\rho_C$};
	\draw[thick] (6.5,0) -- ++(1,0)
    node[pos=0.5, above] {$o$};
  \draw[thick] (7.5,-0.5) rectangle ++(1.5,1)
    node[below] at ++(-0.75,-1) {$\rho_C$};
	\draw[thick] (9,0) -- ++(1,0)
    node[pos=0.5, above] {$t$};
  \node at (10.5,0) {$\cdots$};

  \draw[thick] (11,0) -- ++(1.5,0)
    node[pos=0.5, above] {$u$};
  \draw[thick] (13.5,1) arc (90:270:1)
    node[pos=0, right] {$i$}
    node[pos=1, right] {$j$};

  \draw[dashed,thick] (0,-1) -- ++(0,-0.5) -- ++(13.5,0) -- ++(0,0.5);
  \draw[dashed,thick] (0,1) -- ++(0,0.5) -- ++(13.5,0) -- ++(0,-0.5);

  \def\start{14.5}
\end{tikzpicture}}}\\
&=
\vcenter{\hbox{\begin{tikzpicture}[scale=0.75]
	\draw[thick] (+1,1) arc (90:-90:1)
    node[pos=0, left] {$i$}
    node[pos=1, left] {$j$};
	\draw[thick] (+2,0) -- ++(1.5,0)
    node[pos=0.5, above] {$k$};

  \draw[thick] (+3.5,-0.5) rectangle ++(2.5,1)
    node[below] at ++(-1,-1) {$\rho_C^{m-1}$};

  \draw[thick] (+6,0) -- ++(1.5,0)
    node[pos=0.5, above] {$u$};
  \draw[thick] (+8.5,0+1) arc (90:270:1)
    node[pos=0, right] {$i$}
    node[pos=1, right] {$j$};
  \draw[dashed,thick] (+1,-1) -- ++(0,-0.5) -- ++(7.5,0) -- ++(0,0.5);
  \draw[dashed,thick] (+1,1) -- ++(0,0.5) -- ++(7.5,0) -- ++(0,-0.5);
\end{tikzpicture}}}\ .
\end{aligned}
\ee
Note that the above expression also agrees with the replica partition function for the complementary region $C$ $Z_{AB,m}=\tr \rho_C^m=Z_{C,m}$, consistent with $\ket{\Psi}^{abc}_{ABC}$ being a pure state.

For $m$ being an even integer, it is easy to write down explicitly the state $\ket{\rho_{AB}^{m/2}}$ as,
\be
\begin{tikzpicture}[scale=0.68]
  \draw[<->,thick] (0,1) arc (90:-90:1)
    node[pos=0, left] {$i$}
    node[pos=1, left] {$j$};
  \draw[thick] (1,0) -- ++(1.5,0)
    node[pos=0.5, above] {$k$};
  \draw[thick] (2.5,-0.5) rectangle ++(1.5,1)
    node[below] at ++(-0.75,-1) {$\rho_C$};
	\draw[thick] (4,0) -- ++(1,0)
    node[pos=0.5, above] {$n$};
  \draw[thick] (5,-0.5) rectangle ++(1.5,1)
    node[below] at ++(-0.75,-1) {$\rho_C$};
	\draw[thick] (6.5,0) -- ++(1,0)
    node[pos=0.5, above] {$o$};
  \node at (8,0) {$\cdots$};

  \draw[thick] (8.5,0) -- ++(1.5,0)
    node[pos=0.5, above] {$u$};
  \draw[<->,thick] (11,1) arc (90:270:1)
    node[pos=0, right] {$v$}
    node[pos=1, right] {$w$};

  \def\start{12}

  \node at (\start,0) {$=$};

	\draw[<->,thick] (\start+1,1) arc (90:-90:1)
    node[pos=0, left] {$i$}
    node[pos=1, left] {$j$};
	\draw[thick] (\start+2,0) -- ++(1.5,0)
    node[pos=0.5, above] {$k$};

  \draw[thick] (\start+3.5,-0.5) rectangle ++(2.5,1)
    node[below] at ++(-1,-1) {$\rho_C^{m/2-1}$};

  \draw[thick] (\start+6,0) -- ++(1.5,0)
    node[pos=0.5, above] {$u$};
  \draw[<->,thick] (\start+8.5,1) arc (90:270:1)
    node[pos=0, right] {$v$}
    node[pos=1, right] {$w$};
\end{tikzpicture}
\label{rhom/2}
\ee
where $\ket{i} \in \mathcal{H}_A$, $\ket{j} \in \mathcal{H}_B$, and $\ket{v} \in \mathcal{H}_{A^*}$, $\ket{w} \in \mathcal{H}_{B^*}$. From the right-hand side of the above expression, the analytic continuation is straightforward—we can extend $m$ to generic values without needing the quotient construction in the bulk dual \cite{Dutta:2019gen}. Concretely, we compute the matrix elements of the reduced density matrix $\rho_C$, raise it to the $(m/2 - 1)$-th power, and let it act on the $C$ parts of the two copies of the state $\ket{\Psi}_{ABC}$. This operation is well-defined, since powers of $\rho_C$ remain operators in the algebra associated with region $C$.

In the present setup, we can explicitly compute the reduced density matrix $\rho_C$. The subsystems $AB$ and $C$ form a bipartition of the vacuum state, where $C$ is exactly a \textit{single-interval} region. In this case, the modular Hamiltonian 
\be
H_{\text{mod},C} = -\frac{\ln \rho_C}{2\pi}
\ee
is \textit{local}. In 2D CFT, this locality follows from the fact that after introducing the regulators, the configuration Fig.~\ref{3CONNECTED} can be conformally mapped to the thermofield double state for BCFTs defined on strips $AB$ and $C$ via a conformal transformation $w=f(z)$. Thus, on one side of this thermofield double, the modular Hamiltonian coincides with the BCFT Hamiltonian upto an overall constant normalization\cite{Lauchli:2013jga, Ohmori:2014eia, Cardy:2016fqc},
\be \label{modhambcft}
H_{\text{mod},C} =\mathcal{N} H_{\text{BCFT}}~.
\ee
We will explain below how the constant normalization factor $\mathcal{N}$ is determined from the size of region $C$. Accordingly, the reduced density matrix $\rho_C$ is simply the thermal density matrix of the BCFT defined between boundaries $a$ and $b$. 

We can also explicitly write down the modular Hamiltonian in the original $z$ coordinate as \cite{Cardy:2016fqc},
\be
H_{\text{mod},C} = \int_{C} \frac{T_{00}(x)}{f'(x)}  dx~.
\ee
where $x$ is the coordinate on the original ``time slice'' $C$. It is worth pointing out that the famous Casini–Huerta–Myers result \cite{Casini:2011kv} arises as a special case of this general formula for a spherical subregion in arbitrary dimensions. In two dimensions, however, the enhanced power of the infinite-dimensional conformal group allows this expression to hold in far greater generality.

One might worry that, since the $c$ boundary regulator is introduced between $AB$, the bubble in \eqref{bubbleresistor}—corresponding to a small hole arising from the $c$ boundary in the path integral\footnote{The hole can be made manifest using the fat graph.}—could potentially alter the story. However, as shown in \cite{Chen:2024unp, Geng:2025efs}, in the limit where the hole size shrinks to zero, such bubbles are always contractible in theories with a vacuum state. This is most transparently seen from the open–closed duality: the conformal boundary $c$ is mapped to the conformal boundary state $\ket{c(R)}_{\text{Cardy}}$\cite{Cardy:2004hm}, where $R$ denotes the hole size. In the vanishing-hole limit they universally reduce to the vacuum state,\footnote{More precisely, one first obtains the vacuum Ishibashi state; then, as the hole size goes to zero, the contributions from descendant states are suppressed, leaving only the vacuum.}
\be \label{cardyandvacishi}
\ket{c(R)}_{\text{Cardy}} \approx R^{-\frac{c}{6}} g_c \ket{0}\ .
\ee
After properly normalizing the partition function, such holes leave no imprints, implying that they are indeed shrinkable. Consequently, the reduced density matrix $\rho_C$ coincides with that of a single interval in the vacuum state.

When $m \to 1$, the canonical purification state can be represented as, 

\be \label{purificationCFT}
\ket{\sqrt{\rho_{AB}}}=
\vcenter{\hbox{
\begin{tikzpicture}[scale=0.66]
  \def\start{0}
	\draw[<->,thick] (\start+1,1) arc (90:-90:1)
    node[pos=0, left] {$i$}
    node[pos=1, left] {$j$};
	\draw[thick] (\start+2,0) -- ++(1.5,0)
    node[pos=0.5, above] {$k$};

  \draw[thick] (\start+3.5,-0.5) rectangle ++(2.5,1)
    node[below] at ++(-1,-1) {$\rho_C^{-1/2}$};

  \draw[thick] (\start+6,0) -- ++(1.5,0)
    node[pos=0.5, above] {$u$};
  \draw[<->,thick] (\start+8.5,1) arc (90:270:1)
    node[pos=0, right] {$v$}
    node[pos=1, right] {$w$};

  \def\start{10.5}

  \node at (\start-1,0) {$=$};

	\draw[<->,thick] (\start,1) arc (90:-90:1)
    node[pos=0, left] {$i$}
    node[pos=1, left] {$j$};
	\draw[thick] (\start+1,0) -- ++(1.5,0)
    node[pos=0.5, above] {$k$};

  \draw[thick] (\start+2.5,-0.5) rectangle ++(1.25,1)
    node[below] at ++(-0.5,-1) {$\rho_C^{-1/4}$};

  \draw[thick] (\start+3.75, 0) -- ++(1.5, 0)
    node[pos=0.5, above] {$l$};

  \draw[thick] (\start+5.25,-0.5) rectangle ++(1.25,1)
    node[below] at ++(-0.5,-1) {$\rho_C^{-1/4}$};

  \draw[thick] (\start+6.5,0) -- ++(1.5,0)
    node[pos=0.5, above] {$u$};
  \draw[<->,thick] (\start+9,1) arc (90:270:1)
    node[pos=0, right] {$v$}
    node[pos=1, right] {$w$};
\end{tikzpicture}

}}
\ee
which means we first prepare two copies of the vacuum state $\ket{\Psi}_{ABC}$, then evolve the $C$ subsystem of each copy ``backwards'' using the modular Hamiltonian via $\rho_C^{-1/4}$, and finally glue the two copies together along the surface where this backward evolution terminates, with opposite orientation.

Explicitly, this state can be written as,
\be
\ket{\sqrt{\rho_{AB}}}=\frac{1}{\sqrt{g_a g_b}}\sum_{\text{primaries}, I,J,K,K',L,M} \mathcal{T}_{ijk,IJK}^{abc} \left( \rho_C^{-1/2} \right)_{k,K;u,U}  \mathcal{T}_{vwu,VWU}^{*abc}\ket{i,j,I,J} \ket{v,w,V,W}
\ee
where $\left( \rho_C^{-1/2} \right)_{k,K;u,U}$'s are the matrix elements of $\rho_C^{-1/2}$, spanned in the BCFT basis of states on $C$. We can easily check that this formal expression indeed gives a purification of $\rho_{AB}$: 
\be \label{purificationcheck}
\tr_{A^* B^*} \ket{\sqrt{\rho_{AB}}} \bra{\sqrt{\rho_{AB}}} = 
\vcenter{\hbox{
\begin{tikzpicture}[scale=0.63]
  \def\start{0}

	\draw[<->,thick] (\start,1) arc (90:-90:1)
    node[pos=0, left] {$i$}
    node[pos=1, left] {$j$};
	\draw[thick] (\start+1,0) -- ++(1.5,0)
    node[pos=0.5, above] {$k$};

  \draw[thick] (\start+2.5,-0.5) rectangle ++(2.5,1)
    node[below] at ++(-1,-1) {$\rho_C^{-1/2}$};

  \draw[thick] (\start+5, 0) -- ++(1.5, 0)
    node[pos=0.5, above] {$l$};

  \draw[thick] (\start+7.5,0) circle (1)
	  node[below] at ++(0,1) {$m$}
	  node[above] at ++(0,-1) {$n$};

  \draw[thick] (\start+8.5, 0) -- ++(1.5, 0)
    node[pos=0.5, above] {$o$};

  \draw[thick] (\start+10,-0.5) rectangle ++(2.5,1)
    node[below] at ++(-1,-1) {$\rho_C^{-1/2}$};

  \draw[thick] (\start+12.5,0) -- ++(1.5,0)
    node[pos=0.5, above] {$u$};
  \draw[<->,thick] (\start+15,1) arc (90:270:1)
    node[pos=0, right] {$v$}
    node[pos=1, right] {$w$};

  \def\w{1.5}
  \def\h{1.5}
  \draw[thick, Mulberry, rounded corners=2pt,dashed]
    (7.5-\w, -\h) rectangle ++(2*\w,2*\h);
\end{tikzpicture}
}}~,
\ee
by recalling that the content in the purple box is $\rho_C$. Together with the two factors of $\rho_C^{-1/2}$, it multiplies to $1$, so that after reinstating the OPE coefficients and summing over primaries, this reduced density matrix matches $\rho_{AB}$ \eqref{densitymatrix}.

We now demonstrate that the backward evolution with $\rho_C^{-1/4}$ in (\ref{purificationCFT}) maps the surface $C$ precisely to a special location which, under the hyperbolic slicing \eqref{slicing}, corresponds to the RT surface of $AB$ in the bulk dual.

The strategy of derivation proceeds as follows. We first identify the conformal transformation that maps Fig.~\ref{3CONNECTED} to a rectangular strip, with $C$ on the left and $AB$ on the right. The reduced density matrix $\rho_C$ is then represented by the bracket on the left-hand side of Fig.~\ref{deevolution0}, where the central hole separating $AB$ has already been shrunk. This strip path integral gives the thermal density matrix for $C$, with the modular Hamiltonian equal to the BCFT Hamiltonian up to a normalization constant. Evolution by $\rho_C^{1/4}$ corresponds to traversing only a quarter of the strip, which sends $C$ on the left to the midpoint of the original $ABC$ region (the pink curve). After performing the ensemble average, this pink curve becomes dual to the horizon on the emergent Cauchy slice of the two sided black hole (with branes dual to the $a,b$ boundaries) \cite{Geng:2025efs}, which is the minimal surface between $C$ and $AB$. Mapping back to Fig.~\ref{3CONNECTED} identifies this surface with the RT surface on the original manifold. Consequently, the canonical purification state \eqref{purificationCFT} is realized in the CFT as a path integral on the doubled manifold, similar to Fig.~\ref{crosssection}.

\begin{figure}[h]
\centering
\begin{tikzpicture}
    \def\h{1.5}
    \def\w{2}
    \def\parenShift{0.3}

    \draw[thick] (-\parenShift,\h+\parenShift)
      .. controls (-\parenShift-0.5,\h-0.25) and (-\parenShift-0.5,-\h+0.25) ..
      (-\parenShift,-\h-\parenShift);

    \draw[thick] (4*\w+\parenShift,\h+\parenShift)
      .. controls (4*\w+\parenShift+0.5,\h-0.25) and (4*\w+\parenShift+0.5,-\h+0.25) ..
      (4*\w+\parenShift,-\h-\parenShift)
      node[pos=0, xshift=0.5cm] {$1/4$};

    \draw[thick] (0,-\h) -- ++(4*\w,0)
      node[pos=0.4, below] {$b$};
    \draw[thick] (0,\h) -- ++(4*\w,0)
      node[pos=0.4, above] {$a$};

    \draw[thick] (0,\h) -- ++(0,-2*\h)
      node[pos=0.5, left] {$C$};
    \draw[thick] (4*\w,\h) -- ++(0,-2*\h)
      node[pos=0.5, right] {$C$};

    \draw[thick, dashed] (2*\w,\h) -- ++(0,-2*\h)
      node[pos=0.5, right] {$AB$};

    \draw[thick, \minSurfaceColor] (\w,\h) -- ++(0,-2*\h);

    \node at (4*\w + 1.5, 0) {$=$};

    \def\newx{4*\w + 2.5}

    \draw[thick] (\newx,-\h) -- ++(\w,0)
      node[pos=0.5, below] {$b$};
    \draw[thick] (\newx,\h) -- ++(\w,0)
      node[pos=0.5, above] {$a$};

    \draw[thick] (\newx,\h) -- ++(0,-2*\h)
      node[pos=0.5, left] {$C$};
    \draw[thick] (\newx +\w,\h) -- ++(0,-2*\h)
      node[pos=0.5, right] {$C$};
\end{tikzpicture}
    \caption{\small{After applying the conformal transformations, the path integral preparing the vacuum state on $ABC$ is mapped to the state defined in the square region between $C$ and $AB$, with the tiny hole on $AB$ shrunk. The reduced density matrix associated with $C$ is represented by the entire path integral shown in the bracket. The modular Hamiltonian is simply the BCFT Hamiltonian up to an overall normalization, as is clear from the figure. Taking the $1/4$ power of this density matrix corresponds to a path integral between $C$ and the pink curve.}}
    \label{deevolution0}
\end{figure}

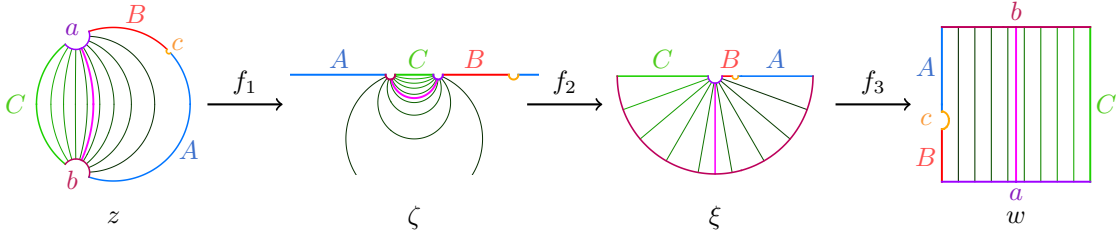
\begin{figure}
    \centering
     \begin{tikzpicture}

  \node (z) at (0,0)
  {\includegraphics[width=2.5cm]{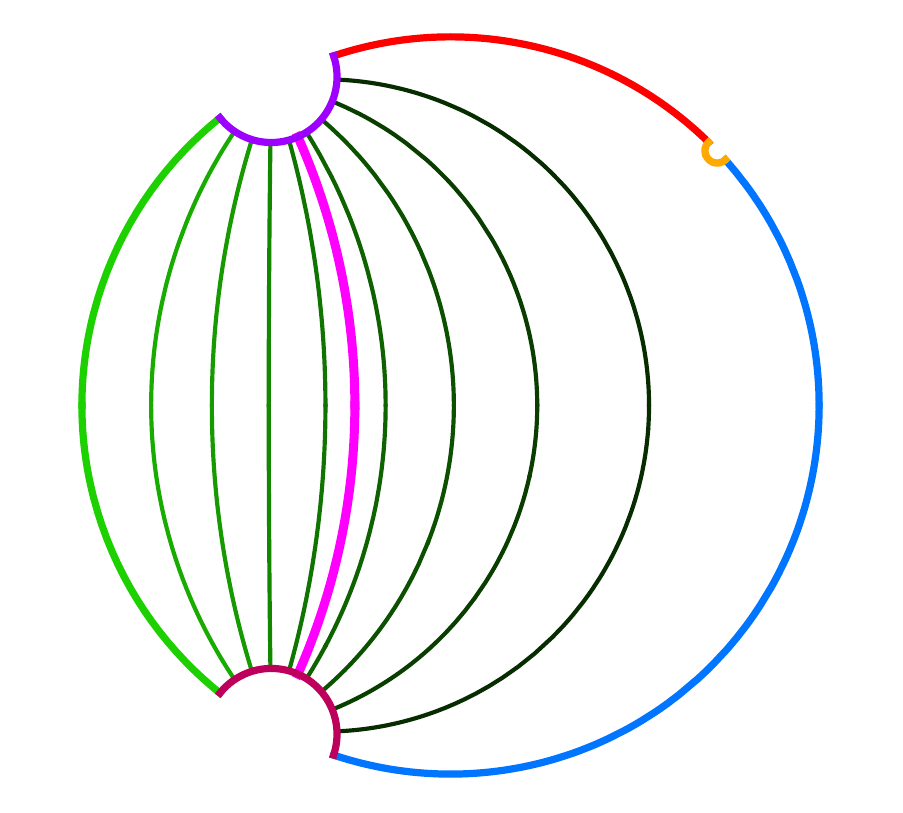}}
  node[\CColor, xshift=-1.3cm, yshift=0cm] {$C$}
  node[\BColor, xshift=0.3cm, yshift=1.2cm] {$B$}
  node[\AColor, xshift=1cm, yshift=-0.6cm] {$A$}
  node[\aColor, xshift=-0.55cm, yshift=1.cm] {$a$}
  node[\bColor, xshift=-0.55cm, yshift=-1.cm] {$b$}
  node[\cColor, xshift=0.85cm, yshift=0.8cm] {$c$};

  \node (zeta) at (4, 0)
  {\includegraphics[width=3.3cm, trim=0 0 0 0.6cm]{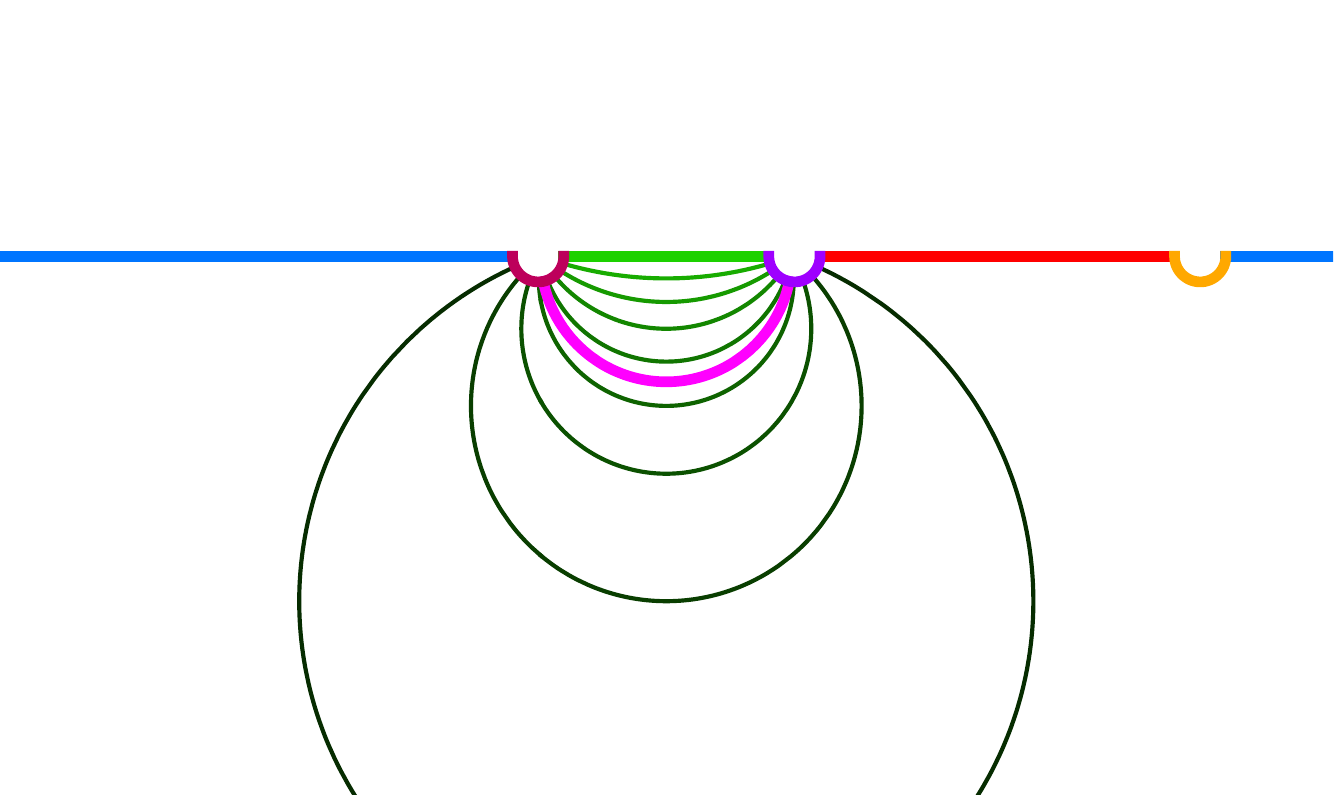}}
    node[\CColor, xshift=4.05cm, yshift=0.6cm] {$C$}
    node[\BColor, xshift=4.8cm, yshift=0.6cm] {$B$}
    node[\AColor, xshift=3cm, yshift=0.6cm] {$A$};

  \node (xi) at (8, 0)
  {\includegraphics[width=3cm, trim= 0 0 0 -2cm]{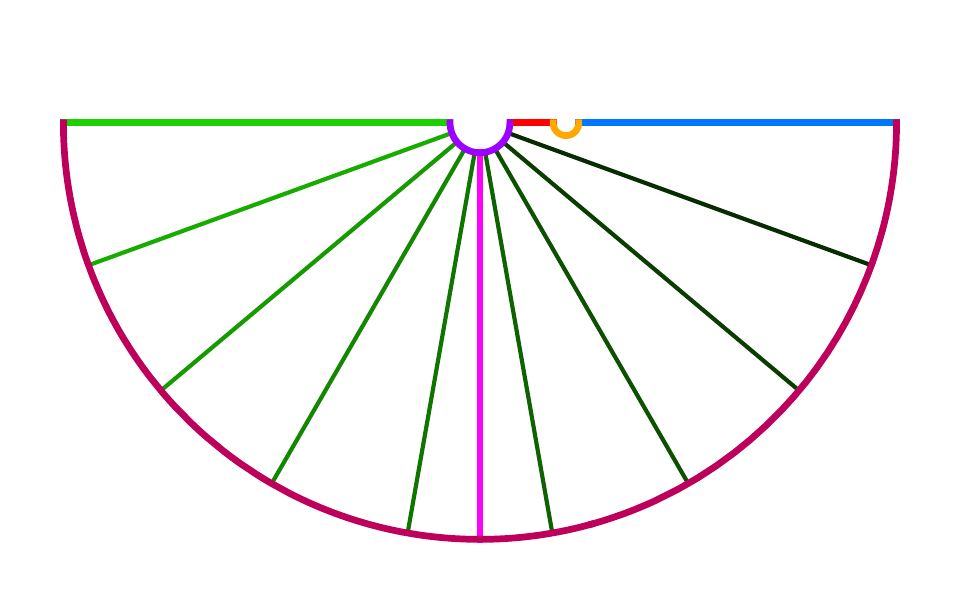}}
    node[\CColor, xshift=7.3cm, yshift=0.6cm] {$C$}
    node[\BColor, xshift=8.2cm, yshift=0.6cm] {$B$}
    node[\AColor, xshift=8.8cm, yshift=0.6cm] {$A$};

  \node (w) at (12, 0)
  {\includegraphics[width=2.6cm]{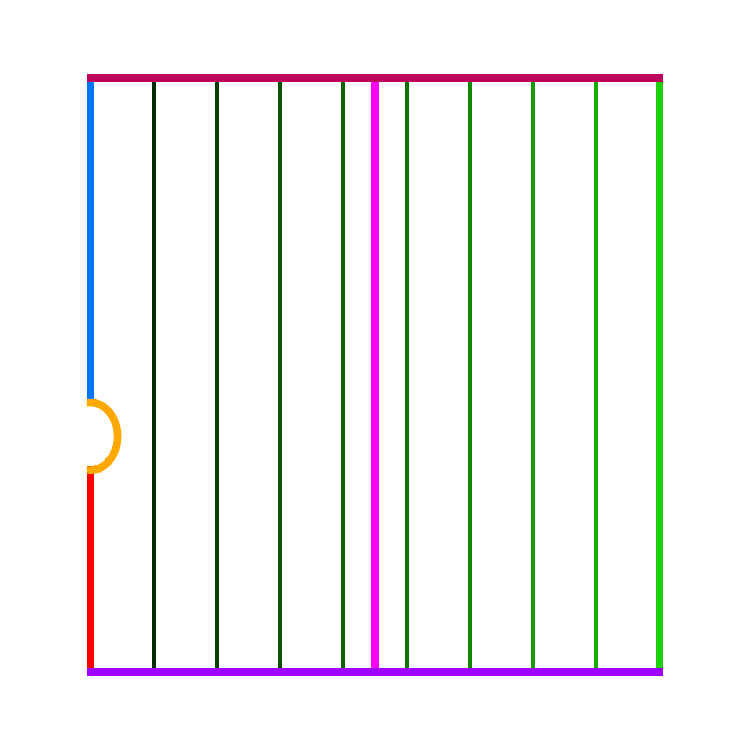}}
    node[\CColor, xshift=1.2cm, yshift=0.0cm] at (w) {$C$}
    node[\BColor, xshift=-1.2cm, yshift=-0.7cm] at (w) {$B$}
    node[\AColor, xshift=-1.2cm, yshift=0.5cm] at (w) {$A$}
    node[\aColor, yshift=-1.2cm] at (w) {$a$}
    node[\bColor, yshift=1.25cm] at (w) {$b$}
    node[\cColor, xshift=-1.2cm, yshift=-0.2cm] at (w) {$c$};

  \node[yshift=-1.5cm] at (z) {$z$};
  \node[yshift=-1.5cm] at (zeta) {$\zeta$};
  \node[yshift=-1.5cm] at (xi) {$\xi$};
  \node[yshift=-1.5cm] at (w) {$w$};

  \draw[->, thick] {(z)++(1.25,0)} -- ++(1,0) node[anchor=south, midway] {$f_1$};
  \draw[->, thick] {(zeta)++(1.5,0)} -- ++(1,0) node[anchor=south, midway] {$f_2$};
  \draw[->, thick] {(xi)++(1.6,0)} -- ++(1,0) node[anchor=south, midway] {$f_3$};

\end{tikzpicture}
    \caption{\small{The entire three-step procedure maps the disk in Fig.~\ref{3CONNECTED} in the $z$ coordinate to the square in the $w$ coordinate.}}
    \label{transforms}
\end{figure}

Let us now explain the sequence of conformal transformations involved in the construction. The entire procedure is summarized in Fig.~\ref{transforms}. Specifically, exploiting the rotational symmetry of the original setup, we place the boundaries of $C$ on the left so that they are symmetric with respect to the horizontal axis. Then we apply a conformal transformation that maps Fig.~\ref{3CONNECTED} to the lower half-plane,
\be
\zeta = f_1(z)=i\frac{z+r}{z-r}
\ee
where $r$ denotes the radius of the disk for the CFT state-preparation manifold. 

The regulators can be introduced either in the $z$ coordinate or in the $\zeta$ coordinate. For simplicity, we choose conformal boundaries of equal size $\epsilon$ in the $\zeta$ coordinate. This is merely a choice of regulator; one could equally well impose equal-size regulators in the $z$ coordinate. While this alters the final value of the reflected entropy—since it is divergent and depends explicitly on the regulator—the derivation itself proceeds in the same way if we make another choice.

Denote the centers of the $a$ and $b$ boundaries in the $\zeta$ coordinate as $(\pm l/2,0)$, and the center of the $c$ boundary as $(x_0,0)$, with the radii of the conformal boundaries taken to be $\epsilon$. Next, we apply the transformation
\be
\xi = f_2(\zeta) = \frac{\zeta - \sqrt{\left(\tfrac{l}{2}\right)^2 - \epsilon^2}}{\zeta+\sqrt{\left(\tfrac{l}{2}\right)^2 - \epsilon^2} }
\ee
which maps $\zeta$ to the lower half-plane, with the two conformal boundaries becoming concentric semi-circles. This mapping is obtained by requiring the four endpoints to the boundaries of two concentric circles \cite{Geng:2025efs}. This differs slightly from the construction in \cite{Cardy:2016fqc}, where the semi-circles become concentric only at leading order in $\epsilon$. In contrast, our mapping is exact for arbitrary $\epsilon$. In the $\xi$ coordinate, the modular flow for $C$ is generated by angular evolution.

Last, we apply the transformation 
\be \label{logmap}
w = f_3(\xi) = i \ln(\xi)
\ee
to map the half-annulus into a strip of horizontal length $\pi$. Since the reduced density matrix is obtained by gluing this geometry to its orientation reversal along $AB$, and the hole in $AB$ can be shrunk, the modular evolution in this coordinate reduces to the standard BCFT time evolution for region $C$. More precisely, the length of $C$ in the $w$ coordinate is given by 
\be
W=\ln \left(f_2(-\frac{l}{2}-\epsilon) \right)-\ln \left(f_2(\frac{l}{2}+\epsilon) \right)=\ln\left(\frac{l \left(\sqrt{l^2-4 \epsilon ^2}+l\right)-2 \epsilon ^2}{2 \epsilon ^2} \right)=2\ln \left(\frac{l}{\epsilon} \right)+\mathcal{O}\left(\frac{\epsilon^2}{l^2} \right) ~.
\ee
Consequently, the normalization factor in \eqref{modhambcft} is $\mathcal{N}=\tfrac{\pi}{W}$, using the fact that the conventional BCFT Hamiltonian is defined on an interval of length $\pi$. The modular Hamiltonian then takes the form
\be
H_{\text{mod},C} = \frac{\pi}{W} H_{\text{BCFT}}
\ee
\begin{figure}
\centering
\begin{tikzpicture}[line cap=rect]
    \node at (-2.5cm,0) {$\rho_{AB} = $};
 
    \def\h{1.75}
    \def\w{1.32}
    \def\delta{0.1}
    \def\tick{0.08}

    \node (strip1) at (0, 0)
    {\includegraphics[width=3.5 cm, trim=0 0 0 0]
    {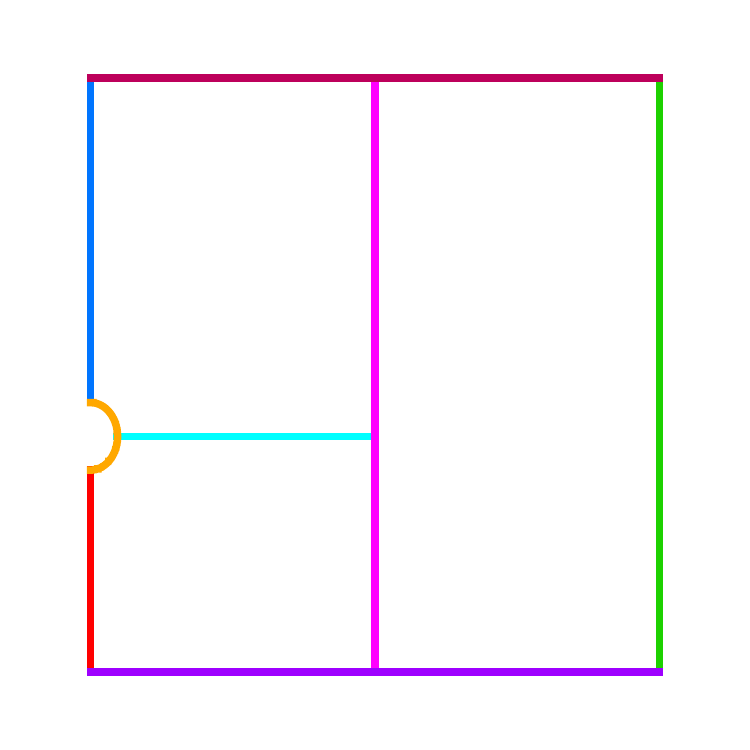}}
    node[\BColor, xshift=-1.6cm, yshift=-0.9cm] at (strip1) {$B$}
    node[\AColor, xshift=-1.6cm, yshift=0.6cm] at (strip1) {$A$}
    node[\aColor, yshift=-1.6cm] at (strip1) {$a$}
    node[\bColor, yshift=1.65cm] at (strip1) {$b$}
    node[\cColor, xshift=-1.6cm, yshift=-0.3cm] at (strip1) {$c$};

  \draw[\CColor,thick]   (\w-\tick,\tick/2) -- (\w+\tick,\tick);
  \draw[\CColor,thick]   (\w-\tick,-\tick/2) -- (\w+\tick,0);

  \draw[black, <->, thick] (-\w,\h+0.25) -- (\w, \h+0.25)
    node[midway, fill=white] {$\pi$}; 

\begin{scope}[shift={(3.5,0)}]
   \node (strip2) at (0, 0)
    {\scalebox{-1}[1]{\includegraphics[width=3.5 cm]
    {./tripartite/state_strip_with_halfway.pdf}}}
    node[\mygreen, xshift=-1.75cm, yshift=0cm] at (strip2) {$C$}
    node[\myred, xshift=1.6cm, yshift=-0.9cm] at (strip2) {$B$}
    node[\myblue, xshift=1.6cm, yshift=0.6cm] at (strip2) {$A$}
    node[\mypurple, yshift=-1.6cm] at (strip2) {$a$}
    node[\mypurple, yshift=1.65cm] at (strip2) {$b$}
    node[\myorange, xshift=1.6cm, yshift=-0.3cm] at (strip2) {$c$};
  \draw[\mygreen,thick]   (-\w-\tick,\tick/2) -- (-\w+\tick,\tick);
  \draw[\mygreen,thick]   (-\w-\tick,-\tick/2) -- (-\w+\tick,0);
  \draw[black, <->, thick] (-\w,\h+0.25) -- (\w, \h+0.25)
    node[midway, fill=white] {$\pi$};
\end{scope}
\end{tikzpicture}

    \caption{\small{Path integral representation of the density matrix $\rho_{AB}$ in the $w$ coordinate.}}
    \label{deevolution1}
\end{figure}

\begin{figure}
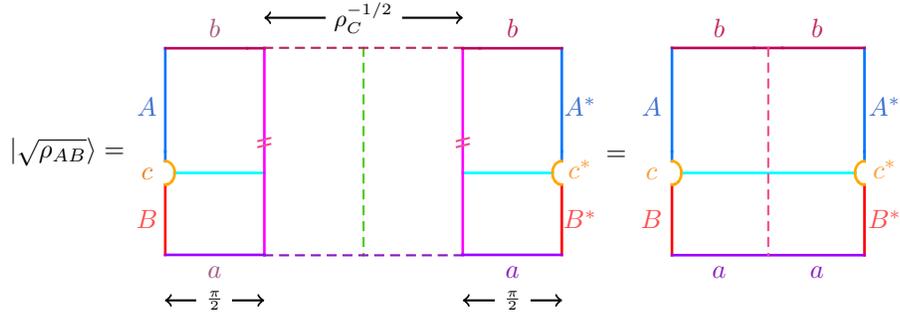

\centering
\begin{tikzpicture}[line cap=rect]
  \def\shift{4}
    \def\h{1.38}
    \def\w{1.32}
    \def\delta{0.1}
    \def\tick{0.08}
    
  \node at (-1.3,0) {$\ket{\sqrt{\rho_{AB}}} = $};

\begin{scope}
  \node[xshift=0.46cm] (strip1) at (0,0) {\includegraphics[width=1.75cm, trim={0 0 6.3cm 0},clip]{./tripartite/state_strip_with_halfway.pdf}}
    node[\myred, xshift=-0.7cm, yshift=-0.9cm] at (strip1) {$B$}
    node[\myblue, xshift=-0.7cm, yshift=0.6cm] at (strip1) {$A$}
    node[\mypurple, xshift=0.2cm, yshift=-1.6cm] at (strip1) {$a$}
    node[\mypurple, xshift=0.2cm, yshift=1.65cm] at (strip1) {$b$}
    node[\myorange, xshift=-0.7cm, yshift=-0.3cm] at (strip1) {$c$};

  \draw[\CColor, thick, dashed]   (2*\w,\h) -- ++(0,-2*\h);
  \draw[\bColor, thick, dashed] (\w,\h) -- ++(2*\w,0);
  \draw[\aColor, thick, dashed] (\w,-\h) -- ++(2*\w,0);
  
  \draw[\minSurfaceColor, thick]  (\w-\tick,\tick/2+0.1) -- (\w+\tick,\tick+0.1);
  \draw[\minSurfaceColor, thick]   (\w-\tick,-\tick/2+0.1) -- (\w+\tick,0.1);
  
  \draw[black, <->, thick] (0,-\h-0.6) -- ++(\w, 0)
    node[midway, fill=white, font=\tiny] {$\frac{\pi}{2}$};
\end{scope}

\begin{scope}[shift={(4*\w,0)}]
    \node[xshift=-0.46cm] (strip2) at (0,0) {\scalebox{-1}[1]
    {\includegraphics[width=1.75cm, trim={0 0 6.3cm 0},clip]{./tripartite/state_strip_with_halfway.pdf}}}
    node[\BColor, xshift=0.7cm, yshift=-0.9cm] at (strip2) {$B^*$}
    node[\AColor, xshift=0.7cm, yshift=0.6cm] at (strip2) {$A^*$}
    node[\aColor, xshift=-0.2cm, yshift=-1.6cm] at (strip2) {$a$}
    node[\bColor, xshift=-0.2cm, yshift=1.65cm] at (strip2) {$b$}
    node[\cColor, xshift=0.7cm, yshift=-0.23cm] at (strip2) {$c^*$};

    \draw[\minSurfaceColor, thick]  (-\w-\tick,\tick/2+0.1) -- ++(2*\tick,\tick/2);
    \draw[\minSurfaceColor, thick]   (-\w-\tick,-\tick/2+0.1) -- ++(2*\tick,\tick/2);

    \draw[black, <->, thick] (0,-\h-0.6) -- ++(-\w, 0)
    node[midway, fill=white, font=\tiny] {$\frac{\pi}{2}$};
\end{scope}

  \draw[black, <->, thick] (\w,\h+0.4) -- ++(2*\w, 0)
    node[midway, fill=white, font=\small] {$\rho_C^{-1/2}$};

\begin{scope}[xshift=6.7cm]
  \node at (-0.7,-0.05) {=};

    \node[xshift=0.47cm] (strip1) at (0,0) 
    {\includegraphics[width=1.7cm, trim={0 0 6.5cm 0},clip]{./tripartite/state_strip_with_halfway.pdf}}
    node[\BColor, xshift=-0.7cm, yshift=-0.9cm] at (strip1) {$B$}
    node[\AColor, xshift=-0.7cm, yshift=0.6cm] at (strip1) {$A$}
    node[\aColor, xshift=0.2cm, yshift=-1.6cm] at (strip1) {$a$}
    node[\bColor, xshift=0.2cm, yshift=1.65cm] at (strip1) {$b$}
    node[\cColor, xshift=-0.7cm, yshift=-0.3cm] at (strip1) {$c$};
    
    \node[xshift=-0.47cm] (strip2) at (2*\w,0) {\scalebox{-1}[1]
    {\includegraphics[width=1.7cm, trim={0 0 6.5cm 0},clip]{./tripartite/state_strip_with_halfway.pdf}}}
    node[\BColor, xshift=0.7cm, yshift=-0.9cm] at (strip2) {$B^*$}
    node[\AColor, xshift=0.7cm, yshift=0.6cm] at (strip2) {$A^*$}
    node[\aColor, xshift=-0.2cm, yshift=-1.6cm] at (strip2) {$a$}
    node[\bColor, xshift=-0.2cm, yshift=1.65cm] at (strip2) {$b$}
    node[\cColor, xshift=0.7cm, yshift=-0.23cm] at (strip2) {$c^*$};

    \draw[thick, \minSurfaceColor, dashed] (\w, \h) -- ++(0, -2*\h); 

\end{scope}
\end{tikzpicture}
    \caption{\small{Path integral representation of the canonical purification state $\ket{\rho_{AB}}$ in the $w$ coordinate.}}
    \label{deevolution2}
\end{figure}

\begin{figure}
    \centering
\begin{tikzpicture}

  \node (strip) at (0,0)
  {\includegraphics[width=3.5cm]{./tripartite/state_strip_with_halfway.pdf}}
    node[\CColor, xshift=1.6cm, yshift=0.0cm] at (strip) {$C$}
    node[\BColor, xshift=-1.6cm, yshift=-0.9cm] at (strip) {$B$}
    node[\AColor, xshift=-1.6cm, yshift=0.6cm] at (strip) {$A$}
    node[\aColor, yshift=-1.6cm] at (strip) {$a$}
    node[\bColor, yshift=1.65cm] at (strip) {$b$}
    node[\cColor, xshift=-1.6cm, yshift=-0.3cm] at (strip) {$c$};

  \node (circle) at (8, 0)
  {\includegraphics[width=4.5cm, trim=0 0 0 0]
    {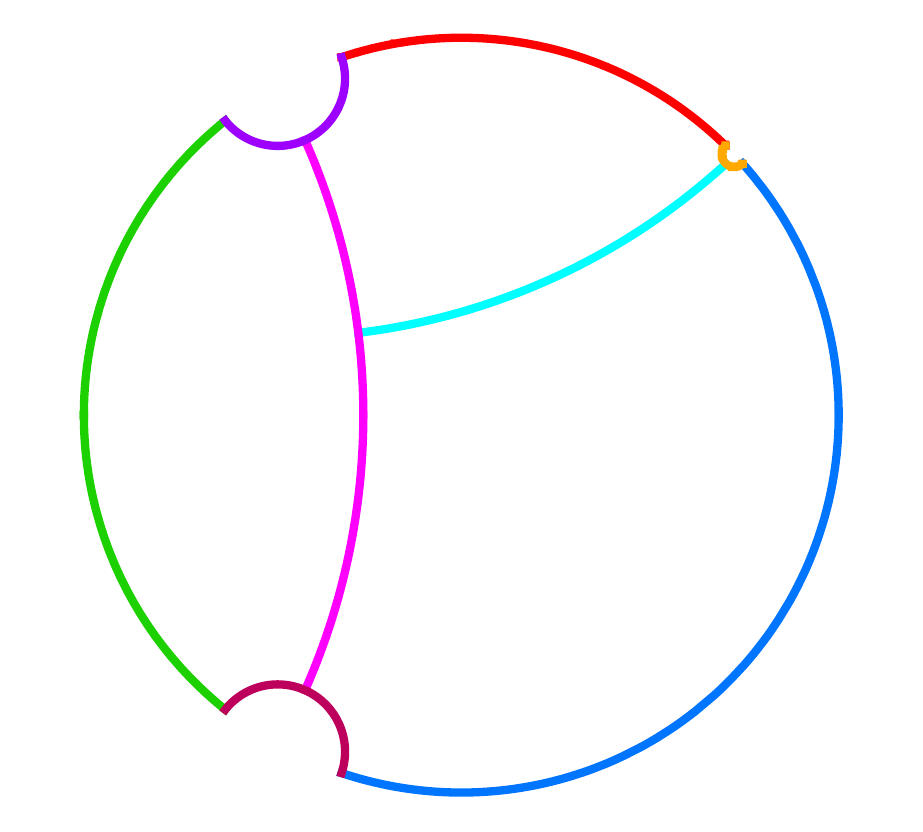}}
    node[\AColor, xshift=1.6cm, yshift=-1.2cm] at (circle) {$A$}
    node[\BColor, xshift=0.5cm, yshift=2cm] at (circle) {$B$}
    node[\CColor, xshift=-2.1cm, yshift=0cm] at (circle) {$C$}
    node[\aColor, xshift=-0.9cm, yshift=1.6cm] at (circle) {$a$}
    node[\bColor, xshift=-0.9cm, yshift=-1.6cm] at (circle) {$b$}
    node[\cColor, xshift=1.45cm, yshift=1.45cm] at (circle) {$c$};

  \draw[->, thick] {(strip)++(2.25,0)} -- ++(3,0) node[above, midway]
    {{$f_1^{-1} \circ f_2^{-1} \circ f_3^{-1}$}};

\end{tikzpicture}
    \caption{\small{After the inverse conformal transformations, the pink curve and the cyan curve get mapped to the positions corresponding to the RT surface and the entanglement wedge cross section on the Cauchy slice in Fig.~\ref{3CONNECTED}.}}
    \label{transforms1}
\end{figure}

The density matrix and the canonical purification state in the $w$ coordinate are illustrated in Fig.~\ref{deevolution1} and Fig.~\ref{deevolution2}. Through the inverse transformations, the modular flow in each coordinate is visualized as a series of green curves, shaded from light to dark in Fig.~\ref{transforms}. At this stage, it is evident that under the hyperbolic slicing \eqref{slicing}, the pink curve at $\tau \to \infty$ maps to the position of the RT surface at $\tau=0$, while the cyan curve maps to the entanglement wedge cross section shown in Fig.~\ref{transforms1}.

\begin{figure}
    \centering
\includegraphics[width=0.2\linewidth]{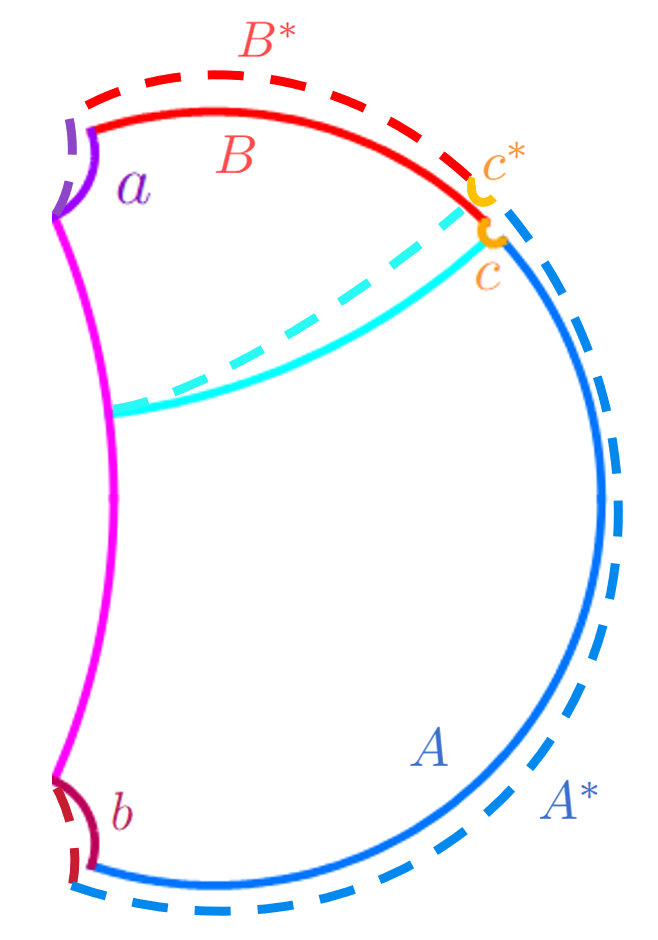}
    \caption{\small{Path integral representation of the canonical purification state $\ket{\rho_{AB}}$ in the $z$ coordinate, with the second copy placed in the back.}}
    \label{taco}
\end{figure}

To summarize, the canonical purification $\ket{\sqrt{\rho_{AB}}}$ can be prepared by the Euclidean BCFT path integral on the doubled manifold shown in Fig.~\ref{taco}. Importantly, we cut and glue along the location corresponding to the minimal surface on the dual 2D hyperbolic geometry. We will rely crucially on this in Sec.~\ref{sec3.3} to relate the emergent geometries in Fig.~\ref{3CONNECTED} and Fig.~\ref{taco}, and to connect the reflected entropy to the entanglement wedge cross section. An upshot is that cutting a 2D hyperbolic manifold along a minimal surface and gluing it to its CPT conjugate yields another smooth 2D hyperbolic geometry. This is possible because minimal surfaces have vanishing extrinsic curvature, as noted by Engelhardt and Wall \cite{Engelhardt:2017aux, Engelhardt:2018kcs}.

\subsubsection{Coarse-Graining BCFT Data in Reflected Entropy} \label{EEsec}
Having constructed the BCFT path-integral representation of the canonical purification \eqref{purificationCFT}, we now proceed to derive the RT formula for the entanglement entropy between $AA^*$ and $BB^*$ (\ref{reflected def}), using the averaging-over-OPE-coefficients procedure of \cite{Bao:2025plr, Geng:2025efs}.

First, as illustrated in Fig.~\ref{deevolution2}, the canonical purification state can be expresssed in terms of the OPE block decomposition in two equivalent ways. One option is to cut the path integral open along the dashed pink lines and insert a complete basis of states; in this case, the two resulting triangles are associated with the state on $A\cup B\cup \text{the pink line}$, and that on $A^*\cup B^*\cup \text{the pink line}$, respectively. Alternatively, one may cut the manifold open along the cyan edge, in which case the two triangles are associated with the states on $A\cup A^*\cup \text{the cyan line}$ and $B\cup B^*\cup \text{the cyan line}$, respectively.

Explicitly, these two equivalent ways give the following expressions:
\begin{equation}
\begin{aligned}
\ket{\sqrt{\rho_{AB}}}&=\sum_{\text{primaries}}  \frac{C^{abc}_{ijk}   C^{*abc}_{lmk}}{\sqrt{g_a g_b}}  \;\mathcal{B}\left[
 \vcenter{\hbox{\begin{tikzpicture}[scale=0.6]
 \draw[<->,thick,black!!40] (0,1) arc (90:-90:1)
    node[pos=0, left] {$i$}
    node[pos=1, left] {$j$};
  \draw[<->,thick,black!!40] (4,1) arc (90:270:1)
  node[pos=0, right] {$l$}
  node[pos=1, right] {$m$};
  \draw[-, thick, black!!40] (1,0) -- ++(2,0)
  node[pos=0.5, above] {$k$};

 \end{tikzpicture}}} \right]\ket{i} \ket{j} \ket{l} \ket{m} \\
 &= 
 \sum_{\text{primaries}}  \frac{C^{cbc}_{inl}   C^{*cac}_{jnm}}{g_c}
 \;\mathcal{B}\left[\vcenter{\hbox{\begin{tikzpicture}[scale=0.6]
	\draw[thick, <->] (1,0) arc (0:-180:1)
    node[pos=0, right] {$l$}
    node[pos=1, left] {$i$};

	\draw[thick] (0,-1) -- ++(0,-1.5)
    node[pos=0.5, right] {$n$};

	\draw[thick, <->] (1,-3.5) arc (0:180:1)
    node[pos=0, right] {$m$}
    node[pos=1, left] {$j$};

 \end{tikzpicture}}}\right]\ket{i} \ket{j} \ket{l} \ket{m}
 ~.
 \end{aligned} \label{OPEstate}
\end{equation}
Notice that the OPE block $\mathcal{B}[\cdots]$ in the first line of \eqref{OPEstate} already incorporates the effect of $\rho_C^{-1/2}$ in (\ref{purificationCFT}), which corresponds to the backward evolution by the BCFT Hamiltonian. It involves less Euclidean evolution than directly gluing $\ket{\Psi}^{abc}_{ABC}$ to its conjugate along $C$, which would instead produce $\ket{\rho_{AB}}$ \eqref{densitymatrix}. In the second line of \eqref{OPEstate}, we invoked crossing symmetry to express the state in the alternative channel of the OPE block decomposition. This step is crucial for extracting the entanglement wedge cross section within the framework of averaging over OPE coefficients \cite{Bao:2025plr, Geng:2025efs}. Roughly speaking, this is because the crossed channel explicitly includes the exchange primary $n$ which propagates through the cyan curve.

We now show that, upon averaging over OPE coefficients in the replica partition function, the entropy of $AA^*$ is determined by the saddle point of the primary integral in the crossed channel. The latter, through its connection to Liouville theory, yields the minimal length of the geodesics on the emergent 2D hyperbolic geometry that are homologous to $AA^*$. 

First, we approximate the replica partition function associated with the reduced density matrix $\rho_{AA^*}$ by the coarse-grained one $\overline{Z_{AA^*,n}}$. This is implemented by performing an ensemble average over the heavy spectrum, which effectively averages over the BCFT data \cite{Hung:2025vgs, Wang:2025bcx, Geng:2025efs}. More explicitly, we have

\be
\label{ZAAn}
Z_{A A^*,n}=\vcenter{\hbox{
\begin{tikzpicture}[scale=0.6, rotate=270]
	\draw[thick] (0,1) arc (90:-90:1)
    node[pos=0, left] {$l$}
    node[pos=1, right] {$i$};
	\draw[thick] (1,0) -- ++(2,0)
    node[pos=0.5, right] {$k$};
  \draw[thick] (4,0) circle (1)
	  node[left] at ++(0,1) {$n$}
	  node[right] at ++(0,-1) {$m$};
	\draw[thick] (5,0) -- ++(2,0)
    node[pos=0.5, right] {$o$};
	\draw[thick] (8,0) circle (1)
	  node[left] at ++(0,1) {$q$}
	  node[right] at ++(0,-1) {$p$};
  \draw[thick] (9,0) -- ++(1,0)
    node[pos=0.6, right] {$r$};

  \node[right] at (10.5,-0.2) {$\vdots$};
  \draw[thick] (11.5,0) -- ++(1,0)
    node[pos=0.3, right] {$u$};
  \draw[thick] (13.5,1) arc (90:270:1)
    node[pos=0, left] {$l$}
    node[pos=1, right] {$i$};

  \def\w{1.4}
  \def\h{2}

  \draw[dashed,thick] (0,-1) -- ++(0,-0.5) -- ++(13.5,0) -- ++(0,0.5);
  \draw[dashed,thick] (0,1) -- ++(0,0.5) -- ++(13.5,0) -- ++(0,-0.5);
\end{tikzpicture}}} \quad  \xrightarrow{\;\text{Coarse-Graining}\;}  \quad \overline{Z_{A A^*,n}}=
\vcenter{\hbox{
\begin{tikzpicture}[scale=0.6, rotate=270]
	\draw[thick] (0,1) arc (90:-90:1)
    node[pos=0, left] {$l$}
    node[pos=1, right] {$i$};
	\draw[thick] (1,0) -- ++(2,0)
    node[pos=0.5, right] {$k$};
  \draw[thick] (4,0) circle (1)
	  node[left] at ++(0,1) {$n$}
	  node[right] at ++(0,-1) {$m$};
	\draw[thick] (5,0) -- ++(2,0)
    node[pos=0.5, right] {$o$};
	\draw[thick] (8,0) circle (1)
	  node[left] at ++(0,1) {$q$}
	  node[right] at ++(0,-1) {$p$};
  \draw[thick] (9,0) -- ++(1,0)
    node[pos=0.6, right] {$r$};

  \node[right] at (10.5,-0.2) {$\vdots$};
  \draw[thick] (11.5,0) -- ++(1,0)
    node[pos=0.3, right] {$u$};
  \draw[thick] (13.5,1) arc (90:270:1)
    node[pos=0, left] {$l$}
    node[pos=1, right] {$i$};

  \def\w{1.4}
  \def\h{2}
  \draw[thick, \myblue, rounded corners=2pt,dashed]
    (0,-\h) -- (\w,-\h) -- (\w,\h) -- (0,\h);
  \draw[thick, \myred, rounded corners=2pt,dashed]
    (4-\w, -\h) rectangle ++(2*\w,2*\h);
  \draw[thick, \myblue, rounded corners=2pt,dashed]
    (8-\w,-2) rectangle ++(2*\w,2*\h);
  \draw[thick, \myblue, rounded corners=2pt,dashed]
    (13.5,-\h) -- ++(-\w,0) -- ++(0,2*\h) -- ++(\w, 0);

  \draw[dashed,thick] (0,-1) -- ++(0,-0.5) -- ++(13.5,0) -- ++(0,0.5);
  \draw[dashed,thick] (0,1) -- ++(0,0.5) -- ++(13.5,0) -- ++(0,-0.5);
\end{tikzpicture}}}\ ,
\ee
where OPE coefficients are again omitted for notational simplicity. The $n$ red and $n$ blue boxes, which carry the propagations of the $AA^*$ and $BB^*$ states, respectively, mark the regions where the averaging is performed. 

Within each box, a Gaussian average\footnote{While non-Gaussian moments can in principle yield dominant contributions in certain settings \cite{Belin:2021ryy, Anous:2021caj, Belin:2023efa}, in the vacuum state with the simplest triangulation employed here, Gaussian moments are sufficient: they reproduce the expected dual geometry, entanglement entropy, Rényi entropies, and reflected entropy. Accordingly, we restrict our attention to Gaussian moments, leaving a better understanding behind their general dominance for future work.} over the OPE coefficients is performed using the formula
\begin{equation}   \overline{C_{ijk}^{cab}C^{*fde}_{lmn}}=\delta_{cf}\delta_{ad}\delta_{be}C_{0}(P_{i},P_{j},P_{k}) (\delta_{il}\delta_{jm}\delta_{kn}+\text{permutations})\,,\label{eq:BCFTC}
\end{equation}
where $C_0(P_i,P_j,P_k)$ denotes the universal OPE coefficient \cite{Collier:2019weq}. It is related to the Liouville DOZZ structure constants \cite{Dorn:1994xn, Zamolodchikov:1995aa} via
\begin{equation} \label{universalZZ}
    C_{0}(P_{i},P_{j},P_{k})=\frac{\hat{C}_{\text{DOZZ}}(P_{i},P_{j},P_{k})}{\sqrt{\rho_{0}(P_{i})\rho_{0}(P_{j})\rho_{0}(P_{k})}}~,
\end{equation}
where
\begin{equation}
\begin{aligned}
    \rho_{0}(P)=4\sqrt{2}\sinh2\pi Pb\sinh2\pi Pb^{-1}\,,
    \quad &c=1+6(b+b^{-1})^{2}~
    \end{aligned}
\end{equation}
is the Cardy density of states \cite{Cardy:1986ie} associated with conformal dimensions $h=\tfrac{c-1}{24}+P^{2}$. This averaging prescription is motivated by the universal statistics of heavy states implied by BCFT bootstrap constraints \cite{Numasawa:2022cni, Hung:2025vgs, Kusuki:2021gpt, Wang:2025bcx, Jafferis:2025yxt}.\footnote{It should also be noted that the dominance of heavy BCFT states is a direct consequence of the tiny regulators introduced, which effectively drive the system into a high-temperature regime. Meanwhile, the reliability of replacing microscopic data with ensemble averages reflects the expectation that black hole microstates exhibit chaotic dynamics.}

In principle, we need to sum over all possible averaging patterns, which correspond to different bulk saddle points through the connection to the Virasoro TQFT \cite{Collier:2023fwi, Belin:2023efa, Jafferis:2025yxt}. The dominant contribution is then selected by choosing the averaging pattern, or equivalently the bulk saddle, that dominates. The candidate leading contributions in the Gaussian average arise from the configuration that imposes the fewest constraints coming from the delta functions in \eqref{eq:BCFTC}. In the present setup, this corresponds to identifying the primary indices associated to the vertical lines in (\ref{ZAAn}), e.g. $k,o,r,u$. Note that the averaging pattern here explicitly preserves the $Z_n$ replica symmetry \cite{Lewkowycz:2013nqa}. 

We also replace the sum over BCFT primary sectors by an integral weighted by the BCFT version of the Cardy density of states \cite{Cardy:1986ie, Hikida:2018khg, Numasawa:2022cni}
\begin{equation} \label{bcftcardy}
\begin{aligned}
    \rho_{ab}(P)=g_{a}g_{b}\rho_{0}(P)~,
    \end{aligned}
\end{equation}
for all $P>0$. To perform the integrals over the primaries $m,n,p,q,i,l,\ldots$, we use the relation between universal BCFT data and the crossing kernels,
\begin{equation} \label{fusionkernel} F_{\mathbb{1}, P_k} \begin{pmatrix} P_i & P_j \\ P_i & P_j \end{pmatrix} =S_{\mathbb{1} P_k} C_0(P_i,P_j,P_k)\,, \quad S_{\mathbb{1} P_k}=\rho_0(P_k) \end{equation}
to convert them into identity-module contributions in certain channels. More explicitly, we have
\be \label{firstphaseZn}
\begin{aligned}
&\int_{0}^\infty d P_m d P_n \rho_0(P_m) \rho_0(P_n) {C}_{0}(P_i,P_m,P_n) \vcenter{\hbox{
	\begin{tikzpicture}[scale=0.75]
	\draw[thick] (0,0) circle (1);
	\draw[thick] (-1,0) -- (-2,0);
	\node[above] at (-2,0) {$k$};
	\node[above] at (0,1) {$n$};
	\node[below] at (0,-1) {$m$};
	\draw[thick] (1,0) -- (2,0);
	\node[above] at (2,0) {$k$};
    \end{tikzpicture}
	}}\\
 &   = \int_{0}^\infty d P_n  \rho_0(P_n)
\vcenter{\hbox{
	\begin{tikzpicture}[scale=0.75]
	\draw[thick] (0,1) circle (1);
	\draw[thick] (0,-1+1) -- (0,-2+1);
	\draw[thick] (0,-2+1) -- (-0.866,-2+1);
	\draw[thick] (0,-2+1) -- (0.866,-2+1);
	\node[left] at (0,-3/2+1) {$\mathbb{1}$};
	\node[left] at (-1.2,0+1) {$n$};
	\node[left] at (-0.866,-2+1) {$k$};
	\node[right] at (0.766,-2+1) {$k$};	
	\end{tikzpicture}
	}} = 
\vcenter{\hbox{
	\begin{tikzpicture}[scale=0.75]
	\draw[thick] (0,1) circle (1);
	\draw[thick] (0,-1+1) -- (0,-2+1);
	\draw[thick] (0,-2+1) -- (-0.866,-2+1);
	\draw[thick] (0,-2+1) -- (0.866,-2+1);
	\node[left] at (0,-3/2+1) {$\mathbb{1}$};
	\node[left] at (-1.2,0+1) {$\mathbb{1}'$};
	\node[left] at (-0.866,-2+1) {$k$};
	\node[right] at (0.766,-2+1) {$k$};	
	\end{tikzpicture}
	}}    ~\,,
\end{aligned}
\ee
where the $'$ indicates that the bubble is in the dual channel. Details of similar computation can be found in \cite{Bao:2025plr, Geng:2025efs}. As a result, (\ref{ZAAn}) turns into the expression
\be \label{integral}
\begin{split}
\overline{Z_{AA^*,n}}=&g_c^2 \left(\frac{g_a^2 g_b^2 g_c^4}{g_a g_b g_c^4} \right)^n\int_0^\infty d P_k \rho_0(P_k) \mathcal{F}_{n}(\mathcal{M}_n,P_k)\\
=&g_a^n g_b^n g_c^2 \int_0^\infty d P_k \rho_0(P_k) \mathcal{F}_{n}(\mathcal{M}_n,P_k)\,,    
\end{split}
\ee
where we have defined the conformal block 
\be\label{eq:F1nold}
\begin{aligned}
\mathcal{F}_{n}(\mathcal{M}_{n},P_k)=
\vcenter{\hbox{
	\begin{tikzpicture}[scale=0.75, rotate=270]
	\draw[thick] (0,1) circle (1);
	\draw[thick] (0,-1+1) -- (0,-2+1);
	\draw[thick] (0,-2+1) -- (-1,-2+1);
	\draw[thick] (0,-2+1) -- (1,-2+1);
	\node[above] at (0,-3/2+1) {$\mathbb{1}$};
	\node[above] at (-1.2,0+1) {$\mathbb{1}'$};

  \draw[thick] (0+3,1) circle (1);
	\draw[thick] (0+3,-1+1) -- (0+3,-2+1);
	\draw[thick] (0+3,-2+1) -- (-2+3,-2+1);
	\draw[thick] (0+3,-2+1) -- (2+3,-2+1);
	\node[above] at (0+3,-3/2+1) {$\mathbb{1}$};
	\node[above] at (-1.2+3.2,0+1) {$\mathbb{1}'$};
  \node at (5.5,0-0.3) {$\vdots$};

  \draw[thick] (0+3+3+2,1) circle (1);
	\draw[thick] (0+3+3+2,-1+1) -- (0+3+3+2,-2+1);
	\draw[thick] (0+3+3+2,-2+1) -- (-2+3+3+2,-2+1);
	\draw[thick] (0+3+3+2,-2+1) -- (2+3+3+1,-2+1);
	\node[above] at (0+3+3+2,-3/2+1) {$\mathbb{1}$};
	\node[above] at (-1.2+3.2+3+2,0+1) {$\mathbb{1}'$};

  \draw [dashed] (-1,-1) -- ++(0,-1.5) -- ++(10, 0) -- ++(0, 1.5);
  \node at (4.5,-1.75) {$k$};
	\end{tikzpicture}~,
	}} 
\end{aligned}
\ee
with $\mathcal{M}_n$ denoting the moduli of the $n$-th replica partition function.

\begin{figure}
    \centering
\includegraphics[width=0.7\linewidth]{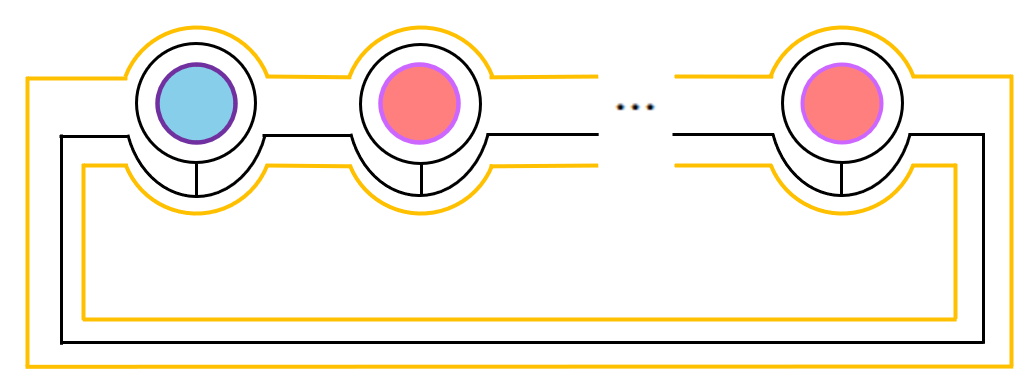}
    \caption{\small{In the bulk dual each boundary condition is capped off by a disk. This dictates that the boundary conditions $a$ and $b$ are each associated with $n$ such disks, while the condition $c$ (highlighted in orange) is associated with two disks (not explicitly shown).}}
    \label{topology}
\end{figure}

As pointed out in \cite{Geng:2025efs}, our careful treatment of the BCFT normalization factors provides a consistency check between the exponents of the $g$-factors and the topology of the emergent dual gravitational solution. For instance, capping off the boundary conditions $a$ and $b$–which correspond to the loops in (\ref{eq:F1nold})—results in $2n$ disks, as illustrated in Fig.~\ref{topology}. These contribute a total Euler characteristic of $\chi = 2n$, with $n$ coming from the disks associated with $a$ and another $n$ from those associated with $b$, which is consistent with the power counting of the $g_a$ and $g_b$ factors in (\ref{integral}). Additionally, there are two disconnected disks capping the boundary condition $c$, contributing an extra $\chi=2$ that accounts for the power of the $g_c$ in (\ref{integral}).

In the large-$c$ limit, the conformal block (\ref{eq:F1nold}) exponentiates  \cite{Zamolodchikov:1987avt}, 
\be
\mathcal{F}_{n}(\mathcal{M}_n,P_k)=e^{-\frac{c}{6} f(\mathcal{M}_n,\gamma_k)}\,,
\ee
where we have substituted $P_{k}=\sqrt{\frac{c}{24}}\gamma_{k}$.
The density of states in this limit is given by \cite{Cardy:1986ie}
\be
\rho_0(P) \to e^{\frac{\pi c \gamma}{6}}~.
\ee
Besides, applying the saddle-point approximation in the large-$c$ limit to the integral \eqref{integral} yields
\be \label{saddleZn}
\overline{Z_{AA^*,n}} \approx g_a^n g_b^n g_c^2 e^{\frac{\pi c \gamma^{*}_n}{6}-\frac{c}{6}f(\mathcal{M}_n,\gamma^{*}_n)  }~,
\ee
where the saddle point $\gamma_n^*$ is determined by the equation
\be \label{saddleforgamman}
\partial_{\gamma}\left(\frac{\pi c \gamma}{6}-\frac{c}{6}f(\mathcal{M}_n,\gamma) \right)\Bigr\rvert_{\gamma=\gamma^{*}_{n}}=\frac{\pi c }{6}-\frac{c }{6}\partial_{\gamma} f(\mathcal{M}_n,\gamma) \Bigr\rvert_{\gamma=\gamma^{*}_{n}}=0~.
\ee

Finally, a useful formula of the von Neumann entropy (\ref{RT}) is given by
\be \label{EEdefinition}
\begin{aligned}
S(A)=-\frac{\partial}{\partial n}\ln \left(\frac{Z_{A,n}}{(Z_{A,1})^n} \right) \Bigr\rvert_{n=1}~.
\end{aligned}
\ee
Plugging \eqref{saddleZn} into it gives the entropy $S(AA^*)$. The result can be further simplified by using the saddle-point equation \eqref{saddleforgamman}, together with the observation of \cite{Bao:2025plr, Geng:2025efs} that, for replica-symmetric configurations such as the one considered above, the function $f(\mathcal{M}_n,\gamma_k)$ is linear in $n$, i.e, $f(\mathcal{M}_n,\gamma_k) = n f(\mathcal{M}_1,\gamma_k)$. These two ingredients both have their dual statements in the holographic dual derivation of the RT formula \cite{Lewkowycz:2013nqa}. Indeed, we see clear CFT indication that, an elongating direction which is contractible in the bulk dual, is precisely what gives rise to the holographic entanglement entropies \cite{Bao:2025plr, Geng:2025efs}.

After carrying out the algebra analogous to that in Sec.~4.1.2 of \cite{Bao:2025plr}, many terms cancel and we ultimately obtain the simple expression,
\be \label{eeinfirstcase}
\text{RE}(A:B)=S(AA^*)=\frac{c}{6} (\pi \gamma_1^*)+2 \ln g_c=\frac{\pi \gamma_1^*}{4G_N}+2 \ln g_c\ ,
\ee
where $\gamma_{1}^{*}$ is the saddle point for the $P_{k}$ integral in the evaluation of $\overline{Z_{AA^*,1}}$, and we have applied the Brown-Henneaux central charge relation $c=3/2G_N$ \cite{1986CMaPh.104..207B}. The $g$-factor provides a constant contribution associated with the tiny regulator we introduced in Fig.~\ref{3CONNECTED}, which renders finite the otherwise divergent length of the cyan curve in the hyperbolic metric \cite{Ohmori:2014eia, Kusuki:2022ozk, Cardy:2016fqc, Geng:2025efs}. This term also has a clear geometrical interpretation in the bulk dual: the $g$-factor encodes the tension of the dual bulk brane \cite{Takayanagi:2011zk, Fujita:2011fp}, whose backreaction increases the length of the RT surface connecting the branes on the Cauchy slice. This produces an additional constant shift in the entanglement entropy compared with the case of tensionless branes with $g=1$, and the extra length induced by the brane deformation is precisely $\ln g$, as shown in \cite{Geng:2021iyq, Geng:2025efs}.

\subsubsection{Emergence of the Entanglement Wedge Cross Section} \label{sec3.3}
We now show how to identify the first term in (\ref{eeinfirstcase}) with $\text{EW}(A:B)$. To this end, let us first show that the $\pi \gamma_1^*$ is precisely the minimal-length geodesic on the hyperbolic metric with asymptotic boundaries $A,A^*,B,B^*$, through its relation to Liouville theory. 

Applying the averages \eqref{eq:BCFTC}, and performing the integral with the density of states \eqref{bcftcardy}, the averaged norm (\ref{ZAAn}) for $n=1$ is given by
\be
\label{ZAA1}
\begin{aligned}
&\overline{Z_{AA^*,1}}=\vcenter{\hbox{\begin{tikzpicture}[scale=0.6]
  \begin{scope}[rotate=270]
	    \draw[thick] (0,1) arc (90:-90:1)
        node[pos=0, left] {$l$}
        node[pos=1, right] {$i$};
	    \draw[thick] (1,0) -- ++(2,0)
        node[pos=0.5, right] {$k$};
      \draw[thick] (4,0) circle (1)
	      node[left] at ++(0,1) {$n$}
	      node[right] at ++(0,-1) {$m$};
	    \draw[thick] (5,0) -- ++(2,0)
        node[pos=0.5, right] {$o$};

      \draw[thick] (8,1) arc (90:270:1)
        node[pos=0, left] {$l$}
        node[pos=1, right] {$i$};

      \def\w{1.4}
      \def\h{2}
      \draw[thick, \myblue, rounded corners=2pt,dashed]
        (0,-\h) -- (\w,-\h) -- (\w,\h) -- (0,\h);
      \draw[thick, \myred, rounded corners=2pt,dashed]
        (4-\w, -\h) rectangle ++(2*\w,2*\h);
      \draw[thick, \myblue, rounded corners=2pt,dashed]
        (8,-\h) -- ++(-\w,0) -- ++(0,2*\h) -- ++(\w, 0);

      \draw[dashed,thick] (0,-1) -- ++(0,-0.5) -- ++(8,0) -- ++(0,0.5);
      \draw[dashed,thick] (0,1) -- ++(0,0.5) -- ++(8,0) -- ++(0,-0.5);
  \end{scope}
    \end{tikzpicture}}}  =g_a g_b g_c^2 \int_0^\infty  dP_k dP_i dP_l dP_m dP_n  \rho_0(P_i) \rho_0(P_l) \rho_0(P_m) \\
    &\times \rho_0(P_n) \rho_0(P_k)C_0(P_i,P_l, P_k) C_0(P_m,P_n,P_k)
    \vcenter{\hbox{
    \begin{tikzpicture}[scale=0.6]
        
    \begin{scope}[xshift=5cm, yshift=-4cm]

    \draw[thick] (0,0) circle (1)
	      node[right] at ++(-1,0) {$i$}
	      node[left] at ++(1,0) {$l$};
    \draw[thick] (3,0) circle (1)
	      node[right] at ++(-1,0) {$m$}
	      node[left] at ++(1,0) {$n$};

    \draw[thick] (0,1) arc (180:0:1.5)
      node[pos=0.5, below] {$k$};
    \draw[thick] (0,-1) arc (180:360:1.5)
      node[pos=0.5, above] {$k$};
  \end{scope} 
  \end{tikzpicture}}}\\
  &=g_a g_b g_c^2 \int_0^\infty  dP_k dP_i dP_l dP_m dP_n  \sqrt{\rho_0(P_i)} \sqrt{\rho_0(P_l)} \sqrt{\rho_0(P_m)}  \sqrt{\rho_0(P_n)}\\
  &\hat C_{\text{DOZZ}}(P_i,P_l, P_k) \hat C_{\text{DOZZ}}(P_m,P_n,P_k) \vcenter{\hbox{
    \begin{tikzpicture}[scale=0.6]
        
    \begin{scope}[xshift=5cm, yshift=-4cm]

    \draw[thick] (0,0) circle (1)
	      node[right] at ++(-1,0) {$i$}
	      node[left] at ++(1,0) {$l$};
    \draw[thick] (3,0) circle (1)
	      node[right] at ++(-1,0) {$m$}
	      node[left] at ++(1,0) {$n$};

    \draw[thick] (0,1) arc (180:0:1.5)
      node[pos=0.5, below] {$k$};
    \draw[thick] (0,-1) arc (180:360:1.5)
      node[pos=0.5, above] {$k$};
  \end{scope} 
  \end{tikzpicture}}}\\
&   \vcenter{\hbox{\begin{tikzpicture}[scale=0.6]
  \begin{scope}[xshift=12cm, yshift=-2.5cm]
    \node at (-2, -1.5) {=  $g_a g_b g_c^2$};
    \draw[thick] (-1,0) arc (180:360:1);
    \draw[thick] (2,0) arc (180:360:1);
    \draw[thick] (0,-1) arc (180:360:1.5);
     \node at (2,0.5) {\text{ZZ}};
     \node at (4,0.5) {\text{ZZ}};
     \node at (1,0.5) {\text{ZZ}};
     \node at (-1,0.5) {\text{ZZ}};
      \node at (-0.65,0) {$i$};
        \node at (0.65,0) {$l$};
          \node at (2.4,0) {$m$};
          \node at (3.6,0) {$n$};
\node at (1.5,-2.2) {$k$};
    \def\capW{0.2}
    \draw[thick] (-1-\capW, 0) -- ++(2*\capW, 0);
    \draw[thick] (1-\capW, 0) -- ++(2*\capW, 0);
    \draw[thick] (2-\capW, 0) -- ++(2*\capW, 0);
    \draw[thick] (4-\capW, 0) -- ++(2*\capW, 0);

    \node at (1.5, -3.5) {Liouville with ZZ boundary conditions};
  \end{scope}
\end{tikzpicture}}}\ .
\end{aligned}
\ee
As shown in \cite{Bao:2025plr, Chua:2023ios, Geng:2025efs}, factors of $\rho_0$ combine precisely into a Liouville CFT partition function with ZZ boundary conditions,\footnote{More specifically, the ZZ boundary states corresponding to the ZZ boundary conditions involve a doubling trick, so the manifold associated to the last line contains only half the second last line. In addition, ZZ boundary states are associated to $\sqrt{\rho_0}$, together with the DOZZ formula on the internal legs, we get the Liouville theory with ZZ boundary conditions in the last line.} for which we introduced the symbolic notation on the last line. Recall that the ZZ boundary condition is given by \eqref{zzbc}, which is imposed at the asymptotic infinity for the 2D metric $e^{\phi} dz d\bar{z}$.

\begin{figure}[h]
    \centering
\includegraphics[width=0.8\linewidth]{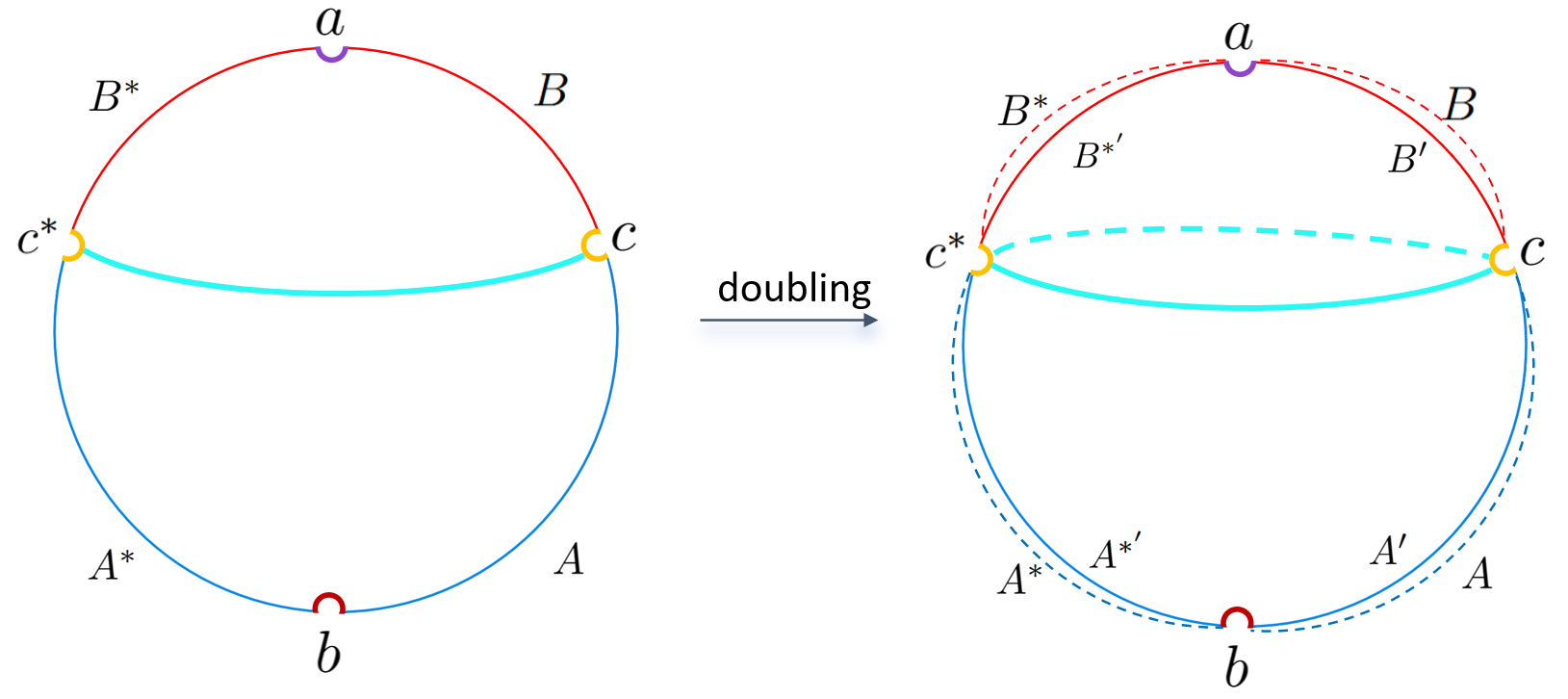}
    \caption{\small{The Liouville CFT is defined on the doubled manifold, constructed by gluing the BCFT manifold to its mirror copy along the conformal boundaries. This construction follows from the BCFT doubling trick.}}
    \label{DOUBLING}
\end{figure}

In this context, the Liouville CFT is defined on the doubled manifold obtained by gluing two copies of the original BCFT manifold with opposite orientation along their conformal boundaries, a procedure known as the “BCFT doubling trick” \cite{Cardy:2004hm, Geng:2025efs}.\footnote{We want to emphasize that this BCFT doubling trick is different from the doubling in the construction of the canonical purification.} As illustrated in Fig.~\ref{DOUBLING}, this construction leads to the multi-boundary black holes\footnote{To avoid confusion with multi-boundary spacetime wormholes, we will refer to these solutions as “multi-boundary black holes.”} analyzed in \cite{Brill:1995jv, Aminneborg:1997pz, Brill:1998pr, Balasubramanian:2014hda}. It is noted in \cite{Geng:2025efs} that this observation establishes a bridge between the RT formula for multiple intervals in the vacuum and that for the multi-boundary black holes considered in \cite{Bao:2025plr}.

Now we use an important result in Liouville CFT \cite{Zamolodchikov:1987avt, Hadasz:2005gk, Harlow:2011ny, Hartman:2013mia, Chandra:2023dgq}, which states that the saddle point $\gamma_1^*$ of the $P_k$ integral in (\ref{ZAA1}) corresponds to the minimal length $L_A$ of the cyan circle (see the right panel of Fig.~\ref{DOUBLING}) on the associated 2D hyperbolic metric. More explicitly, we have the relation
\be
L_A = 2\pi \gamma_1^*\ .
\ee
Since the BCFT manifold constitutes half of the geometry obtained via the doubling trick, $\pi \gamma_1^*$ gives half of $L_A$. It is precisely the length of the entire cyan curve on the left panel of Fig.~\ref{DOUBLING} (or Fig.~\ref{taco}). This establishes the RT formula for the subsystem $AA^*$.

One important feature of our BCFT tensor network framework that we want to point out is that the Renyi entropy does not exhibit a flat spectrum as in random tensor network toy models \cite{Dong:2018seb, Akers:2018fow, Hayden:2016cfa}. Instead, as we will show in upcoming work, it reproduces the full set of Renyi entropies consistent with the bulk duals \cite{Dong:2016fnf}. This is possible due to three crucial facts. First, we include a superposition over all possible primaries propagating along the edges, which can be interpreted as a sum over geometries in the bulk dual. Second, the weights in this superposition are determined by CFT dynamics, thereby encoding the microscopic algebraic data and scaling behaviors of the theory. Third, the bulk entanglement entropy arises from a saddle point of the Virasoro primary integrals. Thus, the \textit{area} is not obtained by a simple counting of cuts in the space we define the tensor network, but rather from the \textit{charge} of the Virasoro algebra that encodes the emergent spacetime structure \cite{Achucarro:1986uwr, Witten:1988hc}. This also guarantees that the outcome is independent of the triangulation chosen for the tensor network, a point emphasized recently in \cite{Akers:2024wab, Chen:2024unp}.\footnote{See also related ideas in \cite{Akers:2024ixq, Donnelly:2016qqt, Wong:2022eiu, Mertens:2022ujr, Dong:2023kyr, Qi:2022lbd, Cheng:2022ori}.}

For completeness, let us also note that there are two additional phases contributing to the computation of $S(AA^*)$ in Fig.~\ref{taco}. The first phase corresponds to a sum over geodesics ending on $a$ and $c$, and a separate geodesic ending on $a$ and $c^*$. The second phase consists of geodesics connecting $b$ to $c$ and $b$ to $c^*$. These two cases also form homologous minimal surfaces to $AA^*$ or $BB^*$. Geometrically, one can see that these contributions are, however, \textit{always larger} than \eqref{eeinfirstcase}, since the corresponding geodesics acquire infinite contributions from the regions near the endpoints $a$ and $b$.\footnote{Equivalently, in the limit of vanishing regulator size, the two geodesics in each of these two phases form a hyperbolic triangle with the cyan curve, so their combined length necessarily exceed that of the third one.}

From the averaging perspective, they arise naturally from two additional averaging patterns in the replica partition function,
\be
\begin{tikzpicture}[scale=0.6, rotate=270]
	\draw[thick] (0,1) arc (90:-90:1)
    node[pos=0, left] {$l$}
    node[pos=1, right] {$i$};
	\draw[thick] (1,0) -- ++(2,0)
    node[pos=0.5, right] {$k$};
  \draw[thick] (4,0) circle (1)
	  node[left] at ++(0,1) {$n$}
	  node[right] at ++(0,-1) {$m$};
	\draw[thick] (5,0) -- ++(2,0)
    node[pos=0.5, right] {$o$};
	\draw[thick] (8,0) circle (1)
	  node[left] at ++(0,1) {$q$}
	  node[right] at ++(0,-1) {$p$};
  \draw[thick] (9,0) -- ++(1,0)
    node[pos=0.6, right] {$r$};

  \node[right] at (10.5,-0.2) {$\vdots$};
  \draw[thick] (11.5,0) -- ++(1,0)
    node[pos=0.3, right] {$u$};
  \draw[thick] (13.5,1) arc (90:270:1)
    node[pos=0, left] {$l$}
    node[pos=1, right] {$i$};

  \def\w{1.4}
  \def\h{2}
  \draw[thick, orange, rounded corners=2pt,dashed]
    (0,-\h) -- ++(5.5*\w,0) -- ++(0,2*\h) -- ++(-5.5*\w,0);
  \draw[thick, orange, rounded corners=2pt,dashed]
    (7*\w,-\h) -- ++(-\w,0) -- ++(0,2*\h) -- ++(\w,0);
  \draw[thick, orange, rounded corners=2pt,dashed]
    (8.2*\w,-\h) -- ++(\w,0) -- ++(0,2*\h) -- ++(-\w,0);
  \draw[thick, orange, rounded corners=2pt,dashed]
    (13.6,-\h) -- ++(-0.3*\w,0) -- ++(0,2*\h) -- ++(0.3*\w, 0);

  \draw[dashed,thick] (0,-1) -- ++(0,-0.5) -- ++(13.5,0) -- ++(0,0.5);
  \draw[dashed,thick] (0,1) -- ++(0,0.5) -- ++(13.5,0) -- ++(0,-0.5);

  \begin{scope}[yshift=7cm]
	  \draw[thick] (0,1) arc (90:-90:1)
      node[pos=0, left] {$l$}
      node[pos=1, right] {$i$};
	  \draw[thick] (1,0) -- ++(2,0)
      node[pos=0.5, right] {$k$};
    \draw[thick] (4,0) circle (1)
	    node[left] at ++(0,1) {$n$}
	    node[right] at ++(0,-1) {$m$};
	  \draw[thick] (5,0) -- ++(2,0)
      node[pos=0.5, right] {$o$};
	  \draw[thick] (8,0) circle (1)
	    node[left] at ++(0,1) {$q$}
	    node[right] at ++(0,-1) {$p$};
    \draw[thick] (9,0) -- ++(1,0)
      node[pos=0.6, right] {$r$};

    \node[right] at (10.5,-0.2) {$\vdots$};
    \draw[thick] (11.5,0) -- ++(1,0)
      node[pos=0.3, right] {$u$};
    \draw[thick] (13.5,1) arc (90:270:1)
      node[pos=0, left] {$l$}
      node[pos=1, right] {$i$};

    \def\w{1.4}
    \def\h{2}
    \draw[thick, orange, rounded corners=2pt,dashed]
      (0,-\h) -- ++(2.5*\w,0) -- ++(0,2*\h) -- ++(-2.5*\w,0);
    \draw[thick, orange, rounded corners=2pt,dashed]
      (7*\w,-\h) -- ++(-3.8*\w,0) -- ++(0,2*\h) -- ++(3.8*\w,0);
    \draw[thick, orange, dashed] (13.6,-\h) -- ++(-1.5*\w,0);
    \draw[thick, orange, dashed] (13.6,\h) -- ++(-1.5*\w,0);

    \draw[dashed,thick] (0,-1) -- ++(0,-0.5) -- ++(13.5,0) -- ++(0,0.5);
    \draw[dashed,thick] (0,1) -- ++(0,0.5) -- ++(13.5,0) -- ++(0,-0.5);
  \end{scope}
\end{tikzpicture}\ , \label{infinitephase}
\ee
where each orange box contains an average over four OPE coefficients, corresponding to the external-leg averaging patterns of \cite{Bao:2025plr, Geng:2025efs}. For instance, in the left contraction pattern above, the left legs on the boundaries of boxes, such as $i,p$, are identified, and similarly for the right legs $l,q,\cdots$, leading to entropy contributions analogous to the mechanism discussed above. However, as discussed above, these two patterns yield only subleading contributions to the entropy, and we shall therefore not elaborate on them here. Further details on these phases can be found in \cite{Bao:2025plr, Geng:2025efs}.

In fact, these three cases exhaust the possibilities in which the replica partition function diagrams can be decomposed into symmetric boxes, allowing consistent Gaussian averaging and leading contributions. They correspond precisely to the three homologous RT surfaces \cite{Bao:2025plr, Geng:2025efs}. A more detailed discussion will be presented in Sec.~\ref{discosection2}, in which the competition between different averaging patterns is shown to generate reflected entropy phase transitions.

In summary, we have shown that the reflected entropy corresponds to the length of the cyan curve measured in the hyperbolic metric of Fig.~\ref{taco}. To complete the proof, it remains to demonstrate that this length is precisely twice the entanglement wedge cross section in Fig.~\ref{3CONNECTED}.

\begin{figure}
    \centering
\includegraphics[width=0.4\linewidth]{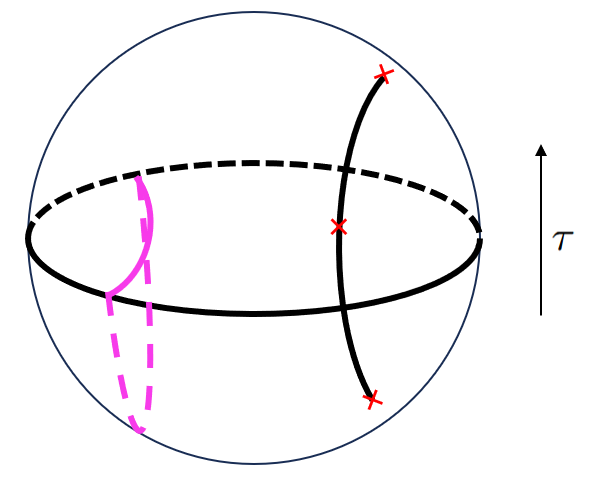}
    \caption{\small{Via the hyperbolic slicing \eqref{slicing}, points on the CFT state-preparation manifold correspond one-to-one to points on the Cauchy slice, as indicated by the red dots. The pink curve representing the RT surface on the Cauchy slice maps to the pink dashed curve on the state-preparation manifold in the limit $\tau \to \infty$.}}
    \label{4CONNECTED0}
\end{figure}

Let us start by noting that the averaging in the 2D CFT norm generally produces the partition function of Liouville CFT with ZZ boundary conditions \cite{Geng:2025efs, Bao:2025plr, Chua:2023ios}, which is also equal to the gravitational partition function on 3D hyperbolic geometries associated with the hyperbolic slicing \eqref{slicing} \cite{Collier:2023fwi}. In this correspondence, the boundary points of the CFT at $\tau \to \pm \infty$ map one-to-one onto points on the Cauchy slice at $\tau=0$ \cite{Goto:2017olq, Bao:2025plr, Geng:2025efs, Verlindetalk, Verlindeunpublished, Chandra:2023dgq, Chua:2023ios}, as illustrated in Fig.~\ref{4CONNECTED0}. 

For the original state $\ket{\Psi}_{ABC}$ in Fig.~\ref{3CONNECTED}, the Liouville field $\phi$ solves the Liouville equation with ZZ boundary conditions on $ABC$. On the other hand, for the canonical purification, the hyperbolic metric comes from the Liouville solution on Fig.~\ref{taco} with ZZ boundary conditions on $ABA^*B^*$. Our goal is to verify that Fig.~\ref{3CONNECTED} and Fig.~\ref{taco} share the same emergent hyperbolic metric on their common region, namely the one bounded by $AB$ and the pink curve. In that case, the length of the cyan edge in Fig.~\ref{taco} is clearly twice that in Fig.~\ref{3CONNECTED}.

This is true for the following reason. First, we solve the Liouville equation on Fig.~\ref{3CONNECTED}, and obtain the Liouville field $\phi$ corresponding to the whole hyperbolic disk $e^{\phi} dz d\bar{z}$. The pink curve marks the location of the minimal surface in this metric. Since a minimal surface has vanishing extrinsic curvature, cutting the disk open along this curve, and gluing it to another copy with reversed orientation using the same Liouville field produces a smooth hyperbolic metric $e^{\phi} dz d\bar{z}$ on Fig.~\ref{taco}. Because the hyperbolic metric is unique, this must be precisely the metric for $ABA^*B^*$.

There is another way to view this from the Liouville CFT path-integral perspective. Using the connection between Liouville theory and minimal-length geodesics in hyperbolic space, what we need to show is that, in the averaged norm computations of $\ket{\Psi}_{ABC}$ and $\ket{\sqrt{\rho_{AB}}}$, the saddle points in the $k$-primary integrals coincide in the following two Liouville theory computations,
\be
 \vcenter{\hbox{\begin{tikzpicture}[scale=0.6]
  \begin{scope}[xshift=12cm, yshift=-2.5cm]
    \draw[thick] (-1,0) arc (180:360:1);
    \draw[thick] (2,0) arc (180:360:1);
    \draw[thick] (0,-1) arc (180:360:1.5);
     \node at (2,0.5) {\text{ZZ}};
     \node at (4,0.5) {\text{ZZ}};
     \node at (1,0.5) {\text{ZZ}};
     \node at (-1,0.5) {\text{ZZ}};
      \node at (-0.65,0) {$i$};
        \node at (0.65,0) {$l$};
          \node at (2.4,0) {$m$};
          \node at (3.6,0) {$n$};
\node at (1.5,-2.2) {$k$};
    \def\capW{0.2}
    \draw[thick] (-1-\capW, 0) -- ++(2*\capW, 0);
    \draw[thick] (1-\capW, 0) -- ++(2*\capW, 0);
    \draw[thick] (2-\capW, 0) -- ++(2*\capW, 0);
    \draw[thick] (4-\capW, 0) -- ++(2*\capW, 0);

    \node at (1.5, -3.5) {Canonical purification manifold};
  \end{scope}
\end{tikzpicture}}}
 \vcenter{\hbox{\begin{tikzpicture}[scale=0.6]
  \begin{scope}[xshift=12cm, yshift=-2.5cm]
    \node at (-2, -1.5) {=  };
    \draw[thick] (-1,0) arc (180:360:1);
    \draw[thick] (0,-1) arc (180:360:1.5);
    \draw[thick] (3,-1) -- ++(0,1);
     \node at (3,0.5) {\text{ZZ}};
     \node at (1,0.5) {\text{ZZ}};
     \node at (-1,0.5) {\text{ZZ}};
      \node at (-0.65,0) {$i$};
        \node at (0.65,0) {$l$};
\node at (1.5,-2.2) {$k$};
    \def\capW{0.2}
    \draw[thick] (-1-\capW, 0) -- ++(2*\capW, 0);
    \draw[thick] (1-\capW, 0) -- ++(2*\capW, 0);
    \draw[thick] (3-\capW, 0) -- ++(2*\capW, 0);

    \node at (1.5, -3.5) {Original manifold};
  \end{scope}
\end{tikzpicture}}}~.
\ee
In fact, this follows directly from our construction of $\ket{\sqrt{\rho_{AB}}}$ (see \eqref{purificationcheck}), and a similar argument using superposition of fixed-area states was presented in \cite{Marolf:2019zoo}.

In this simplest configuration of subregions $ABC$, Fig.~\ref{3CONNECTED}, the only use of the large-$c$ limit is to relate the entropy to the geodesic length, and even this step is not strictly necessary. This case of reflected entropy is in fact analogous to the universal single-interval entanglement entropy, which holds for any 2D CFT. We have nevertheless chosen to present it in this way because it makes the role of all the regulators manifest, and the paradigm of extracting geodesic lengths through the connection to Liouville theory naturally generalizes to the other cases discussed below. By contrast, the examples in Sec.~\ref{discosection} and Sec.~\ref{discosection2} will require essential large-$c$ input and are valid only for large-$c$ holographic CFTs.

Before turning to those examples, in Sec.~\ref{sec3.4} we provide a detour presenting a simpler derivation of this case, which readers may skip on a first pass or if they are only interested in the most general framework.

\subsubsection{Digression: Simpler Derivation of the Universal Reflected Entropy Case} \label{sec3.4}

As mentioned above, for this simplest case, Fig.~\ref{3CONNECTED}, the calculation of the reflected entropy does not necessarily rely on the large-$c$ limit. We now compute in a simpler way. The idea is that after we get the canonical purification, we shrink the small conformal boundaries at $a$ and $b$, thereby reducing the setup to the CFT thermofield double state between $AA^*$ and $BB^*$. The entanglement entropy can then be computed directly, yielding the universal result that matches the cyan geodesic length \cite{Geng:2025efs}.

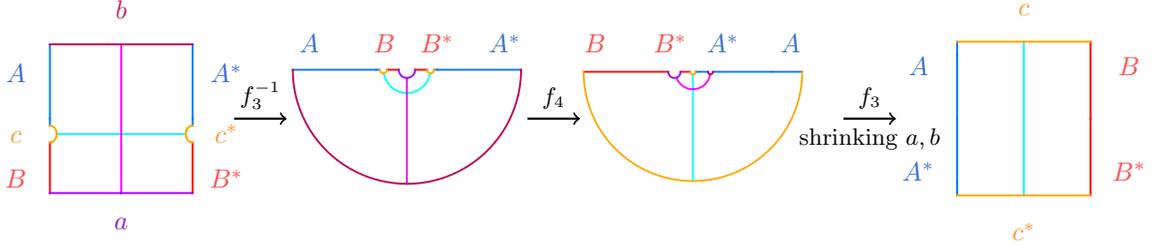
\begin{figure}[h]
    \centering
    \begin{tikzpicture}
    \begin{scope}
    \def\w{1.4}
    \def\Bh{-0.8}
    \def\Ah{0.6}
    \def\ah{-1.4}
    \def\bh{1.45}
    \def\ch{-0.2}
      \node at (0,0)
        {\includegraphics[width=2.5cm]{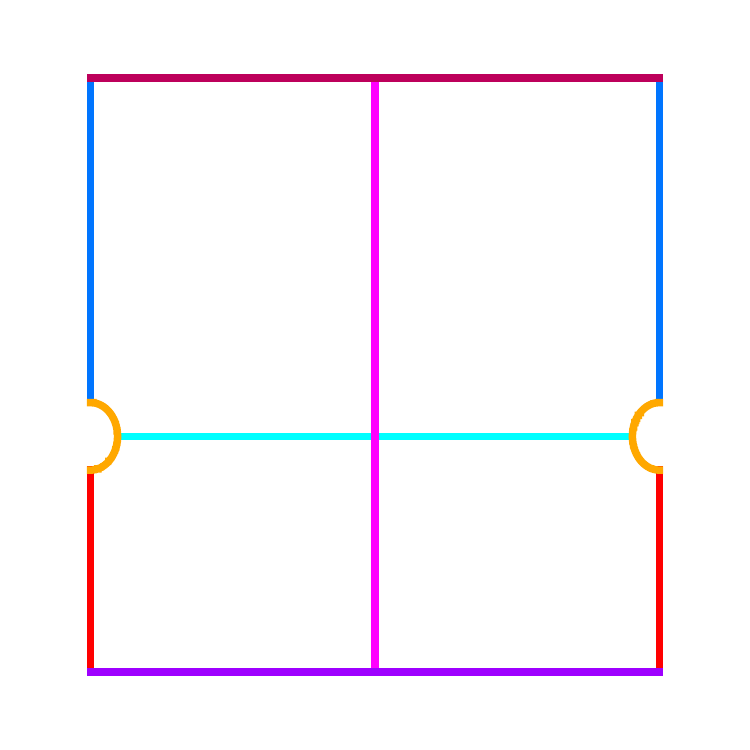}};
        \node[\AColor] at (-\w,\Ah) {$A$};
        \node[\AColor] at (\w, \Ah) {$A^*$};
        \node[\BColor] at (-\w,\Bh) {$B$};
        \node[\BColor] at (\w,\Bh) {$B^*$};
        \node[\aColor] at (0*\w,\ah) {$a$};
        \node[\bColor] at (0*\w,\bh) {$b$};
        \node[\cColor] at (-\w,\ch-0.05) {$c$};
        \node[\cColor] at (\w,\ch) {$c^*$};
    \end{scope}
        
        \begin{scope}[xshift=3.8cm] 
        \def\h{1};
        \def\hp{1.3}
        \def\Aw{1.3}
        \def\Bw{0.3}
        \def\ah{0.5}
        \def\aw{0.6}
        \def\bw{1.1}
        \def\cw{0.8}
            \node at (0,0)
             {\includegraphics[width=3.5cm]{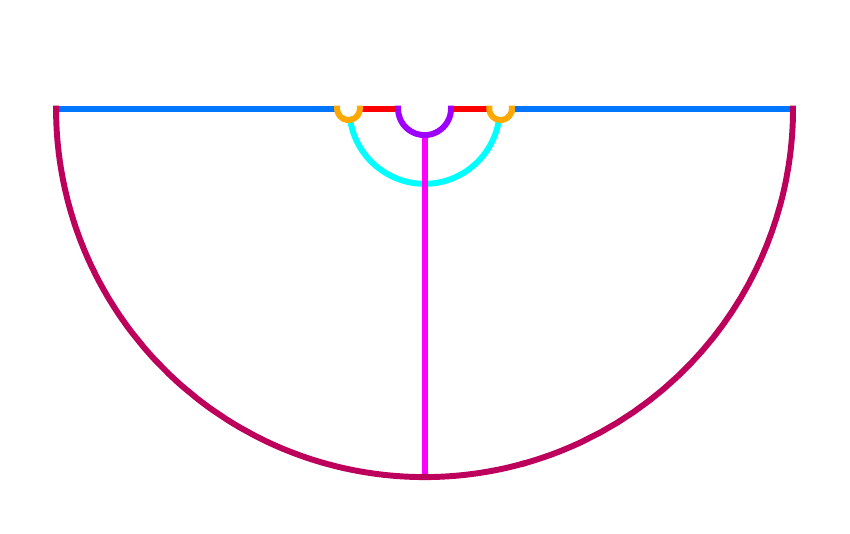}};
             \node[\AColor] at (-\Aw,\h) {$A$};
             \node[\AColor] at (\Aw,\h) {$A^*$};
             \node[\BColor] at (-\Bw,\h) {$B$};
             \node[\BColor] at (\Bw+0.1,\h) {$B^*$};
        \end{scope}

        \begin{scope}[xshift=7.6cm] 
        \def\h{1}
            \node at (0,0)
             {\includegraphics[width=3cm]{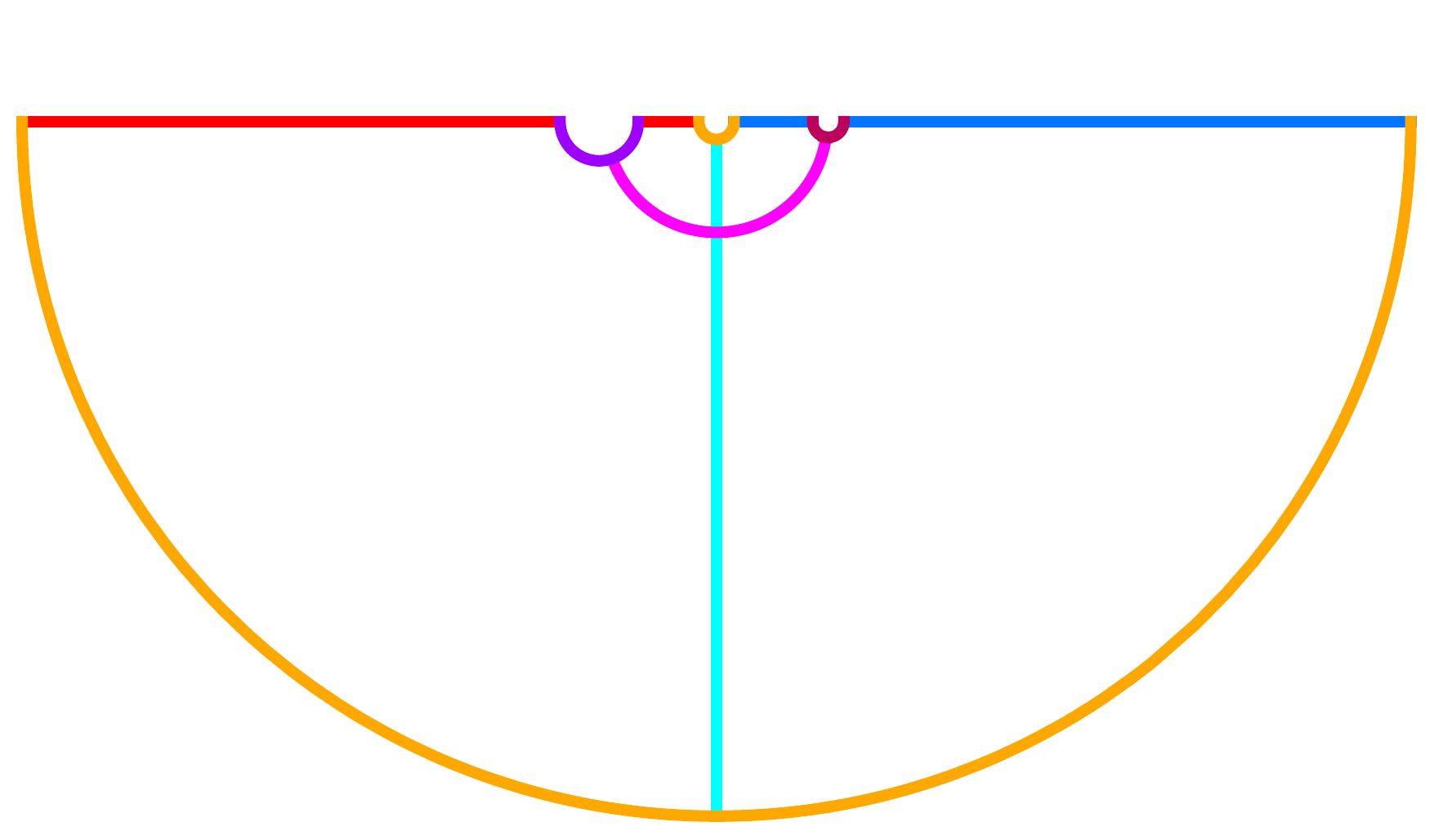}};
            \node[\BColor] at (-1.3,\h) {$B$};
             \node[\AColor] at (1.3,\h) {$A$};
             \node[\BColor] at (-0.3,\h) {$B^*$};
             \node[\AColor] at (0.4,\h) {$A^*$};
        \end{scope}

        \begin{scope}[xshift=12cm] 
        \def\w{1.4}
        \def\ah{0.4}
        \def\bh{\ah}
        \def\ch{1.5}
        \def\h{0.7}
            \node at (0,0)
             {\includegraphics[width=2cm]{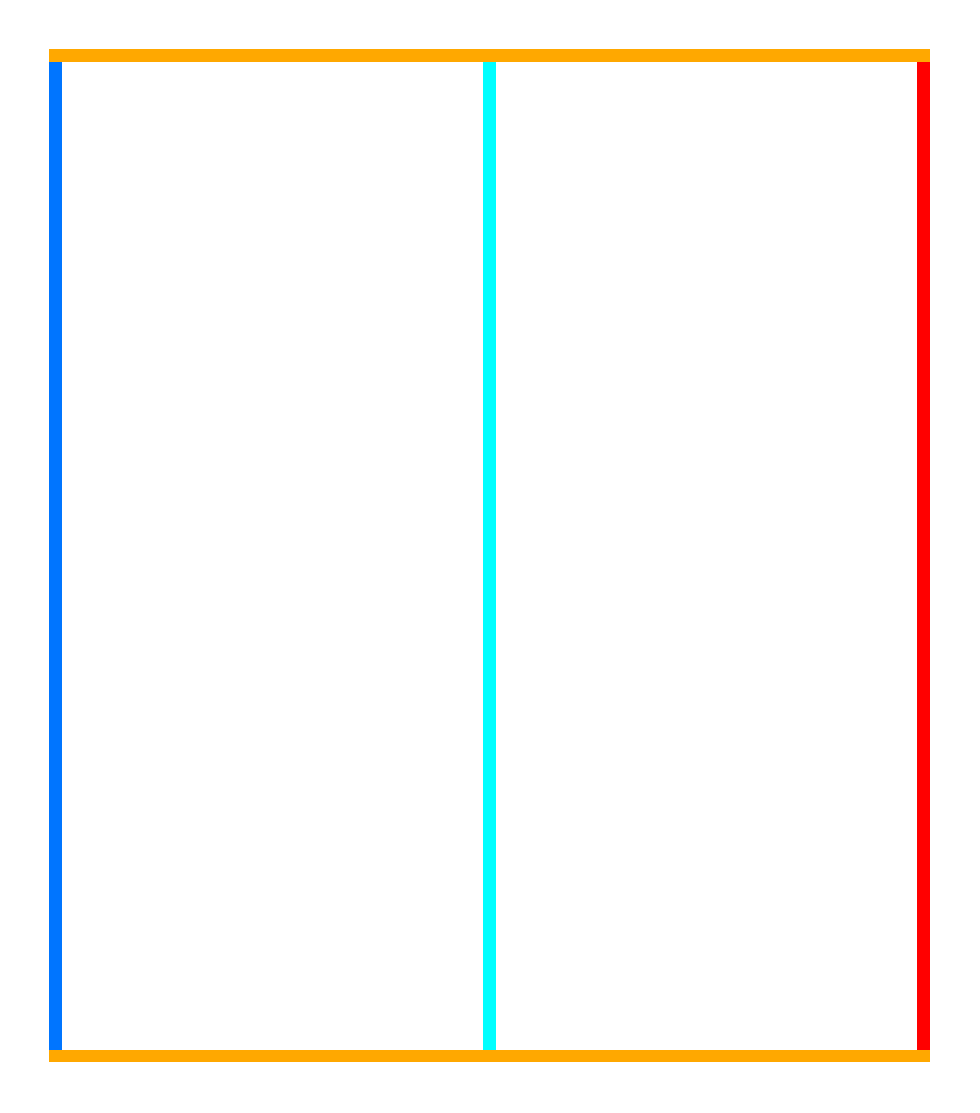}};
             \node[\AColor] at (-\w,\h) {$A$};
             \node[\AColor] at (-\w,-\h) {$A^*$};
            \node[\BColor] at (\w,\h) {$B$};
            \node[\BColor] at (\w,-\h) {$B^*$};
            \node[\cColor] at (0,\ch-0.05) {$c$};
            \node[\cColor] at (0,-\ch) {$c^*$};
        \end{scope}

        \draw[->, thick] (1.5,0) -- ++(0.7,0)
        node[above, pos=0.5] {\small{$f_3^{-1}$}};
        \draw[->, thick] (5.4,0) -- ++(0.7,0)
        node[above, pos=0.5] {\small{$f_4$}};
        \draw[->, thick] (9.6,0) -- ++(0.7,0)
        node[above, pos=0.5] {\small{$f_3$}}
        node[below, pos=0.5] {\small{$\text{shrinking $a,b$}$}};
        
    \end{tikzpicture}
    \caption{Conformal transformations used in the simpler derivation of the universal reflected entropy.}
    \label{inverseDoubled}
\end{figure}

A sequence of conformal transformations is illustrated in Fig.~\ref{inverseDoubled}. We first map the canonical purification state in the $w$ coordinate back to the $\xi$ coordinate using $f_3^{-1}$. In the second figure of Fig.~\ref{inverseDoubled}, the endpoints of the purple $a$ boundaries are located at $(\pm f_2(l/2+\epsilon),0)$, those of the maroon $b$ boundaries at $(\pm f_2(-l/2-\epsilon),0)$, those of the $c$ boundaries at $(-f_2(x_0 \pm \epsilon),0)$, and those of the $c^*$ boundaries at $(f_2(x_0 \pm \epsilon),0)$.

Introducing 
\be
l'=f_2(x_0+ \epsilon)+f_2(x_0- \epsilon), \quad \epsilon'=\frac{1}{2}\left(f_2(x_0- \epsilon)-f_2(x_0+ \epsilon) \right)~,
\ee 
where $l'$ denotes the location of the center of the $c$ boundary and $\epsilon'$ its radius, we apply the conformal transformation
\be
f_4(\xi) = \frac{\xi - \sqrt{\left(\tfrac{l'}{2}\right)^2 - \epsilon'^2}}{\xi+\sqrt{\left(\tfrac{l'}{2}\right)^2 - \epsilon'^2} }
\ee
which maps the $c$ and $c^*$ boundaries to concentric circles, similar to $f_2$ in Fig.~\ref{transforms}. 

Shrinking the conformal boundary regulators at $a$ and $b$ and applying the logarithmic map $f_3$ once more, we obtain the thermofield double state on a strip for $AA^*$ and $BB^*$, whose height is $W'=\ln \left(f_4(-\frac{l'}{2}-\epsilon') \right)-\ln \left(f_4(\frac{l'}{2}+\epsilon') \right)$. This construction reduces to the original BCFT thermofield double setup of \cite{Ohmori:2014eia, Cardy:2016fqc}, where the entanglement entropy can be computed directly, yielding the universal answer,\footnote{Note that $2\ln\tfrac{l'}{\epsilon'}$ is the regularized geodesic length of the cyan curve in the hyperbolic metric, which equals twice the entanglement wedge cross section. Although the CFT need not be holographic, the hyperbolic length still emerges in the result. This parallels the universal single-interval entanglement entropy in 2D CFTs, where the outcome is likewise fixed entirely by conformal symmetry.}
\be
S(AA^*)=2\ln g_c+\frac{c}{6} W'=2\ln g_c+\frac{2\ln\frac{l'}{\epsilon'}}{4G_N}+\mathcal{O}(\frac{\epsilon'}{l'})
\ee

\subsection{Case II: $\text{RE}(A:B)$ for Disjoint $A$ and $B$}\label{discosection}
\subsubsection{$\text{RE}=\text{EW}$ for Connected Entanglement Wedge}

Having detailed the $\text{RE}=\text{EW}$ correspondence for adjacent subregions $A$ and $B$, we now turn to the non-adjacent case. We will first focus on the case with a connected entanglement wedge, as shown in Fig.~\ref{crosssection}, emphasizing the new features that emerge in the large-$c$ limit. 
\begin{figure}
    \centering
\includegraphics[width=0.4\linewidth]{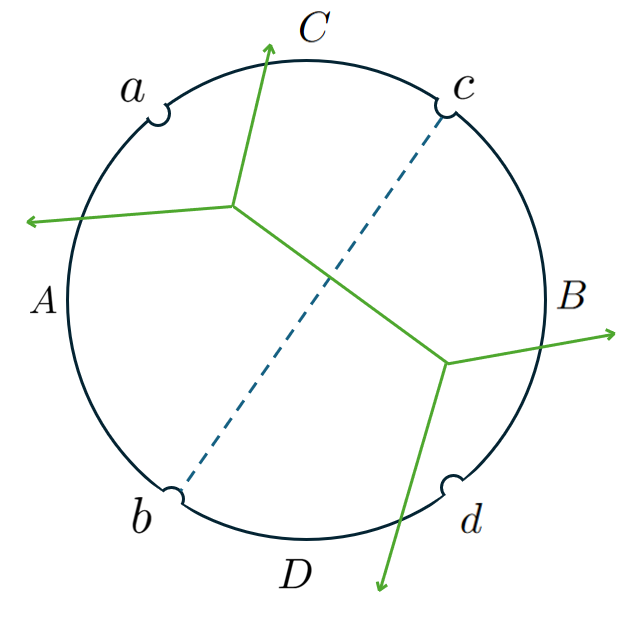}
    \caption{\small{For the vacuum state defined on $ABCD$, we introduce four tiny conformal boundaries and thereby obtain an entangled BCFT state on four intervals. The OPE block decomposition arises from decomposing the path integral into two triangles. The corresponding tensor network is defined on the dual graph, indicated by the green curves.}}
    \label{4CONNECTED}
\end{figure}

\begin{figure}
    \centering
\includegraphics[width=0.8\linewidth]{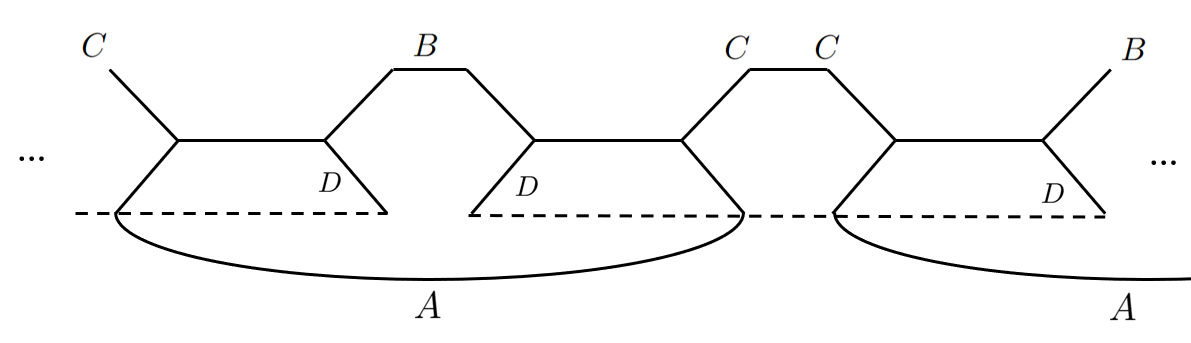}
    \caption{\small{Illustration of the replica partition function $\tr \rho_{AB}^n$ for disjoint $AB$.}}
    \label{4CONNECTEDreplicaZ}
\end{figure}

Introducing the conformal boundary regulators as above, we express the state $\ket{\Psi}_{ABCD}$ as the gluing of two BCFT triangles, as shown in Fig.~\ref{4CONNECTED}. This construction can then be represented in terms of an OPE block, in analogy with the discussion above. The replica partition function associated with $AB$ is depicted in Fig.~\ref{4CONNECTEDreplicaZ}.

\begin{figure}
    \centering
\includegraphics[width=1\linewidth]{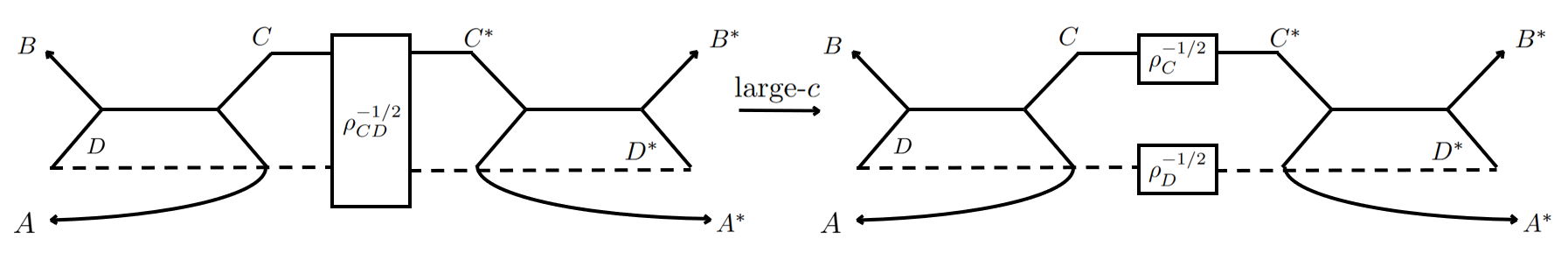}
    \caption{\small{The canonical purification $\ket{\rho_{AB}}$ can be constructed explicitly from the original state together with $\rho_{CD}$. In the large-$c$ limit, $\rho_{CD}$ factorizes $\rho_{C} \otimes{\rho_D}$.}}
    \label{4CONNECTEDcanonical}
\end{figure}

The canonical purification can again be constructed using the reduced density matrix of the complement region, $\rho_{CD}$, as shown on the left panel of Fig.~\ref{4CONNECTEDcanonical}. However, for disjoint $CD$, $\rho_{CD}$ is in general very difficult to compute and, in fact, does not admit a \textit{local} expression. In other words, the modular flow becomes highly nontrivial, in contrast to the previous example where it could be directly identified with the BCFT Hamiltonian. At first sight, therefore, the construction appears to be obstructed at this formal step.

However, the simplifying power of the large-$c$ limit now comes into play. At leading order in large-$c$, when we focus on states whose bulk dual admits a connected entanglement wedge for $AB$, the entanglement wedge associated with $CD$ is completely disconnected. This in turn implies that the density matrix $\rho_{CD}$ factorizes. More explicitly, one can compute the relative entropy between $\rho_{CD}$ and $\rho_C \otimes \rho_D$,
\be \label{relativeent}
\begin{split}
S_{\text{rel}}(\rho_{CD}||\rho_C\otimes\rho_D)\equiv &\Tr{\rho_{CD}\ln \rho_{CD}}-\Tr{\rho_{CD}\ln \left(\rho_C\otimes\rho_D \right)}\\
=&\Tr{\rho_{CD}\ln \rho_{CD}}-\Tr{\rho_{CD}\ln \rho_C}-\Tr{\rho_{CD}\ln \rho_D}\\
=&\Tr{\rho_{CD}\ln \rho_{CD}}-\Tr{\rho_{C}\ln \rho_C}-\Tr{\rho_{D}\ln \rho_D}\\
=&-S(CD)+S(C)+S(D)\ .
\end{split}
\ee
The third line follows from performing partial traces over $D$ and $C$ on the second and third terms of the second line, respectively. The final expression is a linear combination of von Neumann entropies. Thus, this relative entropy can be easily calculated within the CFT—for instance, by averaging over BCFT OPE coefficients as in \cite{Geng:2025efs}—which yields a vanishing result. The vanishing of the relative entropy then implies that the reduced density matrix factorizes,
\be \label{factorize}
\rho_{CD}=\rho_C\otimes\rho_D+\mathcal{O}(1/c)\ .
\ee

\begin{figure}
    \centering
\includegraphics[width=0.9\linewidth]{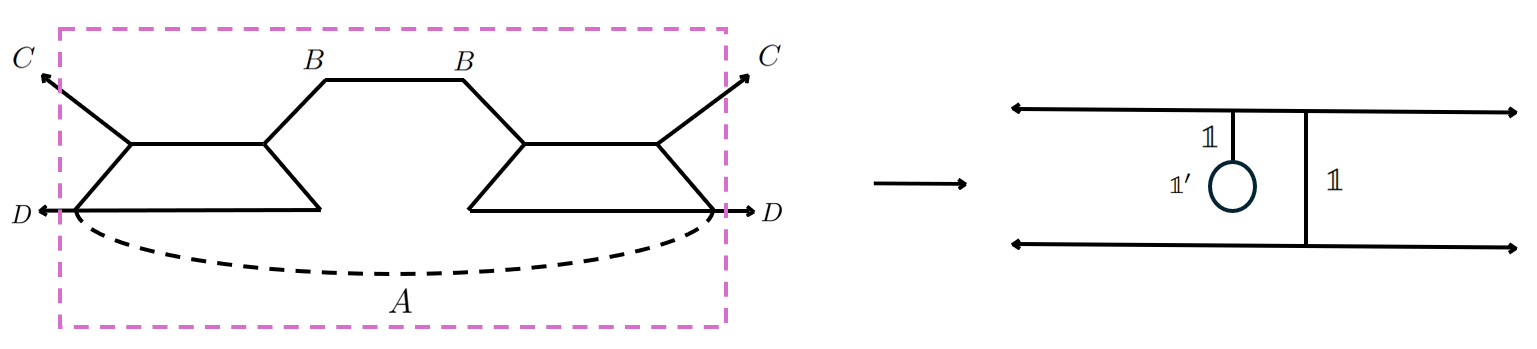}
    \caption{\small{The averaging pattern for the computation of $\rho_{CD}$.}}
    \label{averagerhoCD}
\end{figure}

It is noteworthy that the factorization of $\rho_{CD}$ can in fact be shown directly from the averaging procedure, without passing through the entropy. The averaging pattern that produces the pink RT surfaces in Fig.~\ref{crosssection} corresponds to performing a Gaussian average in the purple box of Fig.~\ref{averagerhoCD},\footnote{This averaging pattern ensures that all primary labels on the $C$ and $D$ legs are identified, leading to entropy contributions from saddles propagating in these channels.} which, after simplification analogous to \eqref{firstphaseZn}, leads to the right panel of Fig.~\ref{averagerhoCD}, yielding a factorizable density matrix at leading order in $c$, in agreement with \eqref{factorize}.

Since the reduced density matrices $\rho_C$ and $\rho_D$ are associated with single intervals in the vacuum state, they are once again \textit{local} and proportional to the BCFT Hamiltonian in appropriate coordinates. Using this result, we can perform the ``backward evolution'' for $C$ and $D$ separately, as illustrated on the right panel of Fig.~\ref{4CONNECTEDcanonical}, moving them precisely to the locations corresponding to the RT surfaces, as discussed in the previous case. This demonstrates that we again obtain the CFT cutting-and-gluing procedure dual to the bulk prescription, as illustrated in Fig.~\ref{crosssection}. In this case, one can shrink the boundary regulators so that the boundaries become two full circles. At leading order in $1/c$, the state reduces to the thermofield double between $AA^*$ and $BB^*$. The entanglement entropy is thus given by the RT surface highlighted in cyan (see Fig.~\ref{crosssection}). We will have more discussion on the finite $c$ corrections in the discussion section.

\begin{figure}
    \centering
\includegraphics[width=0.8\linewidth]{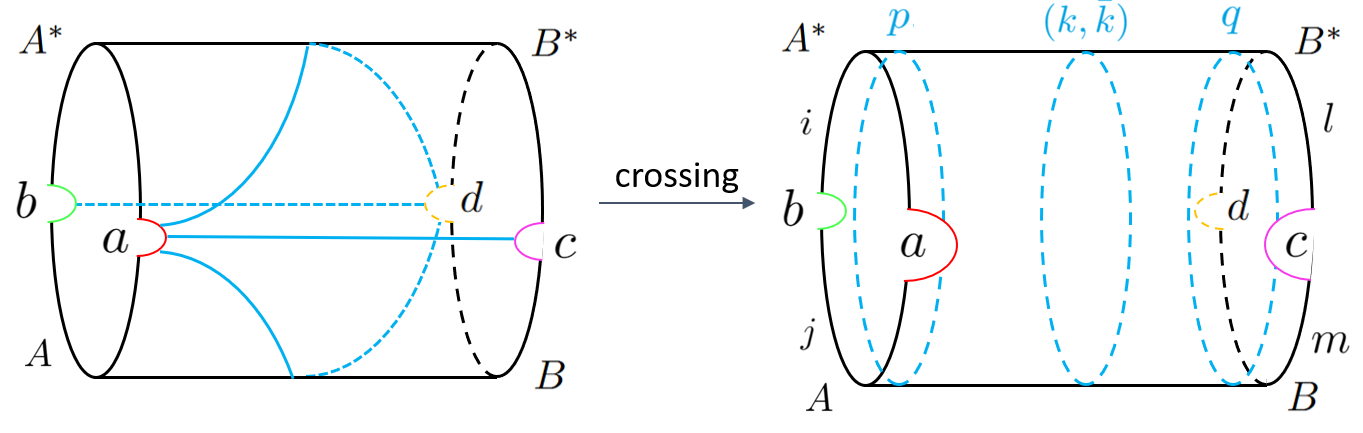}
    \caption{\small{We perform a crossing move that relates the open CFT channel slicings on the left to the closed CFT channel $(k,\bar k)$ slicing on the right, together with two open CFT channels $p$ and $q$.}}
    \label{open-closedcrossing}
\end{figure}

We can also directly compute the entanglement entropy between $AA^*$ and $BB^*$ in the canonical purification Fig.~\ref{4CONNECTEDcanonical} as follows. Similar to the procedure in the previous section, to extract the dominant contribution to the replica partition function under the ensemble average, we first perform a crossing move as depicted in Fig.~\ref{open-closedcrossing}. It corresponds to changing from the slicing with four triangles to the one where an intermediate state lies in the closed CFT channel \cite{Lewellen:1991tb, CARDY1991274, Numasawa:2022cni}. For later convenience, we also indicate the labeling of the primaries propagating in each Hilbert space in the figure.

In the channel corresponding to the right panel of Fig.~\ref{open-closedcrossing}, the OPE block representation of the canonical purification is given by

\begin{equation}
\begin{aligned}
\ket{\sqrt{\rho_{AB}}}=\sum_{\text{primaries}}  C^{aab}_{ijp}  C^a_{k\bar{k}p}  C^{c*}_{k\bar{k}q} C^{ccd*}_{lmq}  \mathcal{B}\left[
 \vcenter{\hbox{\begin{tikzpicture}[scale=0.6]
 \draw[<->,thick,black!!40] (0,1) arc (90:-90:1);
  \draw[<->,thick,black!!40] (6,1) arc (90:270:1);
  \draw[-, thick, black!!40] (1,0) to (2,0);
  \draw[-, thick, black!!40] (4,0) to (5,0);
  \draw[-,thick,black!!40] (3,0) circle (1);
  \node at (-0.2,1) {\textcolor{black}{$i$}};
  \node at (-0.2,-1) {\textcolor{black}{$j$}};
  \node at (3,-0.5) {\textcolor{black}{$k$}};
  \node at (3,0.5) {\textcolor{black}{$\bar{k}$}};
  \node at (1.5,0.25) {\textcolor{black}{$p$}};
  \node at (4.5,0.25) {\textcolor{black}{$q$}};
  \node at (6.2,1) {\textcolor{black}{$l$}};
  \node at (6.2,-1) {\textcolor{black}{$m$}};
 \end{tikzpicture}}} \right] \ket{i} \ket{j} \ket{l}\ket{m}
 ~,
 \end{aligned}\label{eq:2BCFT1CFTchiral1}
\end{equation} 
where the doubling trick \cite{Cardy:2004hm} is applied to the closed state $(k,\bar{k})$, and $C^a_{k\bar{k}p}$ represents the bulk–boundary two-point structure coefficient.

The averaging pattern in the replica partition function that yields the area of the cyan RT surface for $AA^*$ is illustrated in the following diagram

\be
\begin{aligned}
 \vcenter{\hbox{
	\begin{tikzpicture}[scale=0.75]
     \draw[-,thick,black!!40] (0,1) arc (90:-90:1);
	\draw[thick] (1,0) -- (3,0);
	
    \draw[thick] (4,0) circle (1);
	
	\draw[thick] (5,0) -- (7,0);
	\draw[thick] (8,0) circle (1);
    \draw[-,thick,black!!40] (9,0) to (11,0);
    \draw[thick] (12,0) circle (1);
    \draw[-,thick,black!!40] (13,0) to (14,0);
    \node[above] at (15,0-0.3) {$...$};
    \draw[-,thick,black!!40] (16,0) to (17,0);
    \draw[-,thick,black!!40] (18,1) arc (90:270:1);
    \draw[-,dashed,thick,black!!40] (0,-1) to (0,-1.5) to (18,-1.5) to (18,-1);
    \draw[-,dashed,thick,black!!40] (0,1) to (0,1.5) to (18,1.5) to (18,1);
    \draw[-,dashed,thick,blue!!40] (4.3,-2.2) to (12-0.3,-2.2) to (12-0.3,2.2) to (4.3,2.2) to (4.3,-2.2);
    \draw[-,dashed,thick,red!!40] (-0.5,-2.2) to (3.7,-2.2) to (3.7,2.2) to (-0.5,2.2);
    \draw[-,dashed,thick,red!!40] (14.5,-2.2) to (12.3,-2.2) to (12.3,2.2) to (14.5,2.2);
    \draw[-,dashed,thick,red!!40] (15.5,2.2) to (18.6,2.2);
    \draw[-,dashed,thick,red!!40] (15.5,-2.2) to (18.6,-2.2);
	\end{tikzpicture}
	}}~\,,
\end{aligned}
\ee
where, within each box, we perform the average across the $Z_2$-symmetric axes, averaging over the BCFT structure coefficients \eqref{eq:BCFTC} together with the bulk–boundary OPE coefficients \cite{Hung:2025vgs, Wang:2025bcx, Geng:2025efs}
\begin{equation}   
\overline{C_{i \bar{i}j}^{a}C^{*b}_{k \bar{k}l}}=\delta_{ab}\delta_{i k}\delta_{\bar{i} \bar{k}} \delta_{jl}C_{0}(P_{i},\bar{P}_{{i}},P_{j}) \,.\label{eq:BCFTBC}
\end{equation}

Following the same mechanism as in Case I, one can check that the averaged replica partition function and its connection to Liouville theory yield the entropy as \cite{Geng:2025efs}
\be \label{EWfordisconnected}
S(AA^*)=\frac{c}{6} (2\pi \gamma_1^*)=\frac{2\pi \gamma_1^*}{4G_N}\,.
\ee
Here $2\pi \gamma_1^*$ is the minimal geodesic length on the circle\footnote{In this case, the state propagates along a circle rather than an interval in Case I, and includes both chiral and anti-chiral contributions, thereby producing an additional factor of 2 relative to the previous case \cite{Geng:2025efs}.} in the right diagram of Fig.~\ref{crosssection}. This is again twice the length of the entanglement wedge cross section $\text{EW}(A:B)$. Notice that in this case the entanglement wedge cross section does not extend to the asymptotic boundary, and therefore no regulator is required to compute its length. This is reflected in \eqref{EWfordisconnected} by the absence of any $g$-factor contributions. Again, there are two other subleading phases similar to \eqref{infinitephase}, which we will not elaborate on here.

\subsubsection{Connected vs. Disconnected Canonical Purifications}

As noted in \cite{Engelhardt:2022qts}, in the setting of evaporating black holes before and after the Page time, states with the same entropy can exhibit either connected or disconnected canonical purifications.\footnote{We thank Martin Sasieta for suggesting the relevance of this problem to us.} Here, we demonstrate that similar phenomena also exist in vacuum states, and explain their origins from distinct Gaussian averaging patterns.\footnote{As shown in \cite{Geng:2025efs}, the ``replica wormhole'' in black hole contexts \cite{Penington:2019kki, Almheiri:2019qdq} and vacuum entanglement phase transitions are manifestations of this same mechanism.}

Disconnected canonical purifications correspond to disconnected entanglement wedges. By varying subregion lengths, $AB$ can exhibit the same entanglement entropy with either connected or disconnected wedges. For example, interchanging the labels $AB\leftrightarrow CD$ in Fig.\ref{4CONNECTED} yields a disconnected $AB$ phase with the entropy unchanged.

\begin{figure}
    \centering
\includegraphics[width=0.8\linewidth]{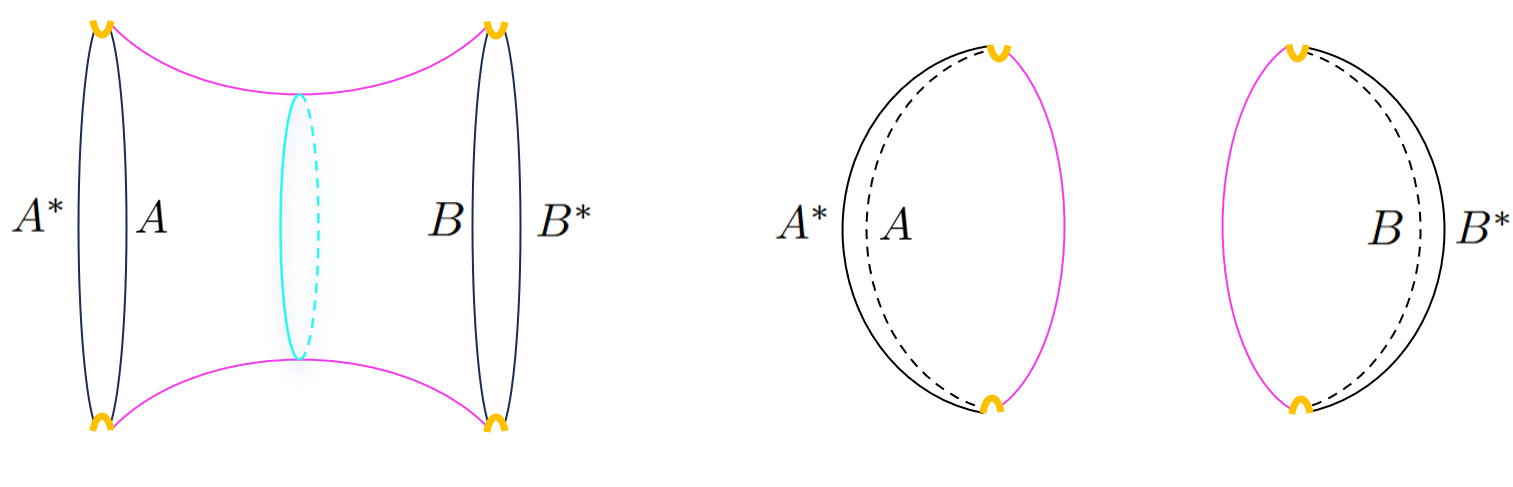}
    \caption{\small{Canonical purifications in the connected and disconnected phases.}}
    \label{disconnected1}
\end{figure}

We have shown above that connected entanglement wedges give rise to connected canonical purifications. The disconnected phase is even simpler: here, $\rho_{AB}$ factorizes as $\rho_A \otimes \rho_B$, with each factor a thermal density matrix for a single interval, similar to \eqref{factorize}. The corresponding canonical purification is then the product of the individual BCFT thermofield double states associated to $\rho_A$ and $\rho_B$. The two cases are summarized in Fig.~\ref{disconnected1}. Note that both types of purifications rely on factorized forms of certain density matrices—one for the $AB$ system itself and the other for the complementary region—and these correspond precisely to the two averaging patterns.

\subsection{Case III: $\text{RE}(A:B_1B_2)$ for Disjoint $A$, $B_1$ and $B_2$}\label{discosection2}
\begin{figure}
    \centering
\includegraphics[width=0.4\linewidth]{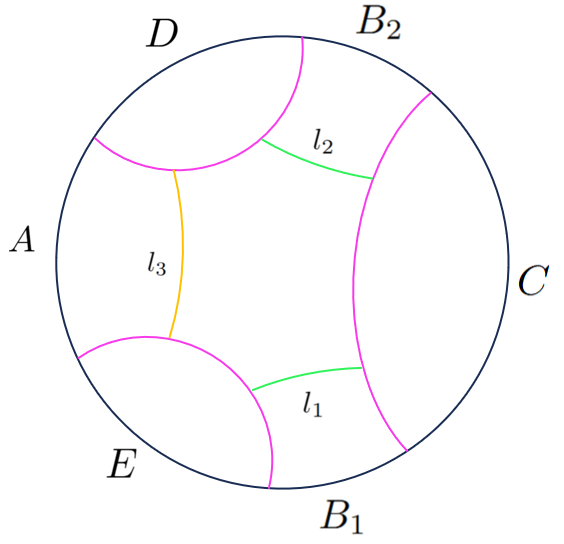}
    \caption{\small{A vacuum state defined on a circle divided into six subregions. Two candidate entanglement wedge cross sections between $A$ and $B_1 \cup B_2$ correspond to $l_1+l_2$ and $l_3$.}}
    \label{disconnected2}
\end{figure}
We now turn to the final representative example, where $B$ consists of two disjoint intervals, $B_1$ and $B_2$, as illustrated in Fig.~\ref{disconnected2}. In the bulk, a distinctive feature of this configuration is the presence of two competing extremal surfaces separating $A$ from $B$, shown in orange and green, respectively. The entanglement wedge cross section $\text{EW}(A:B)$ is defined as the smaller of $l_1+l_2$ and $l_3$. As a consequence, varying the relative sizes of the subregions can induce a phase transition between the two candidate surfaces. Related phase transitions in reflected entropies were first analyzed in \cite{Chandrasekaran:2020qtn, Li:2020ceg}.

Note that phase transitions of this kind also occur in the RT formula. In \cite{Bao:2025plr, Geng:2025efs}, we demonstrated that they arise from \textit{distinct patterns of Gaussian averaging} at leading order in the replica partition function. Moreover, \cite{Geng:2025efs} showed that this is the same mechanism responsible for the appearance of “replica wormholes” in a wide class of models \cite{Penington:2019kki, Almheiri:2019qdq, Rozali:2019day, Geng:2024xpj}. 

Specifically, the canonical purification $\ket{\sqrt{\rho_{AB}}}$ for the configuration in Fig.~\ref{disconnected2} is again constructed by first introducing tiny conformal boundary regulators in the initial pure state $|\Psi\rangle_{AB_1B_2CDE}$, and then acting $\rho_{CDE}^{-1/2}$ on the complementary regions. At leading order in $1/c$, $\rho_{CDE}$ factorizes into $\rho_{C} \otimes \rho_{D} \otimes \rho_{E}$. This can be demonstrated by calculating the relative entropy $S_{\text{rel}}(\rho_{CDE}||\rho_{C} \otimes \rho_{D} \otimes \rho_{E})$, analogous to \eqref{relativeent}:
\be
S_{\text{rel}}(\rho_{CDE}||\rho_{C} \otimes \rho_{D} \otimes \rho_{E})=-S(CDE)+S(C)+S(D)+S(E)\ ,
\ee
which can be easily shown to vanish. The factorization of $\rho_{CDE}$ also admits a clear bulk geometric interpretation: the entanglement wedges of $C$, $D$, and $E$ are disconnected from one another.

Making use of the relation between the single-interval modular Hamiltonian and the BCFT Hamiltonian, the canonical purification can then be obtained by cutting the manifold open along the RT surface of $AB$ and gluing to its CPT conjugate. The resulting manifold has the topology of a pair of pants, with three circular boundaries $AA^*$, $B_1B_1^*$, and $B_2B_2^*$. Each of these boundaries is divided into two intervals by the small conformal boundaries. 

As explained above, one can either work with these intervals within the BCFT framework or, equivalently, shrink the regulators to zero and work directly with the pair of pants. The latter approach reduces the problem to the case dual to a three-boundary black hole \cite{Brill:1995jv, Aminneborg:1997pz, Brill:1998pr, Balasubramanian:2014hda, Bao:2025plr}, as studied in \cite{Bao:2025plr}. Specifically, in terms of OPE blocks, the state can be expressed as
\begin{equation}
\begin{aligned}
\ket{\sqrt{\rho_{AB_1B_2}}}=\sum_{\text{primaries}} C_{ijk} \left|\mathcal{B}\left[
 \vcenter{\hbox{\begin{tikzpicture}[scale=0.6]
  \draw[<->,thick,black!!40] (3.5,1) arc (90:270:1);
  \draw[<-, thick, black!!40] (1,0) to (2.5,0);
  \node at (3.7,-1) {\textcolor{black}{$j$}};
  \node at (3.7,1) {\textcolor{black}{$k$}};
  \node at (1.75,0.25) {\textcolor{black}{$i$}};
 \end{tikzpicture}}} \right]\right|^2 \ket{i} \ket{j}\ket{k}
 ~,
 \end{aligned}\label{eq:1BCFT1CFTchiral}
\end{equation} 
where $\ket{i}, \ket{j}, \ket{k}$ in this case label all the primary states in the CFT on $AA^*,B_1B_1^*,B_2B_2^*$ respectively. We adopt this description to highlight the new ingredients.

The expression (\ref{eq:1BCFT1CFTchiral}) leads to a replica partition function $\overline{Z_{AA^*,n}}$ of the form
\be \label{twophases3bdy}
\begin{aligned}
 \vcenter{\hbox{
	\begin{tikzpicture}[scale=0.73]
	\draw[thick] (0,0) circle (1);
	\draw[thick] (-1,0) -- (-2,0);
	\draw[thick] (1,0) -- (3,0);
    \draw[thick] (4,0) circle (1);
	\draw[thick] (5,0) -- (7,0);
    \node[above] at (8,0-0.3) {$...$};
    \draw[thick] (10,0) circle (1);
	\draw[thick] (11,0) -- (12,0);
      \draw [dashed] (-2,0) -- (-2,-3);
      \draw [dashed] (12,0) -- (12,-3);
       \draw [dashed] (-2,-3) -- (12,-3);
      \draw [red, dashed, thick] (-1.7,1.5) -- (-1.7,-1.5);
      \draw [red, dashed, thick] (1.7,1.5) -- (1.7,-1.5);
       \draw [red, dashed, thick] (-1.7,1.5) -- (1.7,1.5);
     \draw [red, dashed, thick] (-1.7,-1.5) -- (1.7,-1.5);
        \draw [red, dashed, thick] (-1.7+4,1.5) -- (-1.7+4,-1.5);
      \draw [red, dashed, thick] (1.7+4,1.5) -- (1.7+4,-1.5);
       \draw [red, dashed, thick] (-1.7+4,1.5) -- (1.7+4,1.5);
     \draw [red, dashed, thick] (-1.7+4,-1.5) -- (1.7+4,-1.5);
        \draw [red, dashed, thick] (-1.7+10,1.5) -- (-1.7+10,-1.5);
      \draw [red, dashed, thick] (1.7+10,1.5) -- (1.7+10,-1.5);
       \draw [red, dashed, thick] (-1.7+10,1.5) -- (1.7+10,1.5);
     \draw [red, dashed, thick] (-1.7+10,-1.5) -- (1.7+10,-1.5);      
     \draw [blue, dashed, thick] (0.3,2) -- (-0.3+4,2);
      \draw [blue, dashed, thick] (0.3,-2) -- (-0.3+4,-2);
      \draw [blue, dashed, thick] (0.3,2) -- (0.3,-2);
       \draw [blue, dashed, thick] (-0.3+4,2) -- (-0.3+4,-2);
          \draw [blue, dashed, thick] (0.3+4,2) -- (-0.3+4+4,2);
      \draw [blue, dashed, thick] (0.3+4,-2) -- (-0.3+4+4,-2);
      \draw [blue, dashed, thick] (0.3+4,2) -- (0.3+4,-2);
       \draw [blue, dashed, thick] (-0.3+4+4,2) -- (-0.3+4+4,-2);
       \draw [blue, dashed, thick] (+0.3+10,2) -- (0.3+10,-2);
        \draw [blue, dashed, thick] (+0.3+10,2) -- (0.3+10+1.7,2);           \draw [blue, dashed, thick] (+0.3+10,-2) -- (0.3+10+1.7,-2);  
              \draw [blue, dashed, thick] (-0.3,2) -- (-0.3-2,2);
        \draw [blue, dashed, thick] (-0.3,2) -- (-0.3,-2);          
        \draw [blue, dashed, thick] (-0.3,-2) -- (-0.3-2,-2); 
	\end{tikzpicture}
	}}~.
\end{aligned}
\ee
Interestingly, in this replica partition function, two competing distinct types of Gaussian contractions contribute at leading order \cite{Bao:2025plr, Geng:2025efs}, as indicated by the red and blue boxes, respectively. Performing the Gaussian average in the red box and simplifying the result using \eqref{fusionkernel}, we obtain

\be
\overline{Z_{AA^*,n,\text{red}}}=\left|\int_0^\infty d P_i \rho_0(P_i) \mathcal{F}_{\text{red},n}(\mathcal{M}_n,P_i)\right|^2\,,
\ee
where $\mathcal{F}_{\text{red},n}(\mathcal{M}_n,P_i)$ is the conformal block:

\be\label{eq:F1n}
\begin{aligned}
\mathcal{F}_{\text{red},n}(\mathcal{M}_{n},P_i)=
\vcenter{\hbox{
	\begin{tikzpicture}[scale=0.75]
	\draw[thick] (0,1) circle (1);
	\draw[thick] (0,-1+1) -- (0,-2+1);
	\draw[thick] (0,-2+1) -- (-1,-2+1);
	\draw[thick] (0,-2+1) -- (1,-2+1);
	\node[left] at (0,-3/2+1) {$\mathbb{1}$};
	\node[left] at (-1.2,0+1) {$\mathbb{1}'$};
    	\draw[thick] (0+3,1) circle (1);
	\draw[thick] (0+3,-1+1) -- (0+3,-2+1);
	\draw[thick] (0+3,-2+1) -- (-2+3,-2+1);
	\draw[thick] (0+3,-2+1) -- (2+3,-2+1);
	\node[left] at (0+3,-3/2+1) {$\mathbb{1}$};
	\node[left] at (-1.2+3.2,0+1) {$\mathbb{1}'$};
        \node[above] at (5.5,0-0.3) {$...$};
            	\draw[thick] (0+3+3+2,1) circle (1);
	\draw[thick] (0+3+3+2,-1+1) -- (0+3+3+2,-2+1);
	\draw[thick] (0+3+3+2,-2+1) -- (-2+3+3+2,-2+1);
	\draw[thick] (0+3+3+2,-2+1) -- (2+3+3+1,-2+1);
	\node[left] at (0+3+3+2,-3/2+1) {$\mathbb{1}$};
	\node[left] at (-1.2+3.2+3+2,0+1) {$\mathbb{1}'$};
    \draw [dashed] (-1,-1) -- (-1,-2);
      \draw [dashed] (9,-1) -- (9,-2);
       \draw [dashed] (-1,-2) -- (9,-2);
       \node[left] at (4.5,-3/2) {$i, n \beta_i$};
	\end{tikzpicture}
	}}  ~.
\end{aligned}
\ee
For the blue box, we get
\be
\overline{Z_{AA^*,n,\text{blue}}}=\left|\int_0^\infty d P_k d P_j \rho_0(P_k)  \rho_0(P_j) \mathcal{F}_{\text{blue},n}(\mathcal{M}'_n,P_k,P_j) \right|^2\,,
\ee
where
\be
\begin{aligned}
&\mathcal{F}_{\text{blue},n}(\mathcal{M}'_n,P_k,P_j)=\\
&
\qquad \qquad \qquad 
\begin{tikzpicture}[scale=0.75][baseline=(current bounding box.north)]
 \draw[thick] (-3,1) -- (6,1);
  \draw[thick] (8,1) -- (11,1);
    \node[above] at (4,1.1) {$k, n \beta_{k}$};
     \draw[thick] (10,1) -- (10,-1);
     \node[right] at (10,0) {$\mathbb{1}$};
     \draw[thick] (-3,-1) -- (6,-1);
      \draw[thick] (8,-1) -- (11,-1);
    \node[below] at (4,-1.1) {$j, n\beta_j$};
      \node[above] at (7,-0.3) {$...$};
           \draw[thick] (0,1) -- (0,-1);
     \node[right] at (10,0) {$\mathbb{1}$};
       \node[right] at (0,0) {$\mathbb{1}$};
          \draw[thick] (4,1) -- (4,-1);
     \node[right] at (4,0) {$\mathbb{1}$};
       \draw [dashed] (-3,1) -- (-3,2);
      \draw [dashed] (11,1) -- (11,2);
       \draw [dashed] (-3,2) -- (11,2);
      \draw [dashed] (-3,-1) -- (-3,-2);
      \draw [dashed] (11,-1) -- (11,-2);
       \draw [dashed] (-3,-2) -- (11,-2);
\end{tikzpicture}
~.
\end{aligned}
\ee

Following the steps outlined in Sec.~\ref{EEsec}, and using the correspondence with Liouville theory under ZZ boundary conditions, we obtain entropies of $\tfrac{2l_3}{4G_N}$ and $\tfrac{2(l_1+l_2)}{4G_N}$. Further details can be found in \cite{Bao:2025plr}. 

The reflected entropy comes from the dominant contribution in the averaged replica partition functions, which leads to
\be
\text{RE}(A:B)=S(AA^*)=\text{min} \left\{\frac{2l_3}{4G_N}, \frac{2(l_1+l_2)}{4G_N}\right\}\ ,
\ee
and indeed reproduces $\text{EW}(A:B)$.

In fact, there is a guiding principle for identifying all contraction patterns that yield potential leading contributions. Once Gaussian moments are used, we need to find patterns where the replica partition function graphs is decomposable into $Z_2$-symmetric pieces, within each of which the averaging can be applied. As first noted in \cite{Bao:2025plr}, this condition corresponds precisely to the choice of surfaces satisfying the \textit{homology constraint}, since such surfaces split the original graph representing the state into two disconnected components, and break the replica partition function into disconnected $Z_2$-symmetric pieces, further leading to $Z_n$ symmetric configurations. Thus, the averaging patterns giving leading contributions to the entropy computation correspond one-to-one with the homologous entanglement wedge cross sections.

\setlength{\parskip}{1em}

With the three representative examples in hand, we can now see that the mechanism is universal. When we focus on a connected entanglement wedge, the complementary region in AdS$_3$/CFT$_2$ necessarily has a fully disconnected entanglement wedge. At leading order in $c$, this implies that the reduced density matrix of the complementary region factorizes, with each subregion governed by a local modular Hamiltonian equal to the corresponding BCFT Hamiltonian. Evolving backward with the $1/4$ power, this construction consistently reproduces the CFT dual configuration of \cite{Dutta:2019gen}, corresponding to cutting open along the RT surfaces and gluing to the orientation-reversed copy.
\setlength{\parskip}{0em}

As explained above, all candidate choices that give leading contributions correspond precisely to the homologous choices of entanglement wedge cross sections. Incorporating all such possibilities, together with the properties of the universal OPE coefficients and the correspondence to Liouville theory with ZZ boundary conditions, we thus establish the general correspondence between the reflected entropy and entanglement wedge cross section.

\section{Multipartite $\text{RE}=\text{EW}$ from BCFT Random Tensor Networks}\label{multipartitesection}
Much of the existing work on the connection between quantum information theory and holography has focused on bipartite entanglement properties. However, richer information about holographic quantum states is encoded in their multipartite entanglement structures. Therefore, understanding multipartite entanglement and its holographic dual is an important step toward demystifying the emergence of spacetime.

The multipartite generalization of the reflected entropy \cite{Bao:2019zqc, Chu:2019etd} and the dual entanglement wedge cross section \cite{Umemoto:2018jpc} is such an example. In this section, we will show that the construction in earlier sections can be readily generalized to the multipartite case.

The multipartite entanglement wedge cross section is generalized in a simple way. For example, the tripartite one \cite{Umemoto:2018jpc} is defined as
\be
\label{EW3}
\text{EW}_3(A:B:C)\equiv \underset{\partial l=\emptyset}{\min}\Big\{ \text{Area}(l_1)+\text{Area}(l_2)+\text{Area}(l_3)\Big\}\ ,
\ee
where $l = l_1 \cup l_2 \cup l_3$ denotes a collection of surfaces in the entanglement wedge of $ABC$ that are homologous to $A$, $B$, and $C$, respectively. The minimization is over all such collections of surfaces that together form a closed cycle (i.e., $\partial l = \emptyset$), as illustrated in Fig.~\ref{EW}.

The basic idea underlying the derivation of the multipartite $\text{RE}=\text{EW}$ parallels the bipartite case discussed above. In that setting, the canonical purification turns the interval representing the entanglement wedge cross section into a curve on the doubled geometry. The minimal geodesic length of the curve arises from the saddle-point computation in the Liouville CFT describing 2D hyperbolic solutions with ZZ boundary conditions. Because the minimal curve is symmetric across the RT surface where the two copies are glued, its length is equal to twice the entanglement wedge cross section. 

Accordingly, an analogous strategy to the bipartite case can be employed. By applying successive canonical purifications, we replicate the system and map the total length of the three intervals $l_i$ (\ref{EW3}) to the circumference of a circle in the replicated geometry. The minimal value is once again fixed by Liouville theory via the ensemble average over OPE coefficients, yielding a multiple of \eqref{EW3}.

We now turn to the BCFT framework described above to derive this result from the CFT side. As an explicit example, we present the tripartite configuration shown in Fig.~\ref{EW}; the extension to more general situations is straightforward, and the bulk counterpart is discussed in \cite{Bao:2019zqc, Chu:2019etd}. With the bulk–CFT connection in hand, the corresponding dual CFT operation on quantum states can be written down directly.

\begin{figure}
    \centering
\includegraphics[width=0.6\linewidth]{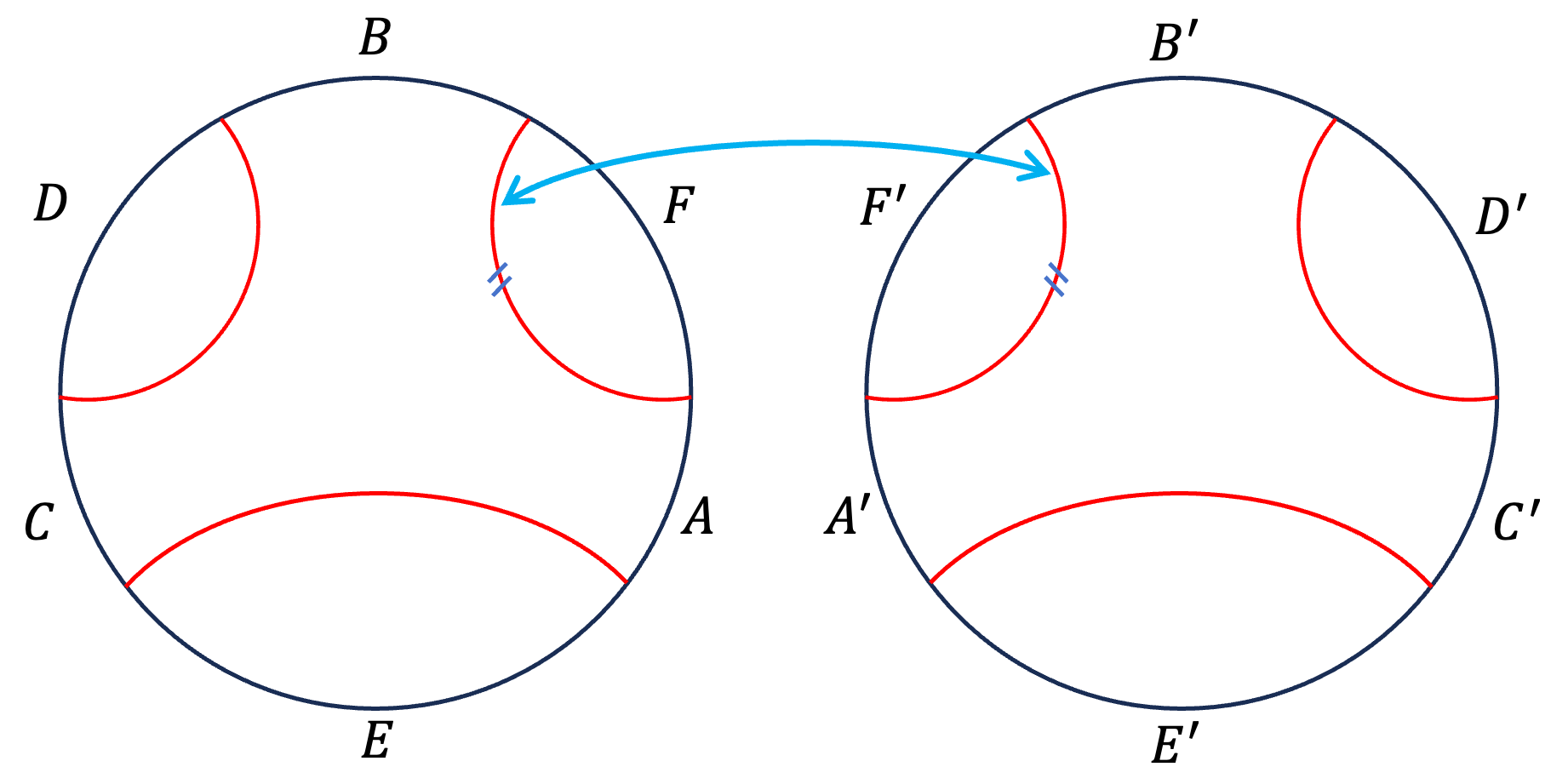}
    \caption{\small{Bulk dual (or state-preparation manifold) of $|\Psi^{(1)}\rangle$ (\ref{Psi1}).}}
    \label{purification1}
\end{figure}

First, let us follow the proposal of \cite{Chu:2019etd} to construct the final pure state for defining the multipartite reflected entropy in several steps. Starting from the initial pure state $\ket{\Psi}$ for $ABCDEF$, the first step is:
\be
\label{Psi1}
\ket{\Psi}\to \ket{\Psi^{(1)}}=\ket{\sqrt{\rho_{ABCDE}}}\ .
\ee
Here, $\rho_{ABCDE}=\tr_F \ket{\Psi} \bra{\Psi}$ denotes the reduced density matrix for the subsystems $ABCDE$. Similar to the bipartite case, the bulk dual of the canonical purification in (\ref{Psi1}) corresponds to gluing two copies of the entanglement wedges for $ABCDE$ along the RT surface, as shown in Fig.~\ref{purification1}. 

\begin{figure}
    \centering
\includegraphics[width=0.6\linewidth]{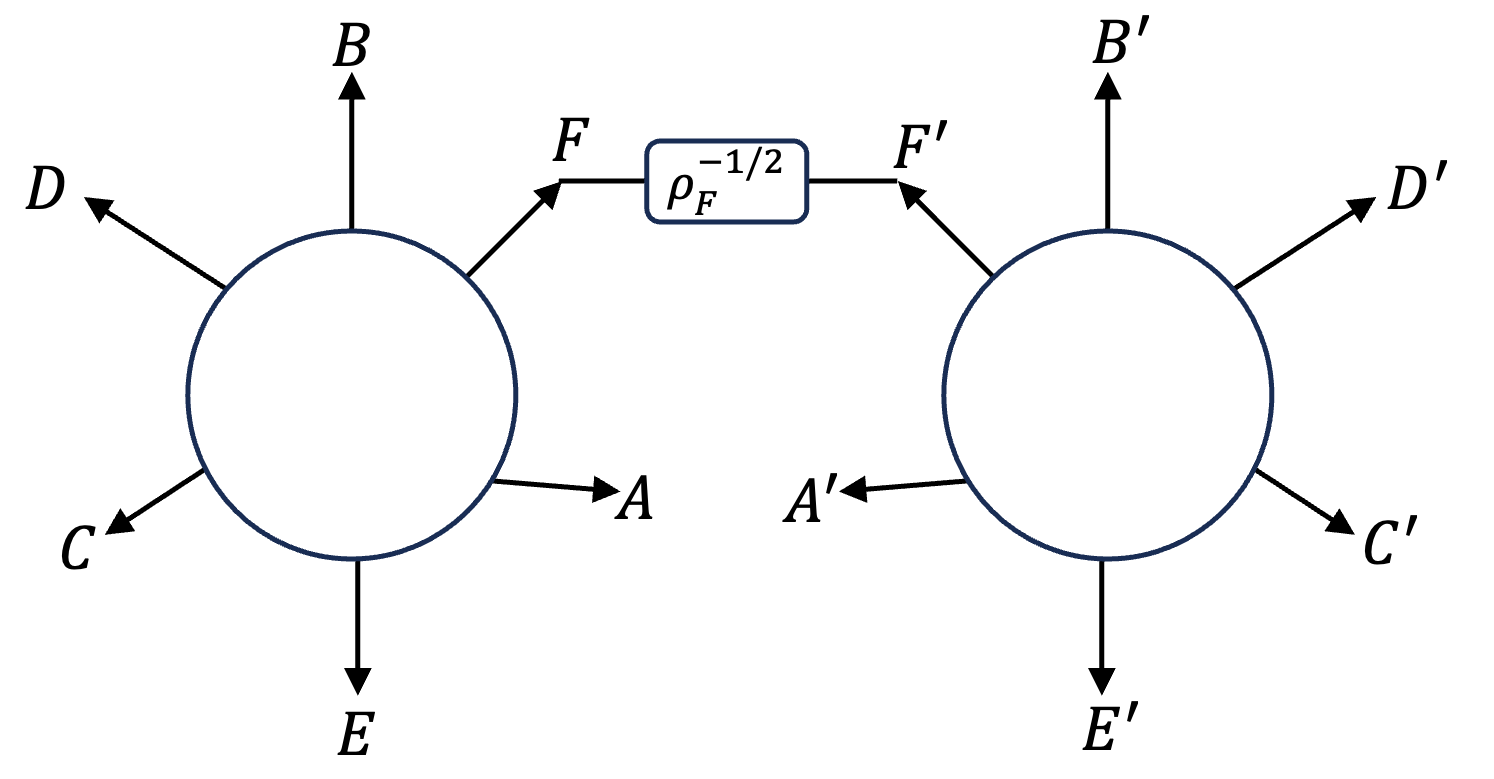}
    \caption{\small{Illustration of the dual BCFT representation realizing the canonical purification (\ref{Psi1}).}}
    \label{purification1CFT}
\end{figure}

On the CFT side, this procedure can be implemented explicitly using BCFT techniques. We introduce small conformal boundaries as regulators so that the path integral is naturally expressed in terms of BCFT, and represent the resulting state schematically in terms of OPE blocks, as shown in Fig.~\ref{purification1CFT}. Each bulb-like diagram denotes the BCFT path integral preparing the state $|\Psi\rangle$ on the six BCFT Hilbert spaces. In principle, this can be decomposed into trivalent vertices corresponding to BCFT operator product expansions, giving rise to the tensor network representation, though for simplicity we leave it schematic. Finally, we glue this CFT path integral to its CPT conjugate along $\mathcal{H}_F$, with an insertion of $\rho_F^{-1/2}$. The outcome is the purified state $\ket{\Psi^{(1)}}$ on ten intervals, where each of the original six intervals is replicated once (e.g. $A\to AA'$) and two of them (namely $F$ and $F'$) are glued together.

\begin{figure}
    \centering
\includegraphics[width=1\linewidth]{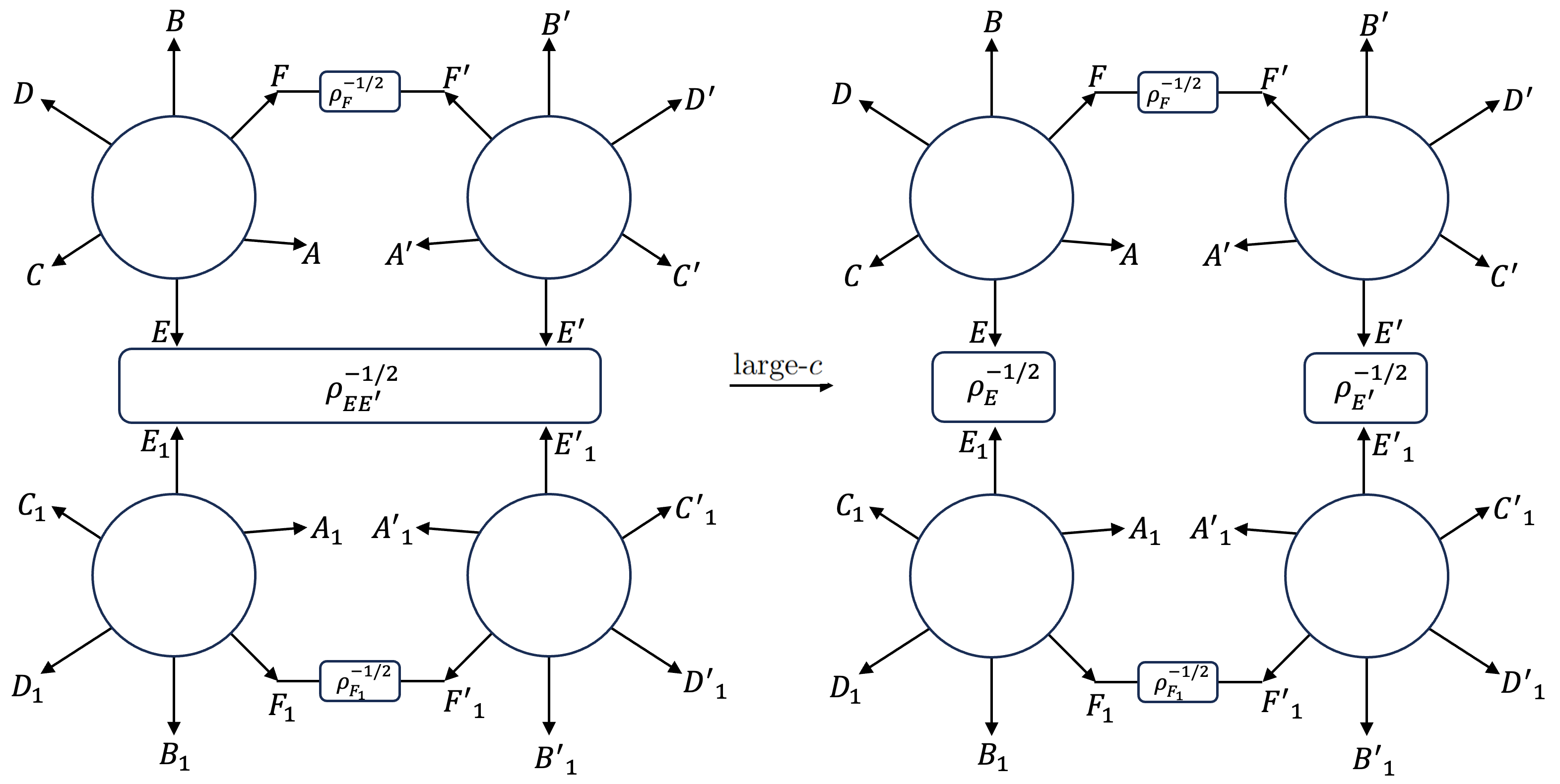}
    \caption{\small{The second step in the first type of construction of the multipartite reflected entropy is to perform the canonical purification obtained by inserting $\rho_{E,E'}^{-1/2}$. In the large-$c$ limit, this reduces to $\rho_{E}^{-1/2} \otimes \rho_{E'}^{-1/2}$.}}
    \label{purification2CFT}
\end{figure}

\begin{figure}
    \centering
\includegraphics[width=1\linewidth]{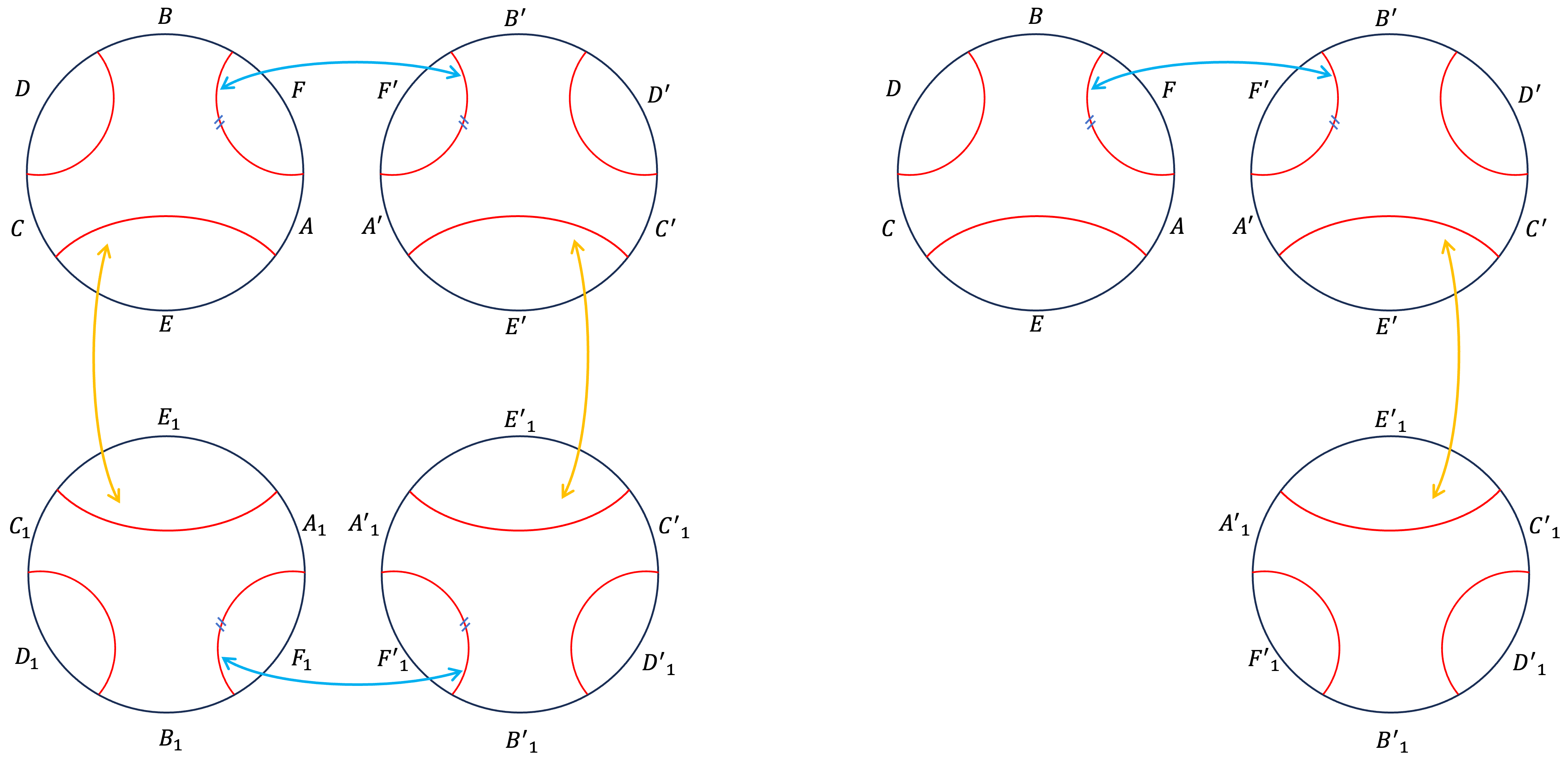}
    \caption{\small{Bulk duals (or state-preparation manifolds) of different proposals for the second step of constructing the tripartite reflected entropy, (\ref{Psi2}).}}
    \label{twoproposals}
\end{figure}

Continuing from (\ref{Psi1}), the second step is to produce the state,
\be
\label{Psi2}
\ket{\Psi^{(1)}} \to \ket{\Psi^{(2)}}= \left|\sqrt{\tr_{E,E'} \ket{\Psi^{(1)}} \bra{\Psi^{(1)}}}\right\rangle\ .
\ee
For its bulk dual, we glue the diagram in Fig.~\ref{purification1} to its CPT conjugate along the RT surfaces of $E$ and $E'$, as illustrated in the left panel of Fig.~\ref{twoproposals}. On the CFT side this canonical purification is constructed by gluing the BCFT state $\ket{\Psi^{(1)}}$ to its CPT conjugate along $\mathcal{H}_E \otimes \mathcal{H}_{E'} $, with an insertion of $\rho_{E,E'}^{-1/2}$. This object is the density matrix for two intervals and, in general, does not admit a simple local expression. In the large-$c$ limit, however, it factorizes as $\rho_{E}^{-1/2} \otimes \rho_{E'}^{-1/2}$, analogous to \eqref{factorize}. Using these factorized local single-interval density matrices, as depicted in the right panel of Fig.~\ref{purification2CFT}, we obtain a local path integral representation of the state (\ref{Psi2}), which now lives in the Hilbert space of sixteen intervals (this time, we will label a copied interval with a subscript ``1'', e.g. $A_1$).

The third step applies the same procedure to the copies of $D$:
\be
\label{Psi3}
\ket{\Psi^{(2)}} \to \ket{\Psi^{(3)}} = \ket{\sqrt{\tr_{D,D',D_1,D_1'} \ket{\Psi_2}\bra{\Psi_2}}} .
\ee
This step is analogous to the two cases discussed above, so we will not elaborate further. After the three steps, the original manifold is replicated $2^3=8$ times, giving a quantum state defined on $3 \times 8 = 24$ intervals, which can be organized into $8$ copies of the Hilbert spaces associated with $ABC$. If we shrink the remaining conformal boundary regulators, the intervals will merge to produce a multi-boundary wormhole geometry with six circular boundaries \cite{Balasubramanian:2014hda, Chu:2019etd, Bao:2019zqc}. At this stage, the resulting quantum state provides a proper purification: tracing over the auxiliary Hilbert spaces reproduces the reduced density matrices of the original state $\ket{\Psi}$, such as $\rho_A$, $\rho_{AB}$, $\rho_{ABC}$, and so on. This follows from the fact that each step in the construction is itself a purification. 
\begin{figure}
    \centering
\includegraphics[width=0.6\linewidth]{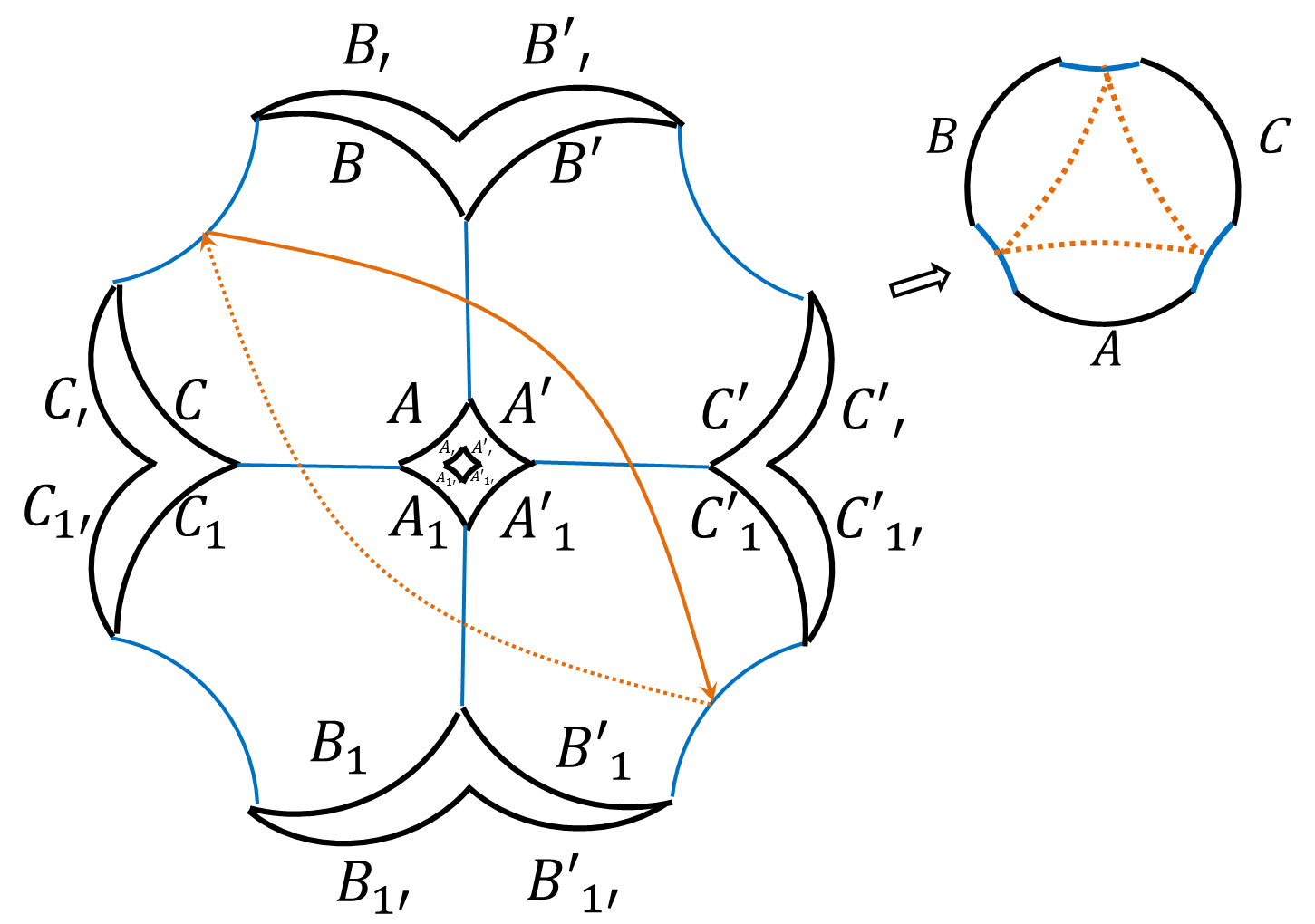}
    \caption{\small{Holographic dual of the tripartite reflected entropy $\text{RE}_3(A:B:C)$. The RT surface for the entanglement entropy involved in the definition \eqref{RE3} is given by the orange loop. This minimal area is twice the area of the tripartite entanglement wedge cross-section $\text{EW}_3(A:B:C)$, which is represented by the dashed orange triangle.}}
    \label{ps}
\end{figure}
Now, labeling the copied Hilbert spaces from the final step in Eq. (\ref{Psi3}) with additional primes in the subscripts, we define the tripartite reflected entropy as
\be
\label{RE3}
\text{RE}_3(A:B:C)\equiv\text{EE}(AA'A_1 A_1' B_1B'_1B_{1'}B'_{1'}CC_{'}C_1C_{1'}:A_{'}{A'}_{'}A_{1'} A_{1'}'BB'B_{'}B'_{'}C'C'_{'}C'_1C'_{1'})\ .
\ee
This expression appears complicated at a first glance. However, the dual gravitational picture, shown in Fig.~\ref{ps}, reveals that the entanglement entropy in (\ref{RE3}) is computed by the RT surface whose area is twice the tripartite entanglement wedge cross section \eqref{EW3}.\footnote{This doubling follows from the $\mathbb{Z}_2$ symmetry inherent in the construction.} This observation led to the tripartite $\text{RE}=\text{EW}$ proposal~\cite{Chu:2019etd}, which states
\be
\label{RE3=2EW3}
\text{RE}_3(A:B:C)=\frac{2\text{EW}_3(A:B:C)}{4G_N}\ .
\ee

Using averaging over OPE coefficients techniques \cite{Bao:2025plr, Geng:2025efs}, analogous to those employed in the bipartite case, we can demonstrate that the relation \eqref{RE3=2EW3} is indeed true. The key point is that, at leading order in $c$, the multi-interval reduced density matrix for disconnected regions—used to construct the canonical purification—factorizes into a product of single-interval contributions, each admitting a simple local expression. This factorization yields precisely the dual bulk prescription proposed in \cite{Chu:2019etd}.

Next we briefly outline the second type of construction proposed in \cite{Cheng:2023kxh}. The spirit of this construction is similar to the previous one, but the details differ. Specifically, after performing the first step of gluing $F$ and $F'$ in Fig.~\ref{purification1} and Fig.~\ref{purification1CFT} to obtain $\ket{\Psi_1}$, we do not replicate the entire diagram that prepares the state—which would otherwise produce four copies of the original disk—but instead introduce only a single additional disk, resulting in three disks in total. In this case, we glue only the second and third copies along the $E'$ edges. See Fig.~\ref{twoproposals} for an illustration on the two different constructions.

We continue this procedure by introducing one additional disk at a time, with the goal of forming a closed loop through successive gluings of the RT surfaces. In the present example, after the previous step it becomes clear that we cannot glue the ${D'}_1$ surface to the original $D$, since the first and third manifolds are not CPT conjugate. Thus, three copies are not sufficient, and more generally this obstruction arises for any odd number $n$. The resolution proposed in \cite{Bao:2019zqc} is to introduce additional $n$ copies and close the loop by gluing the $2n$'th copy to the first one along, e.g. in the $n=3$ case, their common $D$ surfaces. This completes the construction and yields a total of $2n$ copies. We then apply canonical purification to glue all the remaining RT surfaces to their partners on the conjugate manifolds, producing a ring-like geometry. Applying the RT formula in this setup—which can again be justified by averaging over large-$c$ OPE coefficients in the replica partition function—we find that the minimal surface on this manifold spans $4n$ copies. By the same reasoning as in the first proposal, this corresponds to an integer multiple of the entanglement wedge cross section of the original manifold. In contrast with the earlier construction, here each step is not manifestly a canonical purification, but the BCFT representation makes the underlying CFT operation in terms of reduced density matrices transparent.\footnote{For example, to perform the gluing as in the right panel of Fig.~\ref{twoproposals}, one can insert a $\rho_{E'}^{-1/2}$ in the path integral.}

\section{Conclusions and Discussions
} \label{conclusionsec}

In this work, we have established a 2D CFT-based derivation of the correspondence between reflected entropies and entanglement wedge cross sections, both in the bipartite \cite{Dutta:2019gen} and multipartite \cite{Bao:2019zqc, Chu:2019etd} case for the vacuum state. The central idea is to triangulate the state-preparation manifold in the CFT path integral, employ the framework of BCFT tensor networks, and then use universal coarse-grained (B)CFT data in the large-$c$ limit to capture the emergence of 3D hyperbolic geometry \cite{Geng:2025efs}.

Within the BCFT tensor network framework, we first explicitly constructed the canonical purification. Next, combining the factorization of reduced density matrices for disconnected subregions in the large-$c$ limit with the explicit relation between the single-interval modular Hamiltonian and the BCFT Hamiltonian \cite{Ohmori:2014eia, Cardy:2016fqc}, we showed that canonical purifications can be realized directly through CFT cutting-and-gluing operations, which also admits a Euclidean path-integral preparation. 

For the canonical purification state we constructed, we further demonstrated that, upon Gaussian averaging of OPE coefficients in the replica partition functions, the dual hyperbolic bulk geometries naturally emerge via the connection to Liouville theory with ZZ boundary conditions \cite{Chua:2023ios, Bao:2025plr, Geng:2025efs}, and the reflected entropy derived from the averaged replica partition functions precisely matches the areas of entanglement wedge cross sections. 

This work extends our previous derivation of the RT formula based on BCFT random tensor networks \cite{Geng:2025efs} to the case of the ``$\text{RE}=\text{EW}$'' duality, and provides a further microscopic and exact CFT realization of intuitions previously accessible only through toy models or by assuming the AdS/CFT dictionary \cite{Dutta:2019gen, Akers:2021pvd, Akers:2022zxr, Marolf:2019zoo}, now reformulated rigorously in terms of BCFT tensor networks built purely from intrinsic CFT data. Our construction also illustrates the explicit CFT dual operations corresponding to existing proofs based on the gravitational path integral. 

We leave several important and interesting directions for future work. The first is to understand corrections beyond the leading order in large-$c$ limit. As we have seen in the disconnected example for reflected entropy, the construction of canonical purification in general requires density matrices of multiple disjoint regions. Once subleading $1/c$ corrections are included, the factorization property no longer holds, and it will be important to understand how to systematically match the two sides of the holographic correspondence in this regime. A second, related question is to consider more general quantum states, for example by including matter sources in the CFT state-preparation manifold. In this case, subleading $1/c$ corrections naturally require the use of ``quantum extremal surfaces'' \cite{Engelhardt:2014gca} in place of purely geometric minimal surfaces. The CPT gluing in such setups was studied in \cite{Engelhardt:2017aux, Engelhardt:2018kcs, Dutta:2019gen, Parrikar:2023lfr}, and generically leads to discontinuities and shockwaves in the gravitational dual. It would be very interesting to develop a deeper understanding of these phenomena from the BCFT tensor network perspective.

Next, since we have established yet another correspondence between a pair of quantum information quantities and geometrical objects, it is natural to expect that the BCFT tensor network framework is even more powerful, with broader applicability to uncover the mechanisms linking quantum information and holography, and illuminating the general structure of the duality. A wide range of holographic proposals have been inspired or verified by tensor-network models \cite{Miyaji:2015yva, Brown:2015lvg, Brown:2019rox, Akers:2022qdl, Bousso:2022hlz, Kaya:2025vof, Antonini:2025ioh}. It is natural to ask whether BCFT tensor networks can provide a precise CFT-based foundation for many of these conjectures? For instance, can this framework be employed to establish the correspondence between the entanglement of purification and the entanglement wedge cross section \cite{Takayanagi:2017knl, Nguyen:2017yqw}?

Furthermore, our construction of the bipartite canonical purification in the CFT is quite general and applies to arbitrary quantum systems. In the multipartite case, however, it is evident from Sec.~\ref{multipartitesection} that the first approach is not symmetric with respect to the three complement regions $DEF$, once we go beyond the large-$c$ limit where the density matrices factorize. Moreover, both of our constructions in Sec.~\ref{multipartitesection} rely on a specific purification of $\rho_{ABC}$ into $ABCDEF$. In contrast, for the bipartite canonical purification one can show that the result is independent of the choice of purification and is instead an intrinsic property of the density matrix. The current multipartite definitions, however, do not seem to possess this property. 

This naturally raises the question: is there a construction that yields a quantity both symmetric with respect to the purifiers and independent of their choice, valid for general quantum systems beyond large-$c$ holographic CFTs? Given our experience in Sec.~\ref{discosection}, it is tempting to conjecture that such a construction might be realized through a specific combination of gluings, with insertions of $\rho_{D}, \rho_{E}, \rho_{F}, \rho_{DE}, \rho_{EF}, \rho_{DF}$, and $\rho_{DEF}$. Developing such a quantity could provide a genuinely symmetric and purification-independent measure of multipartite correlations for generic quantum systems.

Finally, similar to the case of the RT formula in AdS$_3$/CFT$_2$ \cite{Hartman:2013mia, Faulkner:2013yia}, there exist other proofs of the $\text{RE}=\text{EW}$ proposal based on twist operators and the large-$c$ limit of conformal blocks \cite{Dutta:2019gen, Kusuki:2019evw}. In those approaches, the twist operators encode the gluing of multiple CFT copies, rather than the BCFT construction on a higher-topology manifold as in our case. It would be interesting to clarify the precise relation between this viewpoint and ours. In addition, in the standard treatment of reflected entropy, a double replica trick is employed and a particular order of limits is required to obtain the correct answer \cite{Kusuki:2019evw, Akers:2021pvd}. In contrast, our method constructs the canonical purification first and then performs ensemble averaging of OPE coefficients in the crossed-channel replica partition function. Understanding the precise connection between this approach and the order-of-limits issue in our framework would be very valuable. Moreover, one could also attempt to combine the double replica trick directly with ensemble averaging, followed by analytic continuation, to compute the reflected entropy. In that case, what averaging pattern is required to capture the dominant contribution? Can we see phase transitions explicitly in the double replica trick, similar to \cite{Akers:2021pvd}? What are the roles of the non-replica-symmetric averaging patterns?

\section*{Acknowledgments}

We thank Thomas Faulkner, Keiichiro Furuya, Daniel Jafferis, Hao Geng, Ma-Ke Yuan and Yang Zhou for fruitful discussions. We thank Hao Geng, Song He, Yuya Kusuki, Onkar Parrikar, Pratik Rath, and Martin Sasieta for helpful comments and suggestions on the draft. YJ and JC thank the Kadanoff Center for Theoretical Physics at the University of Chicago for hosting YJ’s talk, which led to the initiation of their collaboration. YJ and NB acknowledge support from the U.S Department of Energy ASCR EXPRESS grant, Novel Quantum Algorithms from Fast Classical Transforms, and Northeastern University. JM is supported by Northeastern University. JC is supported in part by DOE grant 5-29073.

\bibliographystyle{JHEP}
\bibliography{main}
\end{document}